\documentclass[usenatbib,usegraphicx]{mn2e}

\defcitealias{zsfw97-holmberg}{ZSFW}
\defcitealias{brainerd05}{B05}
\defcitealias{sl04}{SL04}
\defcitealias{azpk06}{AZPK}
\defcitealias{appz07}{APPZ}
\defcitealias{RC3}{RC3}
\defcitealias{yang-etal04}{Y04}
\defcitealias{yang-etal06}{Y06}
\defcitealias{ab07}{AB07}

\newcommand{\rout}{{\ensuremath{R_{\mathrm{outer}}}}}
\newcommand{\magout}{{\ensuremath{m_{\mathrm{outer}}}}}
\newcommand{\vout}{{\ensuremath{v_{\mathrm{outer}}}}}
\newcommand{\rin}{{\ensuremath{R_{\mathrm{inner}}}}}
\newcommand{\magin}{{\ensuremath{m_{\mathrm{inner}}}}}
\newcommand{\vin}{{\ensuremath{v_{\mathrm{inner}}}}}
\newcommand{\rsat}{{\ensuremath{R_{\mathrm{sat}}}}}
\newcommand{\magsat}{{\ensuremath{m_{\mathrm{sat}}}}}
\newcommand{\vsat}{{\ensuremath{v_{\mathrm{sat}}}}}
\newcommand{\Nviol}{{\ensuremath{N_{\mathrm{viol}}}}}
\newcommand{\fsatlum}{{\ensuremath{f_{\mathrm{sat}}}}}
\newcommand{\Nsatmax}{{\ensuremath{N_{\mathrm{satmax}}}}}
\newcommand{\kms}{{\ensuremath{\mathrm{km~s^{-1}}}}}

\newcommand{\hikpc}{{\ensuremath{h^{-1}\,\mathrm{kpc}}}}
\newcommand{\hiMpc}{{\ensuremath{h^{-1}\,\mathrm{Mpc}}}}
\newcommand{\fprim}{{\ensuremath{f_{\mathrm{prim}}}}}

\newcommand{\grkc}{{\ensuremath{{}^{0.1}(g-r)}}}
\newcommand{\fnondom}{{\ensuremath{f_{\mathrm{non-dom}}}}}
\newcommand{\cmdfo}{{\ensuremath{\mathrm{CMD}^\mathrm{F}}}}
\newcommand{\cnorm}{{\ensuremath{C_{\mathrm{norm}}}}}

\title[Satellite Galaxy Anisotropy]{The Anisotropic Distribution of
  Satellite Galaxies}

\author[J. Bailin et al.]{Jeremy Bailin$^{1,2}$, Chris Power$^{2,3}$,
  Peder Norberg$^{4,5}$, Dennis Zaritsky$^{6}$, Brad K. Gibson$^{7}$\\
  $^{1}$Department of Physics \&\ Astronomy,
  McMaster University, 1280 Main St. W, Hamilton, ON, L8S 4M1, Canada;
  bailinj@mcmaster.ca\\
  $^{2}$Centre for Astrophysics and Supercomputing,
  Swinburne University of Technology, Mail H39, PO Box 218, Hawthorn, 
  Victoria,\\
  3122, Australia\\
  $^{3}$Theoretical Astrophysics Group, Department of Physics and Astronomy,
  University of Leicester, Leicester, LE1 7RH, UK; cbp1@le.ac.uk\\
  $^{4}$SUPA\thanks{The Scottish Universities Physics Alliance}, Institute for
  Astronomy, University of Edinburgh, Royal Observatory, Blackford Hill,
  Edinburgh, EH9 3HJ, UK; iprn@roe.ac.uk\\
  $^{5}$ETHZ Institut f\"ur Astronomie, HPF G3.1, ETH H\"onggerberg,
  CH-8093 Z\"urich, Switzerland\\
  $^{6}$Steward Observatory, University of Arizona, 933 North Cherry Ave,
  Tucson, AZ, 85721, USA; dzaritsky@as.arizona.edu\\
  $^{7}$Centre for Astrophysics, University of Central Lancashire, Preston,
  PR1 2HE, UK; bkgibson@uclan.ac.uk}

\date{Accepted 2008 August 10. Received 2008 June 30; in original form 2007 June 8}

\begin{document}
  
  \maketitle
  
  \begin{abstract}

We identify satellites of isolated galaxies in the
Sloan Digital Sky Survey and examine their angular distribution.
    Using mock catalogues generated 
    from cosmological $N$-body simulations, we demonstrate that
    the selection criteria used to select isolated galaxies and
    their satellites
    in large galaxy redshift surveys
must be very strict in order to correctly
    identify systems in which the primary galaxy dominates its
    environment. We demonstrate that the criteria used in many previous studies
    instead select predominantly group members.
We refine a set of selection criteria for which the group contamination
is estimated to be less than $7\%$ and present a catalogue of the
resulting sample.

The angular distribution of satellites about their host is biased towards the
major axes for spheroidal galaxies and probably also
for red disc galaxies (the ``intermediate'' class of
Bailin \&\ Harris), but is isotropic for
blue disc galaxies,
i.e.~it is the colour of the host that determines the
distribution of its satellites rather than its morphology.
The similar anisotropy measured in this
study as in studies that were dominated by groups implies that
group-specific processes are not responsible for the angular
distribution.
Satellites that are most likely to have been recently accreted,
late-type galaxies at large projected radii, show a tendancy
to lie along the same axis as the surrounding large scale structure.
The orientations of isolated early and intermediate-type galaxies also
align with the surrounding large scale structures.

We discuss the origin of the anisotropic satellite
    distribution and we consider the implications of our results,
    critically assessing the respective roles played by
    the orientation of the visible galaxy within its dark matter halo;
    anisotropic accretion of satellites 
    from the larger scale environment; and the biased nature of
    satellites as tracers of the underlying dark matter subhalo population.
  \end{abstract}

  \begin{keywords}
    galaxies: haloes ---
    dark matter ---
    galaxies: structure ---
    galaxies: dwarf ---
    galaxies: clusters: general ---
    galaxies: formation
  \end{keywords}
  
\section{Introduction}

The spatial distribution of satellites around isolated galaxies can 
provide important insights into the mass distribution in and around these
galaxies. If dynamical effects could be neglected and if we could
assume that satellite galaxies inhabit an unbiased set of dark matter
subhaloes, then we would expect satellites to cluster preferentially
along the major axis of the host dark matter halo of the primary galaxy, as
in galaxy cluster mass haloes \citep{knebe-etal04}. We could therefore 
determine the orientation of the parent galaxy within its dark matter
halo by using the spatial distribution of its satellite galaxies.

However, there is good reason to believe that dynamical effects will
play an important role in determining the spatial distribution of 
satellites. For example, \citet{hartwick00} has argued that the imprint 
of anisotropic infall of satellites along filaments is evident in the 
orbits of recently accreted systems, and \citet{pkb02} have shown
that the inclination of a satellite's orbit can determine its
dynamical response to the disc or spheroid of its parent galaxy. 
There is also good reason to suspect that the relationship between
satellites and the underlying dark matter subhalo population may not be
straightforward, as has been argued by, for example, \citet{gao-etal04} 
\citep[but see ][]{conroy-etal06a}. Indeed, current galaxy
formation models indicate that satellite galaxies represent a biased 
subset of subhaloes whose spatial distribution is very likely to be 
anisotropic within their parent galaxy's dark matter halo 
\citep[e.g.][]{zentner-etal05,libeskind-etal05},
although current high resolution models suggest that the orientation
of the satellite system with respect to the dark matter halo
is robust \citep{libeskind-etal07}.

These considerations suggest that interpretation of the spatial
distribution of satellites may be more complex than we might 
na\"ively expect. Yet, despite these complexities, satellite galaxies 
represent a powerful observational probe into the mass distribution 
around galaxies. Therefore, it is essential to determine robustly the 
spatial distribution of satellite galaxies, and to establish whether 
or not they show a preferential alignment relative to their parent 
galaxies. This is the aim of the current study.\\

Locally we see strong evidence for the preferential alignment or
\emph{anisotropic distribution} of satellites relative to their
primary galaxies. Both the Milky Way and M31 have satellite populations
that lie in great planes that are highly inclined to their discs. This
has been noted by \citet{lynden-bell76}, \citet{hartwick00} and \citet{ktb05} 
for the Milky Way; by \citet{kg06} and \citet{mi06} for
subsamples of satellites of M31; and by 
\citet{mkj07}, who performed an analysis of the statistical significance
of the planar distribution around both galaxies. The Milky Way and M31 
are the only galaxies for which the sample of satellites is large
enough that their spatial distribution is analysed directly (without 
requiring stacking to obtain a statistical sample). They are also the
only systems for which we are certain that the satellites are physically
associated with their primaries, and for which the three-dimensional 
positions of the satellites with respect to their primaries are
known. In addition, proper motion measurements exist for a number of
Milky Way satellites, which confirm that these systems are on polar orbits
\citep{pmj02}.

\begin{figure}
\includegraphics[scale=0.65]{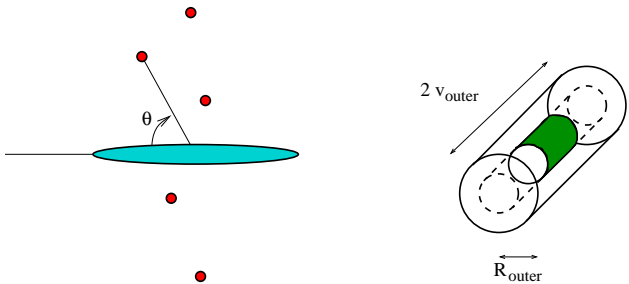}
\caption{%
\textit{(Left)} 
According to the Holmberg Effect, satellites of disc galaxies tend to
lie near the minor axis of the disc. This corresponds to an angle between the
major axis of the galaxy and the satellite (the ``disc angle'', $\theta$)
of greater than $45\degr$. \textit{(Right)} Selection cylinders around
a potential primary galaxy, which lies at the centre of the cylinders.
The outer isolation cylinder is marked with solid lines and has radius 
$\rout$ and length $2\, \vout$. The inner isolation cylinder is marked 
with dashed lines and has radius $\rin$ and length $2\, \vin$. The 
satellites are drawn from the shaded cylinder with radius $\rsat$ and 
length $2\, \vsat$.
\label{selection cylinder figure}}
\end{figure}

Analysing the spatial distribution of satellites in external systems is
generally more challenging because no more than one or two satellites are
detected per primary galaxy, and because the three-dimensional location
of the satellite with respect to its primary is uncertain.
This measurement requires that primaries and
their satellites be stacked to obtain statistical samples from which
the projected angular distribution of satellites about a ``typical'' primary is
determined. Such an analysis was first performed by \citet{holmberg-effect} 
for 58 isolated spiral and lenticular,
or \textit{late-type}, galaxies and their 218 optical 
companions, of which 75 were expected to be physically associated. He
found that satellites at projected radii less than 50~kpc were more
often found near the poles (minor axis) of the primary's disc 
(Figure~\ref{selection cylinder figure}). This preferential polar
distribution, referred to as the ``Holmberg Effect'', was subsequently 
confirmed by \citet{zsfw97-holmberg} (hereafter \citetalias{zsfw97-holmberg}) 
at larger projected radii of 300 -- 500~kpc using a sample of 115 
spectroscopically-confirmed satellite galaxies of 69 isolated late-types.

The advent of large galaxy redshift surveys such as the 2dF Galaxy
Redshift Survey \citep[2dFGRS;][]{colless-etal01} and the Sloan Digital 
Sky Survey \citep[SDSS;][]{sdss-technical-summary} has enabled this 
issue to be revisited. These surveys provide an abundance of galaxies 
with spectroscopic redshifts, and several recent studies have sought 
to use numbers of galaxies far in excess of those that were available to 
\citet{holmberg-effect} and \citetalias{zsfw97-holmberg} to address the 
question of spatial anisotropy with \textit{significantly improved statistics}.
\citet{sl04} (hereafter \citetalias{sl04}) used over 2000 satellites
of almost 1500 primary galaxies in the 2dFGRS, and found that
satellites around blue primaries tended to follow an isotropic
distribution, whereas the locations of low-velocity ($|\Delta v| < 160~\kms$)
satellites around red primaries tended to
align with the \textit{major-axis}. This finding contrasts sharply with 
the polar distribution inferred by both \citet{holmberg-effect} and 
\citetalias{zsfw97-holmberg}\footnote{Note that this is opposite 
to the original claims of \citetalias{sl04}; see the discussion in 
\citetalias{yang-etal06} for further details}. \citet{yang-etal06} 
(hereafter \citetalias{yang-etal06}), \citet{appz07} (hereafter 
\citetalias{appz07}) and \citet{ab07} (hereafter \citetalias{ab07})
detected a similar anisotropy of satellites in the SDSS;
\citetalias{yang-etal06} used over 16000 groups of galaxies selected 
to lie within the same dark matter halo, while \citetalias{appz07}
and \citetalias{ab07} studied satellites of isolated galaxies;
intriguingly, the latter study found that the isotropic distribution around
blue galaxies was composed of a \textit{major-axis} alignment
for satellites at small projected radii and a \textit{minor-axis}
alignment for satellites at large projected radii.
\citet{brainerd05} (hereafter \citetalias{brainerd05}) also selected 
satellites of isolated galaxies in the SDSS (her largest sample
contained approximately 3000 satellites around 2000 primaries) and
found that they exhibited a major-axis distribution.
Similar results were obtained by \citet{faltenbacher-etal07},
and \citet{wang-etal08}, who examined members of tens of thousands
of groups in the SDSS.
Finally, \citet{azpk06} (hereafter 
\citetalias{azpk06}) found no evidence for anisotropy based on a 
smaller sample of 193 satellites of 144 isolated late-type galaxies 
in the SDSS.\\

These results, derived from both 2dFGRS and SDSS, would seem to suggest
that the preferential alignment of satellites about the poles of their 
primaries noted by both \citet{holmberg-effect} and 
\citetalias{zsfw97-holmberg} was a consequence of small number
statistics. However, the spatial anisotropy reported by 
\citetalias{zsfw97-holmberg} was detected with a statistical confidence 
greater than 99\%, suggesting that small number statistics were
unlikely to be a problem. Furthermore, as we have noted already, there
is strong evidence for the preferential alignment of satellites
about the pole of the Milky Way and
evidence for subsamples of satellites lying in inclined discs around M31,
systems for which we can
be certain that the satellites are physically associated
with their primaries and for which we know their spatial distribution.
Both of these observations
raise the spectre that it is systematic rather than random
errors that more strongly affect detection of spatial anisotropies in
the distribution of satellites around external galaxies. This leads us
to the issue of sample selection. 

In the case of both the Milky Way and M31, we have an abundance of 
detailed information that allows us to state with confidence which galaxies can
be considered satellites belonging to these hosts. In the case of
external galaxies, we do not have such detailed information and so we
must employ selection criteria that allow us to identify which faint 
galaxy neighbours in projection are likely to be satellites of the primary.
Differences arising from how satellite galaxies are selected will
affect measurements of the angular distribution of satellites, and
will therefore influence how these data are interpreted.

To illustrate this point, we know that the member galaxies of groups
and clusters tend to cluster about the major axis of the Brightest Group or 
Cluster Galaxy \citep[BGG and BCG respectively; e.g.][]
{binggeli82,west89,mandelbaum06-haloshape}. If we identify these group
or cluster members as satellites of their primary galaxy, the BGG or
BCG, then we would interpret this measurement as showing that group or
cluster members preferentially align with their primary's
major-axis. However, we do not expect groups or clusters to
be dynamically relaxed and so satellites in these systems might not
trace the potential of an equilibrium mass distribution. Therefore, 
the study of satellites around \textit{isolated} galaxies must be
highly successful at excluding such groups and clusters. The ability of 
selection criteria to identify the proper type of system and to
suppress contamination is essential for the result to be considered 
robust and physical.\\

Mock galaxy catalogues constructed from cosmological $N$-body 
simulations provide a powerful method for assessing the reliability 
of selection criteria. Dark matter haloes are populated with mock
galaxies following either a statistical approach based on ``Halo
Occupation Distributions'' \citep[HOD; e.g.][]{bw02}, or a physically
motivated approach based on semi-analytical galaxy formation models
\citep[e.g.][]{cole-etal00}. Both approaches are parameterised and so
statistical properties of the mock galaxy population, such as the 
luminosity function and the 2-point correlation function, are
fine-tuned to recover the observed properties of real galaxies in the 
2dFGRS and the SDSS. This provides an ideal testbed for selection criteria.
We adopt the ``Conditional Luminosity Function'' (CLF) formalism of
\citet{yang-etal03}, a statistical approach based on HODs, to develop
our suite of mock catalogues. The CLF formalism allows us to assign to 
each dark matter halo in a cosmological $N$-body simulation a
probability of hosting $N$ galaxies with a total luminosity $L$, and a
distribution of luminosities drawn from a Schechter function whose
parameters depend on the halo mass $M$. Further details are presented
in \S~\ref{mock catalogues}.

We note that previous studies have used mock galaxy catalogues derived 
from the CLF formalism to investigate the radial distribution 
\citep{vdb-etal05} and kinematics \citep{vdb-etal04}
of satellite galaxies. However, we
use our mock catalogues to establish \textit{optimal selection criteria} that 
preferentially pick out \textit{isolated systems} of primary galaxies 
and their satellites.
We explore the impact of different selection criteria
on the nature of satellite samples (e.g. group or cluster members?)
and to establish robust criteria that minimise the influence of
interlopers and a primary's larger-scale environment. This allows 
us to assess selection criteria adopted in previous studies and to
quantify the ability of these criteria to identify isolated systems.\\

By taking such care in establishing the criteria by which our 
isolated systems are selected, we are able to address a number of important
outstanding physical questions using a \textit{robust} sample of galaxies taken
from the SDSS. In particular, we revisit the question of whether or not
the angular distribution of satellite galaxies about their primary is 
anisotropic, and whether or not anisotropy is linked to the
morphological type of the primary. Such a dependence would present an 
exciting possibility to study the connection between subhaloes, haloes, 
and galaxy morphology.
In \S~\ref{summary discussion section} we discuss the
physical significance of our results in this context.

We note also natural overlaps with studies that seek to measure the 
flattening of dark matter haloes. There is good evidence to suggest
that the haloes of early-type galaxies are flattened along the minor 
axis of the light -- based on studies of X-ray isophotes 
\citep[e.g][]{buote-etal02} and weak lensing \citep[e.g][]{hyg04} -- 
whereas the results for disc galaxies are inconclusive.
For example, the tidal 
stream of the Sagittarius Dwarf has been used to infer that the Milky 
Way's halo is spherical \citep{ibata-etal01,fellhauer-etal06},
flattened in the same 
sense as the disc \citep{martinez-delgado-etal04,jlm05}, or flattened 
in the opposite sense as the disc \citep{helmi04b}, depending on which part 
of the stream is modelled \citep{ljm05}. We discuss the implications 
of our results for halo flattening in \S~\ref{summary discussion section}.\\

The structure of this paper is as follows: in \S~\ref{sample selection section}
we discuss the sample selection criteria, and present our sample of satellites
of isolated galaxies. In \S~\ref{results section} we present
our results for the anisotropic distribution of satellite galaxies 
with respect to their primary galaxy, with respect to large scale filaments,
and the relative alignment of isolated galaxies with large scale filaments.
In \S~\ref{comparison with previous results section} we
compare our results to previous studies and analyse
how different selection criteria provide constraints on the origin
of satellite anisotropy.
Finally, we discuss the interpretation of the results in terms
of the shapes of dark matter haloes, anisotropic infall,
and the formation of galactic discs in 
\S~\ref{summary discussion section}.

\section{Sample Selection}\label{sample selection section}
The most critical part of the analysis is the process of selecting satellite
galaxies of isolated primaries. We must simultaneously:
\begin{enumerate}
 \item Minimize the number of primaries that are not isolated, i.e. which
do not dominate the dynamics of their environment.
 \item Minimize the number of ``interlopers'', or satellite-primary pairs
that do not represent physical satellites of the primary galaxy.
 \item Maximize the sample size, subject to the above constraints.
\end{enumerate}

We use mock catalogues generated from cosmological simulations to
refine our criteria to fulfil the above requirements and to
critically examine the selection criteria that have been used in previous
studies.

\subsection{Observational Data}
\subsubsection{Sloan Digital Sky Survey}
Our sample is drawn from the Sloan Digital Sky Survey Data Release 6
(SDSS DR6) \citep{sdss-dr6}. All primary survey objects classified as 
galaxies in the imaging data that satisfy the spectroscopic targeting 
algorithm of either the Main galaxy sample \citep{sdss-maingalaxysample}
or the Luminous Red Galaxy sample \citep{sdss-lrgsample} are
considered. Only galaxies with valid spectroscopic redshifts
($\mathrm{sciencePrimary}=1$ and $\mathrm{zConf}>0.85$)
that are also spectroscopically classified as galaxies are considered 
as primary or satellite galaxies; however, galaxies without such
spectra or which are spectroscopically classified as stars
(which often indicates that a foreground star appears near the
centre of a galaxy in projection)
are also used when evaluating the isolation criteria.
We have excluded all objects with unrealistic colours
(differences between successive bands of at least $5$ magnitudes),
as this always indicate a spurious detection 
at the magnitudes we consider.
Petrosian magnitudes are used throughout, and are dereddened using 
the corrections in \citet{sfd98} and k-corrected using KCORRECT 
v4\_1\_4 \citep{kcorrect} to the ${}^{0.1}r$ band\footnote{The $r$ 
band redshifted to $z=0.1$; for simplicity we use the notation $M_r$
to refer to $M_{{}^{0.1}r}$.}
to minimise the effect of errors in the 
k-correction. The nominal survey limit for spectroscopic targets in the 
SDSS Main galaxy sample is $r \le 17.77$; however, the actual limit 
varies across the sky. Therefore, we conservatively
treat it as only being complete to $r \le 17.5$, although we make use
of galaxies as faint as $r = 17.77$ when they are available.
When we require photometric redshifts, we use
the D1 neural network photometric redshifts available in the SDSS DR6 catalogue
\citep{oyaizu-etal07},
which are the most accurate available for the $r<18$ galaxies that
we consider.
Angular diameter distances and distance moduli are calculated
assuming $\Omega_m=0.3$ and $\Omega_{\Lambda}=0.7$ and are quoted
in $h$-independent units, where $H_0 = h\,100\,\mathrm{km\,s^{-1}\,kpc^{-1}}$.

\subsubsection{Galaxy Classification}
\label{galaxy classification section}

It is important to separate spheroid-dominated early-type galaxies
from disc-dominated late-type galaxies. The major axes of spheroidal
galaxies are determined by their anisotropic velocity dispersions while
those of disc galaxies are determined by their angular momentum;
therefore, their orientation with respect to the dark matter halo may
be different. Furthermore, the dynamical effects of discs versus
spheroids on satellite orbits may be different.
Galactic orientations may also depend on their history,
which is probed by their stellar populations.
Indeed, many previous studies have found that
the satellites around red and blue galaxies are distributed differently.

We adopt the
galaxy classification scheme of \citet{bailinharris08-classification},
which is based on the global concentration of the light profile and the
location of the galaxy on the colour-magnitude diagram. This method
explicitly accounts for inclination effects, and has been validated
using high-quality imaging data from the Millennium Galaxy Catalogue
\citep{mgc-imaging}. Galaxies congregate in three distinct regions
of parameter space: Early-type galaxies are red, highly concentrated,
and ellipsoidal; Intermediate-type galaxies are red, have intermediate
concentrations, and contain discs; and Late-type galaxies are blue,
have low concentrations, and are disc-dominated.
Determining whether the satellite systems of Intermediate-type galaxies
bear closer relation to those of Early- vs. Late-type galaxies will
provide useful insights into their nature.

An analysis of the results when other classification schemes are adopted
is given in Appendix~\ref{appendix:classification}. We find that our
qualitative results are unchanged for any reasonable method of splitting
the sample, although the numerical magnitude of the effect can vary by
$\sim 1\sigma$ depending on the classification method.

\subsection{Definition of Selection Criteria}
The format of our selection criteria is based on \citet{nfc08}
and is similar to that used in previous
studies; however, the details differ in several important ways.
To be considered isolated, primaries must not have any
comparably bright neighbours within a large surrounding region,
and must be much brighter than all potential satellites in
the immediate vicinity. We define three cylinders around each potential 
primary (see Figure~\ref{selection cylinder figure}):
\begin{itemize}
 \item Outer isolation cylinder: All galaxies within a projected 
   separation of 
    $\Delta R \le \rout$ and a velocity difference of
    $|\Delta v| \le \vout$ must be at least $\magout$ magnitudes
    fainter in $r$.
 \item Inner isolation cylinder: All galaxies within a projected 
   separation of
    $\Delta R \le \rin$ and a velocity difference of
    $|\Delta v| \le \vin$ must be at least $\magin$ magnitudes
    fainter in $r$ ($\magin > \magout$).
 \item Satellite cylinder: Satellites are galaxies within a projected 
   separation of
    $\Delta R \le \rsat$ and a velocity difference of
    $|\Delta v| \le \vsat$. Satellites must be at least
    $\magsat$ magnitudes fainter in $r$. For our criteria, we enforce
    $\rsat = \rin$, $\magsat=\magin$, and $\vsat = \frac{1}{2} \vin$.
\end{itemize}
The adopted values of the parameters are given in
Table~\ref{selection criteria parameters table}.
To ensure that satellites are not associated with more than
one primary (a situation we refer to as a ``multi-homed'' satellite)
and that there are no near neighbours too luminous to be
satellites, it is important that
$\rin \le \frac{1}{2} \rout$, $\vin = \vout$, $\rsat \le \rin$,
$\vsat \le \frac{1}{2} \vin$, and $\magsat = \magin$.
These sanity checks are not fulfilled by many of the criteria that 
have been used in previous studies.

\begin{table*}
\begin{minipage}{160mm}
\caption{Parameters of Selection Criteria%
\label{selection criteria parameters table}}
\begin{tabular}{lcccccccccc}
\hline
{Parameter} & {This Work} &
 {\citetalias{zsfw97-holmberg}} &
 {\citetalias{sl04}} &
\multicolumn{3}{c}{\citetalias{brainerd05}} &
 {\citetalias{azpk06}} &
\multicolumn{2}{c}{\citetalias{appz07}} & \citetalias{ab07}\\
 & & & & S1 & S2 & S3 & & S1 & S2\\
\hline
\multicolumn{7}{l}{Outer isolation cylinder:}\\
\hspace{1em}$\rout\>(\hikpc)$ & $1000$ & $750$ & ... &
  ... & ... & $700$ & ... & ... & ... & ...\\
\hspace{1em}$\vout\>(\kms)$ & $1500$ & $1000$ & ... &
  ... & ... & $1000$ & ... & ... & ... & ...\\
\hspace{1em}$\magout$ & $0.7$ & $0.7$ & ... &
  ... & ... & $0.7$ & ... & ... & ... & ...\\
\multicolumn{7}{l}{Inner isolation cylinder:}\\
\hspace{1em}$\rin\>(\hikpc)$ & $500$ & $375$ & $700$ &
  $490$ & $2000$ & $350$ & $500$ & $490$ & $500$ & $511$\\
\hspace{1em}$\vin\>(\kms)$ & $1500$ & $1000$ & $1000$ &
  $1000$ & $1000$ & $1000$ & $1000$ & $1000$ & $1000$ & $1000$\\
\hspace{1em}$\magin$ & $2.0$ & $2.2$ & $1.0$ &
  $1.0$ & $0.7$ & $2.2$ & $2.0$ & $1.0$ & $2.0$ & $1.0$\\
\multicolumn{7}{l}{Satellite cylinder:}\\
\hspace{1em}$\rsat\>(\hikpc)$ & $500$ & $375$ & $500$ &
  $350$ & $350$ & $350$ & $350$ & $350$ & $350$ & $365$\\
\hspace{1em}$\vsat\>(\kms)$ & $750$ & $500$ & $500$ &
  $500$ & $1000$ & $500$ & $500$ & $500$ & $500$ & $500$\\
\hspace{1em}$\magsat$ & $2.0$ & $2.2$ & $2.0$ &
  $2.0$ & $1.5$ & $2.2$ & $2.0$ & $2.0$ & $2.0$ & $2.0$\\
$\Nviol$ & NED+photo-z$^{e}$ & 0$^{e}$ & ... &
  ... & ... & ... & ... & ... & ... & ...\\
$\fsatlum$ & $0.2$ & ... & ... &
  $1.0$ & $1.0$ & $1.0$ & ... & ... & ... & $1.0$\\
$\Nsatmax$ & $4$ & ... & $4$ & 
  ... & ... & ... & ... & ... & ... & $9$\\
\multicolumn{7}{l}{Sanity checks:}\\
\hspace{1em}Forbids multi-homed satellites$^{a}$ & Yes & Yes & No &
  No & No & Yes & Yes & No & Yes & No\\
\hspace{1em}Forbids nearby non-satellites$^{b}$ & Yes & Yes & No &
  No & No & Yes & Yes & No & Yes & No\\
\hspace{1em}Avoids survey edge$^{c}$ & Yes & N/A & No &
  No & No & No & Yes$^{e}$ & No & No & Yes\\
\hspace{1em}Avoids survey magnitude limit$^{d}$ & Yes & Yes & No &
  No & No & No & Yes & No & Yes & No\\
\hline
\end{tabular}
$^{a}$The criteria do not permit satellites to belong to more
than one primary galaxy.\\
$^{b}$The criteria do not permit there to be bright (non-satellite)
galaxies within the satellite cylinder.\\
$^{c}$The criteria do not permit primaries so near the edge
of the survey that potential bright neighbours would lie outside the survey region.\\
$^{d}$The criteria do not permit primaries faint enough that
potential bright neighbours fall below the local survey magnitude limit.\\
$^{e}$See text for further clarification.\\
\end{minipage}
\end{table*}

We also apply the following additional criteria:
\begin{enumerate}
  \item All primaries must be at least \magin\ magnitudes brighter than
$r=17.5$ to ensure that all potential
bright neighbours are brighter than the local survey limit.
  \item The projected distance from the primary to the nearest edge
of the photometric survey footprint
must be at least \rout\ to ensure
that any potential bright neighbours have been observed
photometrically. The projected distance from the primary
to the nearest edge of the spectroscopic survey footprint
must be at least \rsat\ to ensure that all potential satellites
are equally likely to have been observed spectroscopically and
therefore that the survey edge does not impose an angular bias
in the selected satellite population.
  \item Because of spectral incompleteness, $\sim10$\%\ of galaxies that fulfil
the requirements of the SDSS Main galaxy targeting criteria do not have
observed redshifts.
Therefore, there are potential primaries that would not be
considered isolated if it turned out that their non-spectroscopic
neighbours are at the same redshift.
This issue is particularly important because the limited number of
fibres per tile causes
the fractional completeness to be lower in regions of high galactic density.
There are a number of ways of treating such galaxies
(hereafter referred to as ``violators''): one could
assume that most do not lie at the same redshift as
the primary and simply ignore their presence
\citepalias{sl04,brainerd05,azpk06,appz07,yang-etal06,ab07},
one could
establish a threshold such that if there are more than a number \Nviol\ 
then the chances that at least one is at the same redshift
as the primary is high and therefore eliminate those primaries
\citep{nfc08,herbertfort-etal08},
or one could eliminate all such primaries on the grounds that there
is a chance that they are not isolated (equivalent to setting $\Nviol=0$;
\citetalias{zsfw97-holmberg}).
The existence of photometric redshifts in the SDSS DR6 catalogue allows
us to use a more sophisticated method of determining whether the violators
are likely to be at the same redshift as the primary.
We first query the NASA/IPAC Extragalactic Database
(NED\footnote{http://nedwww.ipac.caltech.edu/}) for
literature spectroscopic redshifts of the violators of all
primaries that would otherwise be included in our sample.
If no spectroscopic redshift is available then
we consider the photometric redshift
$z_{\mathrm{viol,photo}}$;
if it is within $2 \sigma_{\mathrm{viol,photo}}$
of the spectroscopic redshift of the primary, where
$\sigma_{\mathrm{viol,photo}}$ is the estimated error on the photometric
redshift in the catalogue, 
then we eliminate the primary.
We ignore the presence of violators that do not
exist in the photometric redshift database, as these are mostly
galaxies that are not detected in one or more bands; such galaxies are
highly unlikely to be truly bright physical neighbours of the relatively
luminous nearby galaxies that constitute our sample of primary galaxies.
We have confirmed that excluding primaries with such neighbours does not
alter our conclusions.
  \item The
total luminosity of the satellites must not be more than $\fsatlum$ times the
luminosity of the primary, and systems with more than $\Nsatmax$ satellites are
discarded. This ensures that the primary galaxy dominates the satellite
system.
  \item Postage stamp images of each potential primary were examined by eye.
11 objects were removed, 4 of which suffered from 
catastrophically bad background subtraction
due to nearby bright objects and 7 of which are major mergers
in progress.
\end{enumerate}

Our choices for the relevant parameters are motivated by
analysis using mock catalogues as described in
\S~\ref{tests of selection criteria section}.
Where applicable, the values 
previous authors have used for the selection parameters are
given in Table~\ref{selection criteria parameters table}.
For those criteria that use only one isolation cylinder,
we characterize it as an ``Inner'' cylinder.
We summarize the selection criteria used by each previous study below:

\citetalias{zsfw97-holmberg} used somewhat thinner and shorter cylinders
than those used in this work, but 
the criteria fulfil the above sanity checks.
They did not have redshifts and magnitudes for every galaxy
in the field, but eliminated all primaries that appeared by eye to
have potentially criteria-violating neighbours
(i.e.~$\Nviol=0$).
There was no formal edge of the surveyed area,
and primaries were chosen to be at least 2.5~mag brighter than
the POSS magnitude limit,
so all potential bright neighbours were considered.
No cut on the number or luminosity fraction of satellites was imposed,
and only morphological late-types were included as primaries. 

\citetalias{sl04}
used only one isolation cylinder, 
which did not satisfy $\magsat \le \magout$.
Therefore, relatively bright galaxies are allowed to be in the
satellite region.
Because of this definition, satellites may be multi-homed.
Only primary galaxies with absolute magnitude
$M_{\rm b_J} - 5 \log h < -18$ were used.
They did not impose any
constraints on proximity to the survey magnitude limit,
proximity to the survey edge, \Nviol, or \fsatlum,
but used $\Nsatmax=4$.

\citetalias{brainerd05} tested three selection criteria; her
Sample~1 (S1) used isolation criteria based on \citetalias{sl04}
(though adopting a different value of the Hubble constant);
her Sample~2 (S2) used one very wide isolation cylinder,
which permits bright galaxies to be in the satellite region and
satellites to be multi-homed; and her Sample~3 (S3)
used isolation criteria based on \citetalias{zsfw97-holmberg}
(though adopting a different value of the Hubble constant).
There was no cut on proximity to the survey edge, \Nviol,
or \Nsatmax\ in any of the samples. There was a cut on \fsatlum\ of $1.0$.

\citetalias{azpk06} used Sample~2 of \citet{prada-etal03}, but 
restricted
the primaries to be morphological late-types, adopted a redshift
limit of $cz \le 11000~\kms$, and only examined primaries with
$-20.5 \le M_B \le -19.5$. These criteria
are similar to \citetalias{sl04}, except that because $\magsat = \magout$,
the satellite cylinder is not permitted to contain bright galaxies and
satellites can only belong to a single primary.
Although they did not formally adopt a cut on proximity to the survey
edge, they searched for bright neighbours in \citet{RC3} (\citetalias{RC3}),
which fulfils the same purpose.
The narrow range of absolute magnitudes and redshift limit serve the
same purpose as our cut on galaxies near the survey magnitude limit.
They did not make any cut on \Nviol, \Nsatmax, or \fsatlum.

\citetalias{appz07} tested two sets of criteria. Their Sample~1 (S1)
uses the same criteria as \citetalias{brainerd05}~S1 but without the cut
on \fsatlum. Their Sample~2 (S2) is identical to 
\citetalias{azpk06} except that a wider range of primary
luminosities, $-23 \le M_r \le -21$ is used, the redshift limit
is extended to $cz \le 30000~\kms$, and no check for bright neighbours
outside of the survey boundary is performed.

\citetalias{ab07} used criteria very similar to \citetalias{brainerd05}~S1,
but they adopted a slightly different value for the Hubble constant,
restricted \Nsatmax\ to $9$, and ensured that their primary galaxies
were not near the edge of the spectroscopic survey.

\citet{holmberg-effect} did not have redshifts for any of his galaxies,
making it difficult to directly compare our selection.
It is also difficult to compare our selection with \citetalias{yang-etal06}
or \citet{faltenbacher-etal07},
who used an iterated percolation algorithm rather than isolation criteria.
Their selections were tuned
using mock catalogues to minimize the number of interlopers; however, they
were not designed to find isolated galaxies and most of their systems should
be considered groups or clusters rather than satellite systems of isolated
galaxies.

Our criteria are more restrictive than other
criteria that have been used to select isolated galaxies.
The advantage of using such a large sample as SDSS DR6 is less
the ability to boost the statistics than the ability
to be extremely conservative with our selection criteria
and still retain an acceptable number of galaxies.
Given that the disagreement between previous results is more
likely due to systematic errors than statistics, such a rigorous
treatment is essential to disentangling the nature of the
disagreement and determining the true distribution.

\subsection{Mock Catalogues}\label{mock catalogues}
\label{mock catalogue section}

\subsubsection{The Conditional Luminosity Function $\Phi(L|M)$}

We construct our mock galaxy catalogues following the prescription
presented in \citet{yang-etal04} (hereafter \citetalias{yang-etal04}), 
which is based on earlier studies by \citet{yang-etal03} and 
\citet{vdb-etal03}. This approach allows us to assign to each dark 
matter halo of mass $M$ a probability of hosting a population of $N$ 
galaxies of total luminosity $L$. This probability is governed by the 
``Conditional Luminosity Function'', $\Phi(L|M)$, which is
parameterised by the Schechter function,
\begin{equation}
  \label{eq:phi}
	{ \Phi(L|M)dL =
	  \frac{\tilde{\Phi}^{\ast}}{\tilde{L}^{\ast}}\left(\frac{L}{\tilde{L}^{\ast}}\right)^{\tilde{\alpha}}
	  \exp(-L/\tilde{L}^{\ast})dL .}
\end{equation}

\noindent
The normalisation $\tilde{\Phi}^{\ast}$, characteristic luminosity 
$\tilde{L}^{\ast}$, and faint-end slope $\tilde{\alpha}$ are all
functions of halo mass $M$; appropriate expressions for these
quantities are in Appendix~\ref{appendix:details}.\\

Given $\Phi(L|M)$, we compute various ``observable'' properties of 
the galaxy population associated with an average dark matter halo 
of mass $M$:
\begin{itemize}
\item the mean number of galaxies $\left<N\right>(M)$ (see equation~\ref{eq:nav});
\item the luminosities of the central galaxy $L_{\rm cen}$ and
  satellite galaxies $L_{\rm sat}$ (see equation~\ref{eq:l1});
\item the morphological type of each galaxy (i.e. early- versus late-
  type; see Appendix~\ref{appendix:mocks}). 
\end{itemize}
All of these properties can be recovered using analytic dark matter 
halo mass functions, but a cosmological $N$-body simulation 
is required to assign phase space coordinates to mock galaxies. In
what follows, we describe briefly the main steps involved in
constructing the mock catalogues. A more detailed description is provided
in Appendix~\ref{appendix:mocks}.\\

\subsubsection{Populating Dark Matter Haloes with Galaxies}

We perform a series of cosmological $N$-body simulations following
the formation of structure in a $\Lambda$CDM model from $z$=50 to
$z$=0. Each simulation contains $256^3$ particles. We adopt
cosmological parameters of $\Omega_m$=0.3, $\Omega_{\Lambda}$=0.7 
and $h$=0.7, and the power spectrum of initial density perturbations 
\citep[generated using CMBFAST;][]{cmbfast} is normalised assuming a 
mass variance of $\sigma_8$=0.9. 
Details of the runs are presented in Table~\ref{table:simulations}.
We quote results using the five $100~h^{-1}~\mathrm{Mpc}$ boxes
(labelled ``A'' through ``E''), but have verified using the smaller
boxes that this resolution is sufficient to reproduce the
statistical properties of the mock catalogues.

\begin{table}
  \caption{\label{simulations}\label{table:simulations}%
    Cosmological $N$-body Simulations. $L_{\rm box}$ refers to the
    length of the simulation box; $m_{\rm part}$ is the particle mass,
    which depends on both $L_{\rm box}$ and the number of particles
    $N_{\rm part}$; $\epsilon$ is the gravitational force softening; 
    $M_{\rm min}$ is a (conservative) estimate of minimum halo mass
    that can be reliably resolved, based on convergence of the mass
    function; and $L_{\rm min}$ is the minimum luminosity.}
  \begin{tabular}{cccccc}
    \hline
    {$L_{\rm box}$} & {$m_{\rm part}$} & {$\epsilon$} & 
    {$M_{\rm min}$} & {$L_{\rm min}$}\\
    {[$h^{-1} {\rm Mpc}$]} & [$h^{-1} {\rm M_{\odot}}$] &
    [$h^{-1} {\rm kpc}$] & [$h^{-1} {\rm M_{\odot}}$] & 
    [$h^{-2} {\rm L_{\odot}}$]\\
    \hline
  35 & $0.21 \times 10^{9}$ & 2.7 & $10^{10}$           & $1.1 \times 10^{8}$\\
  50 & $0.62 \times 10^{9}$ & 3.9 & $3 \times 10^{10}$  & $1.1 \times 10^{8}$\\
  70 & $1.7 \times 10^{9}$  & 5.5 & $9 \times 10^{10}$  & $1.1 \times 10^{9}$\\
  100& $4.96 \times 10^{9}$ & 7.8 & $25 \times 10^{10}$ & $1.1 \times 10^{9}$\\
\hline
\end{tabular}
\end{table}

These simulations provide the dark matter host haloes that we populate 
with mock galaxies. Haloes are identified at $z$=0 using the 
friends-of-friends (FOF) algorithm with a linking length of $b$=0.2 
times the mean interparticle separation. For each halo we compute its 
virial mass and radius, and we record also the coordinates of its 
most bound particle and a list of all particles that reside within its 
virial radius. This is (in principle) all the information we need to
construct our mock galaxy catalogues.\\

We note that $\tilde{\Phi}^{\ast}$, $\tilde{L}^{\ast}$, and
$\tilde{\alpha}$ are functions of halo mass $M$, but they also depend 
on the cosmological parameters. \citet{yang-etal03} presented a number
of different CLFs for different cosmologies and different assumptions 
regarding the free parameters, and we adopt those used by 
\citetalias{yang-etal04}, who assumed a flat $\Lambda$CDM cosmology 
with $\Omega_m$=0.3, $\Omega_{\Lambda}$=0.7, $h$=0.7 and a
normalisation $\sigma_8$=0.9. Precise values for the CLF parameters 
are given in Appendix~\ref{appendix:details}. 

These parameters were chosen to recover the observed luminosity 
functions and correlation lengths of galaxies in the 2dFGRS (both 
as a function of their luminosity and their type), but we find that 
the clustering and luminosity functions of galaxies in our mock
catalogues are in very good agreement with the corresponding SDSS
correlation and luminosity functions.\\

It is important to estimate the minimum reliably resolved halo mass 
$M_{\rm min}$; below this threshold the number density of haloes tends 
to be suppressed relative to the number density they would have in the 
limit of infinite numerical resolution. In this work we assume that 
the mass function is converged for haloes containing 50 particles or
more (see discussion in Appendix~\ref{appendix:mocks}); this gives 
$M_{\rm min}$=50$\,m_{\rm part}$ in Table~\ref{table:simulations}.

Knowing $M_{\rm min}$ allows us to estimate $L_{\rm min}$, the minimum 
luminosity that we can assign to a mock galaxy, using the``conditional 
probability distribution'' $P(M|L)$. $L_{\rm min}$ fixes the halo
occupation number, the average number of galaxies per halo of mass $M$,
$\left<N\right>(M)$ (equation~\ref{eq:nav}); as $M_{\rm min}$ and 
therefore $L_{\rm min}$ increases (decreases), $\left<N\right>(M)$ 
increases (decreases). The mass resolution of our simulations means
that we adopt $L_{\rm min}=1.1 \times 10^{9}h^{-2}\,{\rm L_{\odot}}$ in
our mock catalogues A to E, upon which our analysis is based.

We assume that the number of galaxies $N$ in a dark matter halo is 
Poisson distributed about $\left<N\right>(M)$ \citep{yang-etal03}, and
that each galaxy is assigned a luminosity drawn from $\Phi(L|M)$. The 
central galaxy is defined to be the brightest galaxy in the halo and
has a luminosity $L>L_1$, where $L_1$ satisfies the condition that
$\left<N\right>(M)=1$ (see equation~\ref{eq:l1}). The remaining $N$-1 
galaxies have luminosities that are drawn randomly from the luminosity
function in the range $L_{\rm min} < L < L_1$.

The morphological type of each galaxy is obtained from $f_{\rm late}(L,M)$ 
(equation~\ref{eq:late}), the probability that a galaxy of luminosity $L$ 
hosted by a dark matter halo of mass $M$ is late-type. Finally, the
position and velocity of the central galaxy are associated 
with those of the most bound particle in the halo, while the positions 
and velocities of the remaining $N$-1 galaxies are obtained by randomly 
sampling the particles in the FOF group.

A more detailed description of our approach to populating dark matter
haloes with galaxies is given in Appendix~\ref{appendix:mocks}.

\subsubsection{Constructing a Mock Galaxy Redshift Survey}
\label{constructing mock grs section}

At this point we depart from \citetalias{yang-etal04}, who wished to study a 
mock 2dFGRS and stacked simulation boxes to recover the survey's median 
redshift. Our needs are more modest -- we wish to evaluate the
reliability of our sample selection criteria. To transform our raw mock 
galaxy distribution into a mock galaxy catalogue,

\begin{itemize}

\item We select a single simulation box, recalling that each box has
  periodic boundary conditions, and we replicate it 3 times along
  each dimension, producing a stack of 3$\times$3$\times$3 boxes. We
  then centre the stacked boxes on the median redshift of the SDSS 
  $z_{\rm med}$=0.11. 

\item We place a virtual observer at $z$=0 and we define an 
  $(\alpha,\delta)$ coordinate frame with respect to the centre of 
  the stacked box.
  
\item We compute a redshift $z$ for each galaxy as seen by the virtual
  observer from the recessional velocity $cz=Hr+\vec{v}\cdot
  \vec{r}/|r|$, where the galaxy is at $\vec{r}$ with respect to the 
  virtual observer, and $\vec{v}\cdot\vec{r}/|r|$ is its line-of-sight 
  velocity. We account for
  observational velocity uncertainties by adding a random velocity 
  drawn from a Gaussian of width $30~\kms$. We also
  compute the apparent magnitude according to its luminosity and
  distance, to which we add an RMS error of $0.02~\mathrm{mag}$, in 
  accordance with the SDSS internal estimates of the redshift and 
  photometric errors for galaxies.
  
\item We remove the redshifts of all galaxies that fall below the
  magnitude limit $r=17.77$ of the SDSS Main galaxy sample and those
  with declinations less than $+30\degr$,
  for which the depth of the stacked simulation boxes is insufficient
  and which we define as the edge of the mock survey boundary
  
\end{itemize}

There are a number of attributes of the observed survey that are not 
accurately reproduced by the mock survey. The true survey boundary is 
much more complicated than the boundary of our mock catalogue,
and hence a much larger fraction of galaxies lie near an edge and may 
be near an unseen bright galaxy.
Also, the true survey is not spectroscopically complete,
and is less complete in regions of higher density due to the lower
availability of fibres. Therefore, our analysis using
the mock catalogues would underestimate the importance of excluding 
primaries near the survey edge, excluding primaries near the magnitude 
limit of the survey, and implementing a cut based on \Nviol. As these 
issues cannot be addressed well using the mock catalogues, we do not 
implement them in our analysis of the mock catalogue. Their effects 
on the sample are investigated empirically in \S~\ref{holmberg effect 
results section}.

\subsubsection{Tests of our Selection Criteria}%
\label{tests of selection criteria section}

Using the mock catalogues, we quantify the degree to which current
and previous selection criteria accurately identify physical satellites 
of isolated galaxies. We quote the results of the ``Mock A'' catalogue
(see Table~\ref{mock anisotropy stats table}), but the results are
consistent with those from the other mock catalogues.

Interlopers are identified in the mock catalogues as satellites that 
do not belong to the same halo as their selected primary galaxy.%
\footnote{Note that many previous works refer to all unwanted satellites
as ``interlopers'', regardless of the reason for wanting to exclude them
from the sample. We prefer to designate only satellites that are not 
physically associated with their selected primary as ``interlopers'' to 
distinguish them from satellites that are physically associated with
unwanted primaries.} We do not consider primary galaxies as isolated if:
 \begin{enumerate}
 \item they are not the central galaxy of their halo, or%
   \label{non-central primary enum}%
 \item they are not sufficiently more massive than other galaxies 
   within the halo.%
   \label{non-dominant primary enum}%
\end{enumerate}
 
\begin{figure}
\includegraphics[scale=0.5]{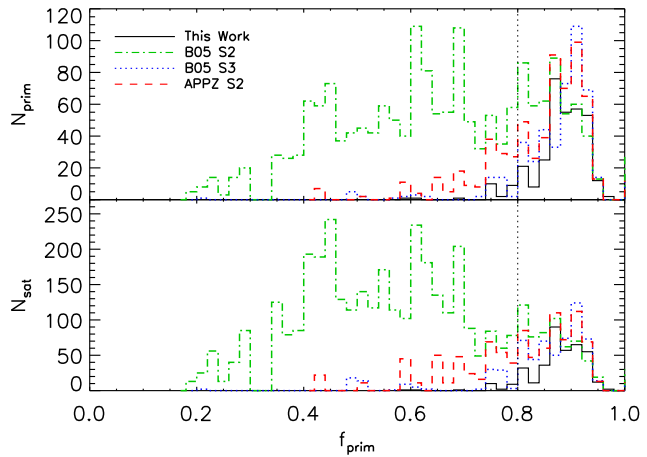}
\caption{%
\textit{(Top)} Histogram of the number of primaries selected
from the mock catalogue as a function of \fprim,
the fraction of the true halo
luminosity that comes from the primary galaxy. The solid lines
represent systems chosen using
our adopted selection criteria,
while the other line styles indicate other representative criteria
(dashed, dotted and dot-dashed for \citetalias{brainerd05}~S2,
\citetalias{brainerd05}~S3 and \citetalias{appz07}~S2).
\textit{(Bottom)} As above, but weighted per satellite in the sample.
The vertical dotted lines denote the \fprim\ below which primaries
are considered non-dominant.%
\label{primary halo luminosity fraction plot}}
\end{figure}

\noindent
Point~\ref{non-central primary enum} is easily determined from the mock
catalogues because we know which galaxy is the central galaxy of each halo.
Point~\ref{non-dominant primary enum} is more difficult to determine
because the CLF formalism assigns luminosities and not masses to
individual galaxies. However, we can estimate the degree to which the 
primary galaxy dominates its halo by calculating the fraction of the 
total halo luminosity that is contributed by the primary galaxy,
\begin{equation}
  \fprim \equiv L_{\mathrm{prim}}/L_{\mathrm{tot}}.
\end{equation}
In the top panel of Figure~\ref{primary halo luminosity fraction plot}
we plot a histogram of the number of selected primaries as a function 
of \fprim\ for four sample sets of criteria (ours, 
\citetalias{brainerd05}~S2, \citetalias{brainerd05}~S3
and \citetalias{appz07}~S2) that span the range of observed
behaviours. In the bottom panel, we weight each primary by its number 
of selected satellites to demonstrate its influence on the observed sample.
Because primaries that contribute less to the luminosity of their halo 
have more satellites, the tail to low fractions is exacerbated.
The histograms are characterised by a symmetric peak centered at
$\fprim = 0.9$ that extends down to 0.8 that we identify as truly
isolated primaries, and a long tail to low values
that we wish to eliminate. Based on examination of these histograms, we 
consider a primary to be ``non-dominant'' if $\fprim < 0.8$ (denoted by 
the vertical line in Figure~\ref{primary halo luminosity fraction plot}).

Table~\ref{selections on mocks table} lists the number of
primaries and satellites selected from the mock catalogue
that pass and fail the above interloper and isolation criteria.
Row 1 indicates the number of primaries selected,
row 2 indicates the fraction of those primaries that are not
the central galaxy of their halo (category \ref{non-central primary enum}
above),
row 3 indicates the fraction of primaries that do not dominate
the dynamics of their halo (category \ref{non-dominant primary enum} above),
and row 4 indicates the fraction of selected primaries that
are truly isolated.
Row 5 indicates the number of satellites selected,
row 6 indicates the fraction of those satellites that are
``multi-homed'', i.e. selected as satellites of more than one primary,
row 7 indicates the fraction of satellites that do not belong
to the same halo as their selected primary (``interlopers''),
row 8 indicates the fraction of satellites that are physically
associated with non-dominant primaries (``\fnondom''),
and row 9 indicates the fraction of selected satellites that
are selected correctly, i.e. they
are physically associated with isolated primaries.

When evaluating criteria based on different bandpasses, we use the
following simple transformations:
for the \citetalias{azpk06} criteria,
which are based on \citetalias{RC3} $B_T$ magnitudes,
we assume a constant $B-{}^{0.1}r=0.6$,
typical of the late-type galaxies they studied;
for the \citetalias{appz07} criteria,
we assume a constant difference between the $z=0$ and
$z=0.1$ $r$-band k-corrections of $0.23~\mathrm{mag}$;
and for the \citetalias{sl04} criteria,
which are based on UK Schmidt $b_J$ magnitudes,
we assume a constant $b_J-{}^{0.1}r=0.7$, intermediate
between the typical values for early and late type
galaxies.

The use of the mock
catalogues to evaluate selection criteria for studies that did not use SDSS
data is not as accurate
as for those studies based on SDSS
because the parameters of the mock catalogues,
such as the typical photometric and redshift error, are specifically
tuned to mimic SDSS.
The photometric errors in \citetalias{zsfw97-holmberg} and the
velocity errors in \citetalias{sl04} are significantly larger.

\begin{table*}
\begin{minipage}{175mm}
\caption{\label{selections on mocks table}%
Results of Applying Selection Criteria to Mock Catalogues}
\begin{tabular}{lcccccccccc}
\hline
{Parameter} & {This Work} &
 {\citetalias{zsfw97-holmberg}} &
 {\citetalias{sl04}} &
\multicolumn{3}{c}{\citetalias{brainerd05}} &
 {\citetalias{azpk06}} &
\multicolumn{2}{c}{\citetalias{appz07}} & \citetalias{ab07}\\
 & & & & S1 & S2 & S3 & & S1 & S2\\
\hline
Selected primaries & $337$ & $135$ & $1516$ &
   $1828$ & $1778$ & $404$ & $114$ & $1595$ & $636$ & $589$\\
\hspace{1em}Non-central primary fraction & $0.000$ & $0.007$ & $0.001$ &
   $0.002$ & $0.002$ & $0.002$ & $0.000$ & $0.003$ & $0.000$ & $0.005$\\
\hspace{1em}Non-dominant primary fraction & $0.071$ & $0.059$ & $0.600$ &
   $0.640$ & $0.724$ & $0.087$ & $0.123$ & $0.656$ & $0.272$ & $0.626$\\
Isolated primary fraction & $0.929$ & $0.941$ & $0.400$ &
  $0.360$ & $0.276$ & $0.913$ & $0.877$ & $0.344$ & $0.728$ & $0.374$\\
\\
Selected satellites & $388$ & $187$ & $2396$ &
   $3461$ & $3852$ & $563$ & $133$ & $3081$ & $980$ & $1112$\\
\hspace{1em}Multi-homed fraction & $0.000$ & $0.000$ & $0.000$ &
  $0.000$ & $0.001$ & $0.000$ & $0.000$ & $0.000$ & $0.000$ & $0.000$\\
\hspace{1em}Interloper fraction & $0.046$ & $0.048$ & $0.037$ &
   $0.036$ & $0.038$ & $0.025$ & $0.023$ & $0.042$ & $0.036$ & $0.049$\\
\hspace{1em}\fnondom & $0.062$ & $0.102$ & $0.691$ &
   $0.767$ & $0.844$ & $0.151$ & $0.218$ & $0.778$ & $0.409$ & $0.754$\\
Correctly-selected satellite fraction & $0.892$ & $0.861$ & $0.290$ &
   $0.216$ & $0.138$ & $0.824$ & $0.759$ & $0.203$ & $0.563$ & $0.224$\\
\hline
\end{tabular}
\end{minipage}
\end{table*}

All sets of criteria do an adequate job of
selecting central galaxies as primaries and of minimizing the fraction 
of interlopers and multi-homed satellites,
with each source of contamination contributing less 
than 5\%\ to each sample\footnote{Although it is in principle possible
to select multi-homed satellites using many of the sets of criteria,
only for \citetalias{brainerd05}~S2 does this situation ever occur
in practice.}.
However, the fraction of the sample that
lies around non-dominant primaries (\fnondom) extends from a low of 6\%\ for 
our adopted criteria to almost 85\%\ in the case of
\citetalias{brainerd05}~S2.
We have examined by eye the fields surrounding a subset of the primary galaxies
in our SDSS sample in order to confirm that this is an accurate measure
of the degree of isolation of the sample and estimate
that $7\%$ of our primaries
are members of groups, in very good agreement with the value
that we derive from the mock catalogues.

The greatest single predictor of the magnitude of \fnondom\ 
is whether non-satellites are permitted to lie within the satellite cylinder.
In other words, criteria with $\magsat > \magin$ 
generally fail to select isolated
primaries.  Although we do not evaluate the effects of \Nviol\ using the mock
catalogues, neglecting to account for spectroscopic incompleteness
is the other factor that can result in galaxies larger than satellites
lying within the satellite cylinder; we therefore expect that this
also has a significant effect on \fnondom.
The impact of non-dominant primaries
on the results will be discussed in detail in
\S~\ref{comparison with previous results section}.

One might ask how sensitive these conclusions are to the CLF method
used to assign luminous galaxies to dark matter halos versus, for example,
a semi-analytic model.
Motivated by the halo luminosity function found in a particularly
discrepant semi-analytic model of
\citet[their figure 5]{eke-etal04-luminosity},
we have tested a mock catalogue where we arbitrarily doubled the
luminosity of the central galaxy.
The resulting \fnondom\ changes by less than $0.1$ for the vast majority of
selection criteria tested.
Given the insensitivity
of \fnondom\ to such relatively dramatic departures from our method
of assigning luminous galaxies to dark matter halos,
we feel confident that
our conclusions regarding the selection criteria used in previous
studies are robust.

We have optimized the parameters we use for the selection criteria
using the mock catalogues to maximise both the size of the sample
(Row 5) and the fraction of the sample that passes all of the checks 
(Row 9). In particular, all parameters in
Table~\ref{selection criteria parameters table} were varied in turn
and the new value kept if it increased the sample size without
a correspondingly large increase in either the incorrectly-selected
fraction (the inverse of the value in Row~9),
the interloper fraction, or the fraction of satellites
around non-dominant primaries; or, conversely, if it
decreased the incorrectly-selected fraction without a
correspondingly large decrease in sample size. When in doubt, we
erred on the side of more restrictive criteria. This process was
repeated until convergence.

\subsection{The Sample}%
\label{the sample section}

The following quality cuts are imposed on the sample:
\begin{enumerate}
 \item Satellites within
$35\,\hikpc$ of the primary are removed due to the known sky subtraction problem
around bright sources 
\citep{mandelbaum05-systematicerrors,mandelbaum06-haloshape} and the
possibility of bright knots in the outer regions of a galaxy being
mistakenly deblended as separate galaxies.
 \item As the interloper fraction rises at large projected radius,
and the radius at which interlopers dominate increases with halo mass,
we eliminate satellites of
less luminous ($M_r - 5 \log h > -21.1$, i.e.~fainter than the
median) intermediate- and early-type primaries
and of all late-type primaries which lie
beyond a projected radius of $345~\hikpc$; this choice of parameters
is justified below.
\end{enumerate}

The resulting sample of primary and satellite galaxies is given in
Table~\ref{sample table}.
The full sample contains $866$ satellites of $722$ primaries;
$311$/$138$/$273$ of the primaries are classified as
early/intermediate/late-type
hosting $378$/$167$/$321$ satellites.

\begin{table*}
 \vbox to220mm{\vfil Landscape table to go here.
 \caption{\label{sample table}}\vfil}
\end{table*}

The following cuts are further imposed on systems used to measure
the anisotropy around primary galaxies:
\begin{enumerate}
 \item Primaries that do not have a measured position angle (PA)
in the SDSS DR6 database are excluded; this cut excludes 11 primaries.
 \item Primaries with isophotal axis ratios $b/a > 0.8$ are excluded.
Galaxies with nearly circular isophotes have poorly-constrained
PAs. In addition, any anisotropy that exists in three dimensions
gets washed out in projection as the system is viewed close to
its axis of symmetry. The numerical choice of $b/a > 0.8$ for
this cutoff is motivated
in \S~\ref{holmberg effect results section} and the effects of
changing this value are discussed.
\end{enumerate}
Unless otherwise specified, this subsample is the sample referred to for
the remainder of the paper, and
contains $440$ satellites of $372$ primaries. 
In \S~\ref{lss angle section}, we analyze the distribution of
satellites with respect to the large scale structure (LSS).
For those purposes, the following further cuts are imposed instead:
\begin{enumerate}
 \item Systems within a projected radius $3000~\hikpc$ of the
edge of the spectroscopic survey footprint are excluded in order to
ensure that the survey boundary does not introduce a bias in the
derived LSS axis.
 \item Primaries for which the LSS axis is undefined because there
are no galaxies within the cylinder used to define the axis or for which the
LSS axis ratio is greater than $0.9$ are excluded.
\end{enumerate}
This subsample contains $572$ satellites of $480$ primaries.

\begin{figure*}
\includegraphics[scale=0.4]{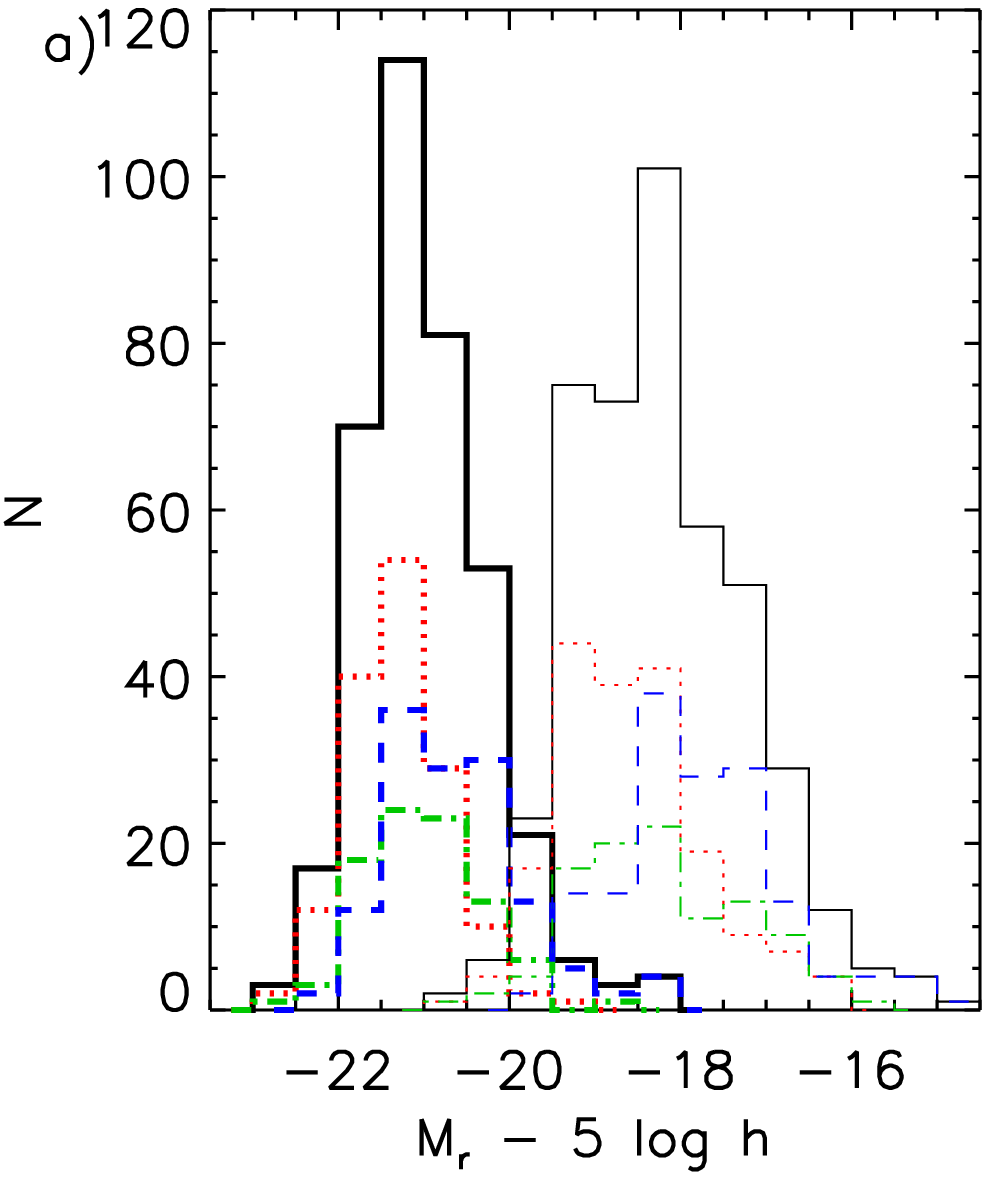}%
\includegraphics[scale=0.4]{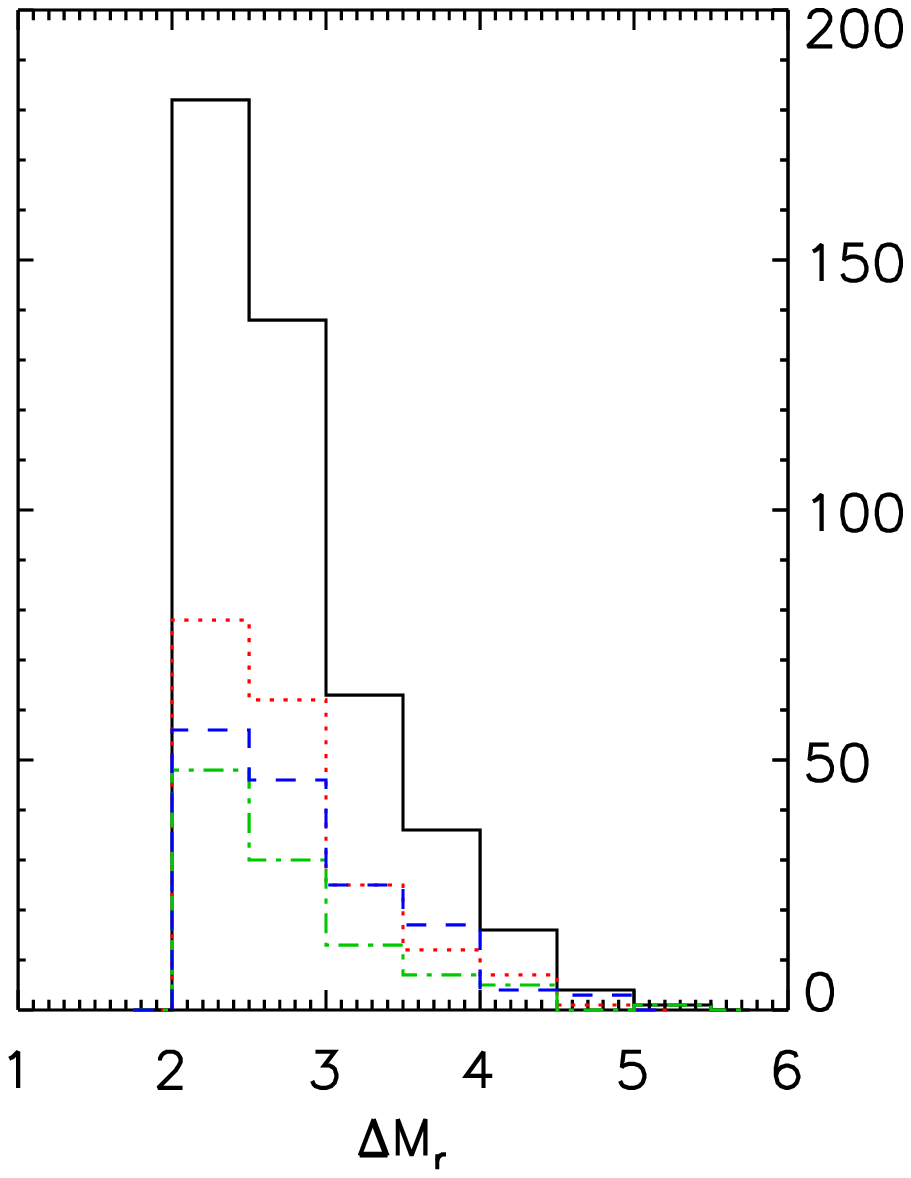} 
\scalebox{0.46}[0.43]{\includegraphics{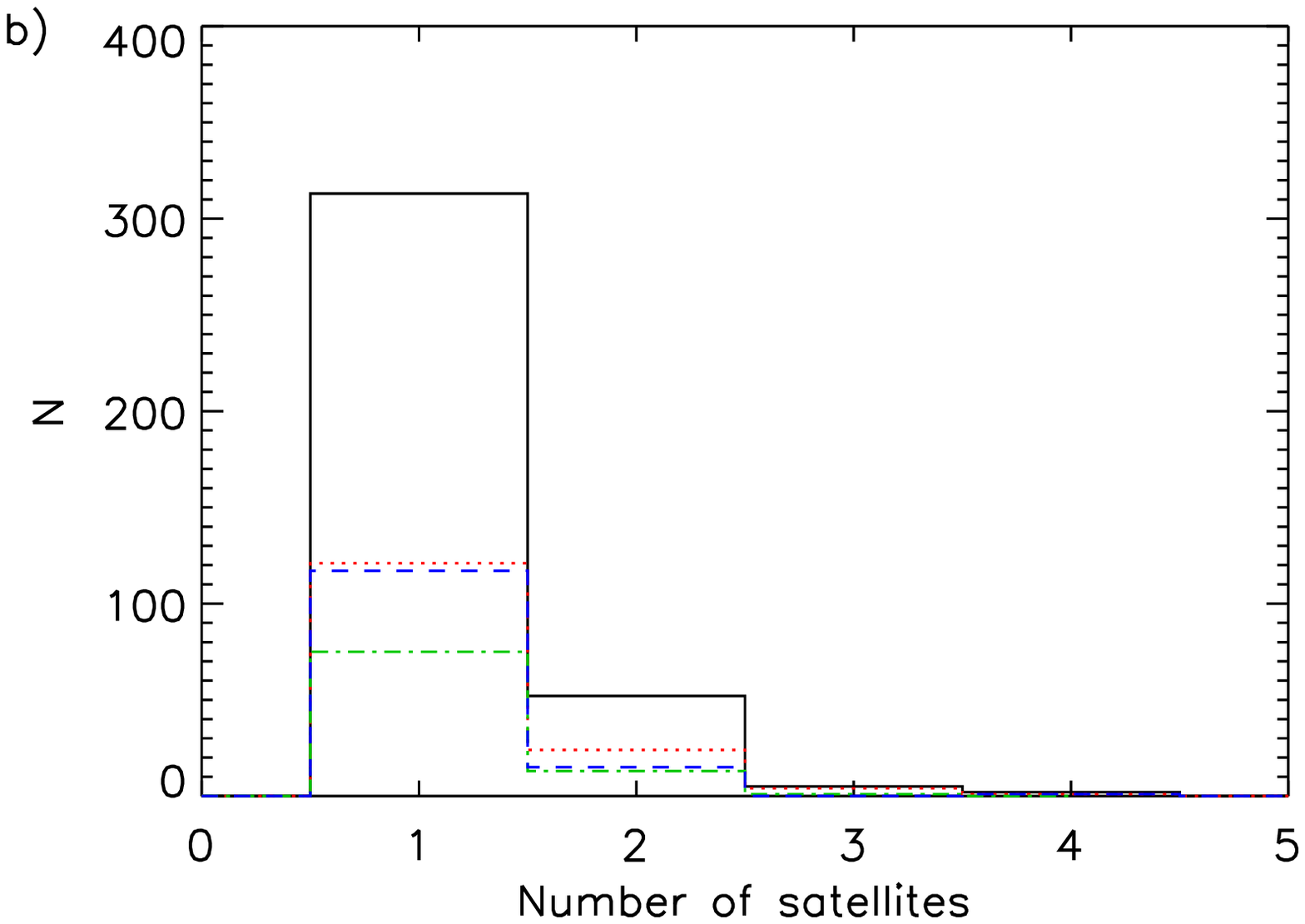}}\\
\includegraphics[scale=0.45]{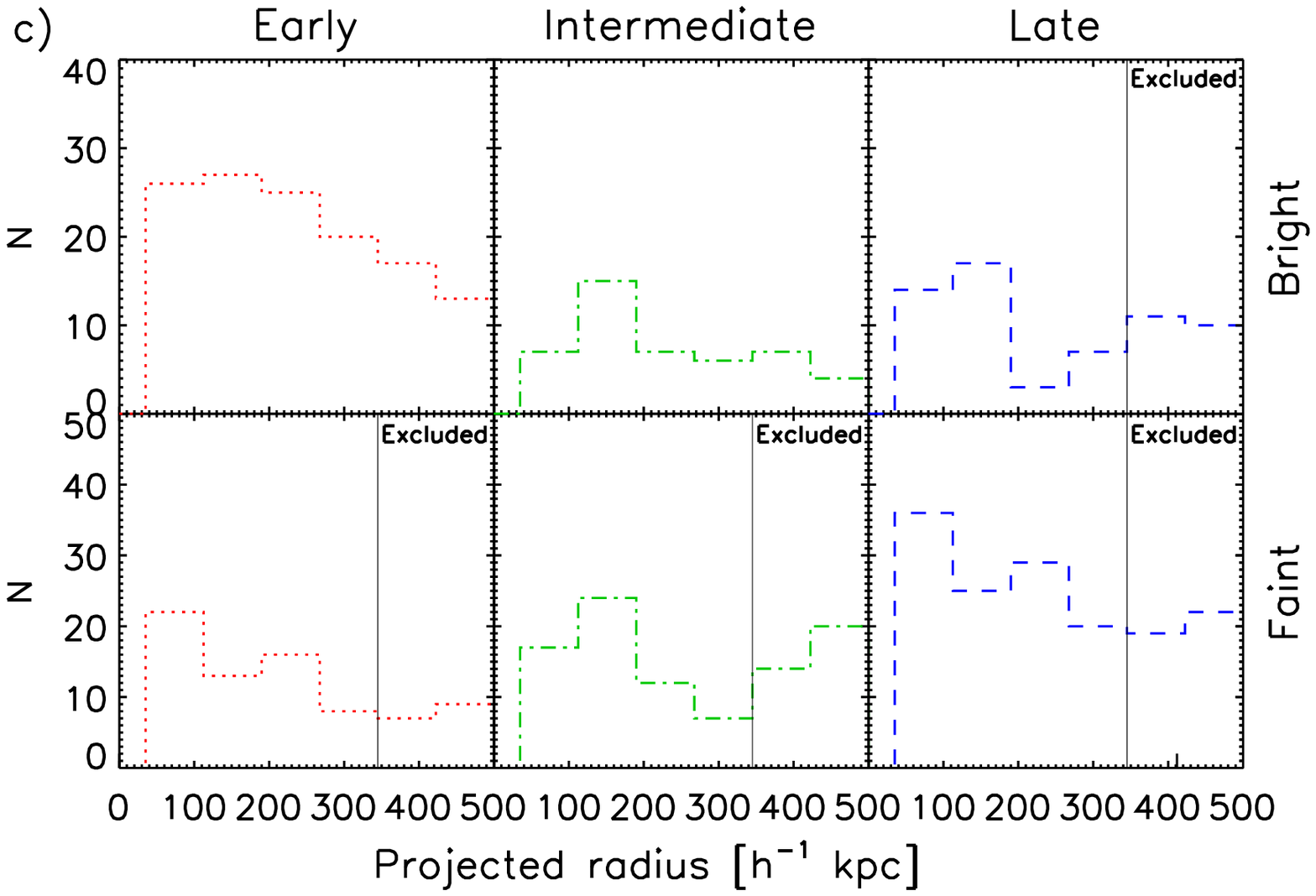} 
\includegraphics[scale=0.45]{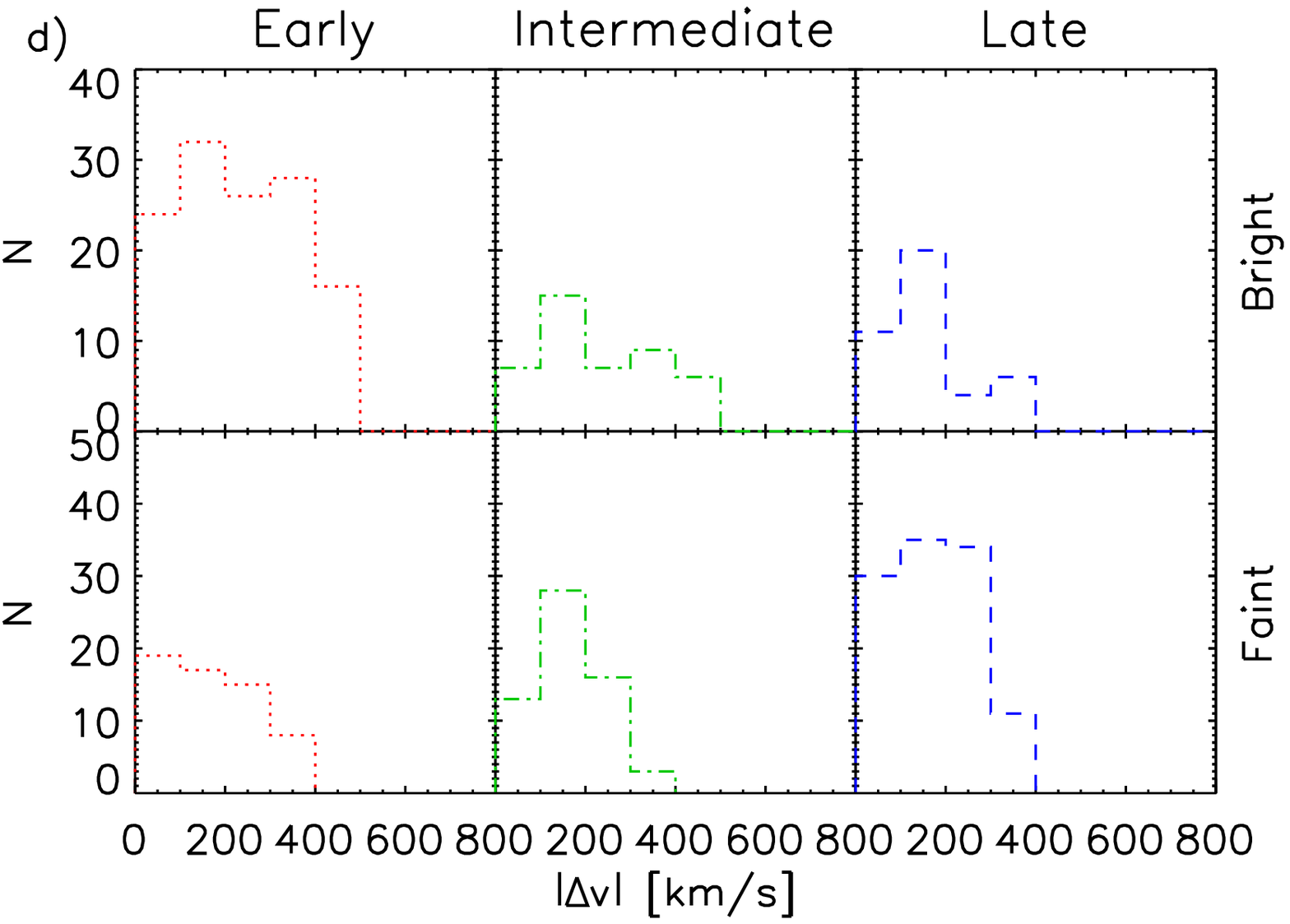}
\caption{%
\textit{(a)} 
Absolute magnitude distributions of primary galaxies (thick lines)
and of satellites (thin lines) (left panel) and the difference
between the magnitude of each satellite and its primary (right panel).
The red/dotted lines shows the distribution for early-type primaries,
the green/dot-dashed lines shows the distribution for intermediate-type
primaries,
and blue/dashed lines refer to late-type primaries. Note that the satellites
are also separated according to the type of their primary, not according to
their own type.
\textit{(b)} Distribution of number of satellites per primary.
Line colours/styles indicate the classification of the
primary galaxy as in (a).
\textit{(c)}
Distribution of projected radial separation of satellites from
their primary.
Left panels show
satellites of early-type galaxies,
middle panels show satellites of intermediate-type galaxies,
and right panels show
satellites of late-type galaxies. Top panels show satellites of
primary galaxies more luminous than the median
($M_r - 5 \log h < -21.1$) while bottom
panels show satellites of the less-luminous primary galaxies
($M_r - 5 \log h > -21.1$).
The vertical lines mark the maximum radial extent for satellites.
These satellites are not included in the remaining plots.
\textit{(d)} Distribution of velocity differences between 
satellites and their primary. Panels are as in (c).%
\label{sample distribution figure}}
\end{figure*}

The distribution of luminosities, number of satellites per primary,
radial separations and velocity differences are shown in
Figure~\ref{sample distribution figure}.
In Figure~\ref{sample distribution figure}a we plot the absolute
magnitude distributions of the primary and satellite galaxies,
along with the magnitude difference.
The median absolute magnitude is $M_r - 5 \log h = -21.1$.
The typical primary has
a luminosity similar to the Milky Way, while the typical satellite
is $\sim2.5$~magnitudes fainter, slightly brighter than the LMC.
Figure~\ref{sample distribution figure}b shows the number
of satellites per primary. Most primaries are surrounded by
only one satellite.
Figure~\ref{sample distribution figure}c presents the distribution
of radial separation between satellites and primary galaxies for
samples split by the type and luminosity of the primary
(the median absolute magnitude, $-21.1$ is used as the cutoff
between the ``bright'' and ``faint'' subsamples).
In this panel only, we include the distant satellites of faint
primaries.
The radial distribution of true satellites declines with
radius, while the distribution of interlopers increases with
radius \citep{chen-etal06}. The crossover between these regimes
occurs at the edge of the dark matter halo, and therefore depends
on galaxy mass. Figure~\ref{sample distribution figure}c shows
no evidence for an increase in the number of outer satellites
due to interlopers around bright early- and intermediate-type
primaries. However, such an upturn
is evident around faint primaries of both morphologies and
all late-type primaries beyond $\Delta R > 345~\hikpc$.
Therefore, around these primaries we exclude satellites at projected radii
$\Delta R > 345~\hikpc$ in all other panels and remaining plots;
however, 
we have confirmed that if we include these satellites, our results
are qualitatively unchanged.
Figure~\ref{sample distribution figure}d presents the distribution
of velocity differences between satellites and their primaries.
The velocities of selected satellites cluster strongly about the
velocity of the primary indicating that they are indeed physically
associated. The velocity dispersion of satellites around bright
primaries is higher than around faint primaries due to their
larger mass
\citep[see, e.g.][and references therein]{conroy-etal07,nfc08}.

\section{Results}\label{results section}
\subsection{Distribution About The Primary Galaxy}%
\label{holmberg effect results section}

\begin{figure*}
\includegraphics[scale=0.9]{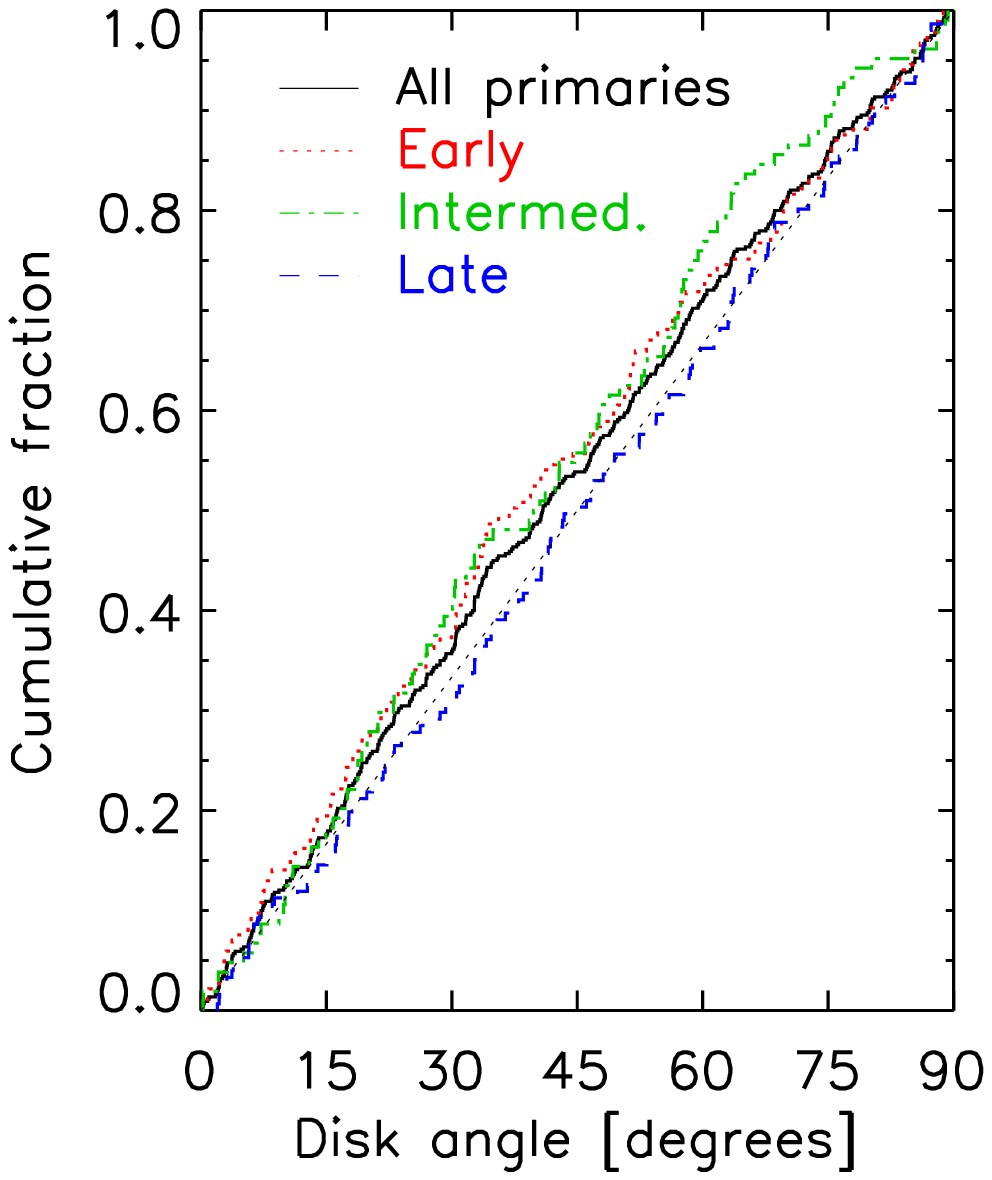}%
\includegraphics[scale=0.9]{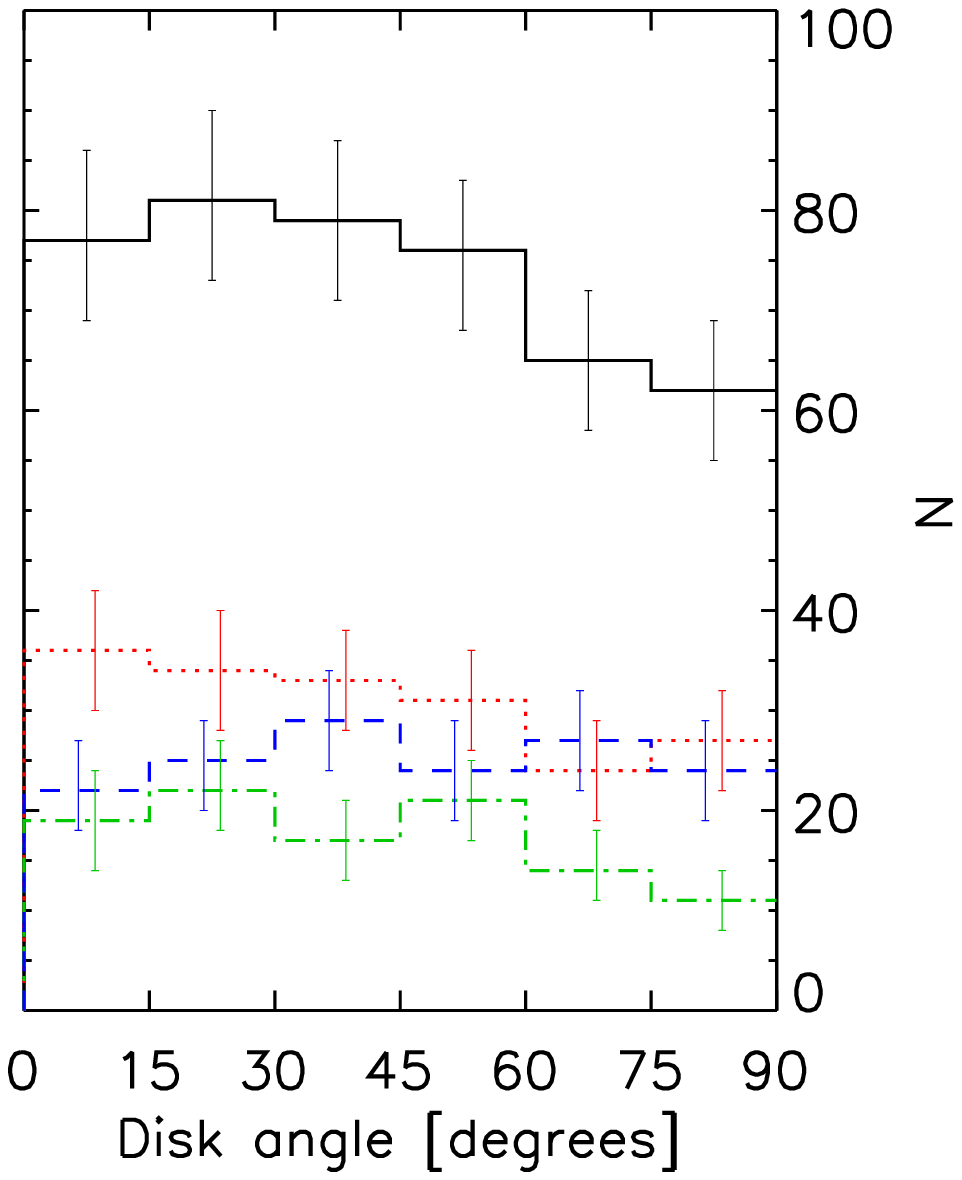}
\caption{%
\textit{(Left)} Cumulative distribution of angle between
the major axis of the primary and the location of the satellite
(``disc angle''). The thick
black/solid, red/dotted, green/dot-dashed, and blue/dashed lines refer to the
distribution of satellites
around all primaries, early-type primaries, intermediate-type primaries,
and late-type primaries
respectively. The thin dotted line shows the distribution expected
if satellites were isotropically distributed.
\textit{(Right)} Differential distribution of the disc angles.
The error bars are determined by bootstrap resampling
of all primary galaxies.%
\label{disc angle distribution figure}}
\end{figure*}

We define the ``disc angle'' as the angle
between the $r$-band isophotal position angle of the major axis
of the primary and the
PA of the great circle between the primary and satellite
(see Figure~\ref{selection cylinder figure}); we fold this angle
into the range $0$--$90\degr$.
If satellites are distributed isotropically around the primaries then
the distribution of disc angles is uniform with a mean of $45\degr$.

In Figure~\ref{disc angle distribution figure} we plot the cumulative
and differential disc angle distributions. The full sample shows a
tendency to lie at small disc angles, i.e.~for the satellites to lie
near the major axis of their parent galaxy. The hypothesis that the
angles are distributed randomly is ruled out at a greater than
$2\sigma$ level:
according to the Kolmogorov-Smirnov (KS) test, the probability is
$0.06$ with a mean disc angle of $42\fdg5\pm1\fdg2$.

The satellites surrounding different types of primaries show
different angular distributions:
there is a clear excess concentration of satellites along the major
axis for early-types, hints of a major-axis excess
around intermediate-types (the measured
magnitude of the anisotropy is in fact larger
than around early-types, but is detected at less than $2\sigma$),
and no detectable anisotropy around late-types.
The mean disc angles are $41\fdg3\pm0\fdg9$, $40\fdg5\pm2\fdg5$
and $45\fdg5\pm2\fdg1$ around
early, intermediate and late-types respectively,
with KS test probabilities of being
drawn from a random distribution of $0.03$, $0.10$, and $0.99$.
When the samples are combined these
effects counteract each other, with the isotropic satellites of
late-types diluting the major-axis alignment seen around the
early-types.
The KS test results, the mean and median disc angles,
and the polar fraction (the fraction of satellites with disc angles
larger than $45\degr$) are listed in
Table~\ref{satellite anisotropy stats table}.
The quoted uncertainties are determined using bootstrap resampling of
the primaries and represent 68\%\ confidence intervals.
All statistics reinforce the same conclusion:
satellites of spheroidal early-type galaxies tend to lie near the
long axis of the spheroid, while satellites of disc galaxies
are isotropically distributed around blue late-types but show hints
of lying near the disc around red intermediate-types.
We have verified using the mock catalogues, which have
intrinsically isotropically-distributed satellites, that we
could not have measured this level of anisotropy around
the early-type primaries if it were not physically
present; see Appendix~\ref{appendix:mockdist}.

\begin{table*}
\caption{\label{satellite anisotropy stats table}%
Anisotropy of Satellite and Primary Distributions}
\begin{tabular}{lcccc}
\hline
{Parameter} & {Full sample} & {Early-type primaries} &
 {Intermediate-type primaries} & {Late-type primaries}\\
\hline
\multicolumn{4}{l}{Disc angle:}\\
\hspace{1em}KS probability & $0.06$ & $0.03$ & $0.10$ & $0.99$\\
\hspace{1em}Mean disc angle [\degr] & $42.5\pm1.2$ &
  $41.3\pm0.9$ & $40.5\pm2.5$ & $45.5\pm2.1$\\
\hspace{1em}Median disc angle [\degr] & $40.9^{+1.9}_{-1.6}$ &
  $37.1^{+3.8}_{-3.5}$ & $40.5^{+4.1}_{-7.8}$ & $43.6^{+5.6}_{-2.1}$\\
\hspace{1em}Polar fraction & $0.46\pm0.02$ & $0.44\pm0.03$ &
  $0.44\pm0.05$ & $0.50\pm0.04$\\
\hspace{1em}Early-type satellite mean & $42.5^{+4.5}_{-4.3}$ &
  $36.6^{+6.0}_{-6.7}$ & $49.5^{+6.6}_{-7.4}$ & $47.7^{+11.7}_{-10.6}$\\
\hspace{1em}Intermediate-type satellite mean & $40.8^{+3.3}_{-3.7}$ &
  $42.1^{+4.0}_{-4.5}$ & $43.0^{+7.9}_{-7.6}$ & $28.7^{+7.3}_{-6.9}$\\
\hspace{1em}Late-type satellite mean & $42.8\pm1.3$ &
  $41.6\pm2.3$ & $38.9^{+2.6}_{-2.8}$ & $46.2\pm2.1$\\
\multicolumn{4}{l}{LSS angle:}\\
\hspace{1em}KS probability & $0.56$ & $0.76$ & $0.49$ & $0.40$\\
\hspace{1em}Mean LSS angle [\degr] & $44.6\pm1.2$ & $44.0^{+1.5}_{-1.7}$ &
  $43.0^{+3.0}_{-3.2}$ & $46.4\pm2.2$\\
\multicolumn{4}{l}{Disc/LSS angle:}\\
\hspace{1em}With satellites -- KS probability & $0.06$ & $0.4$ & $0.05$ & $0.10$\\
\hspace{2em}-- mean & $41.8\pm1.4$ &
  $43.4\pm2.2$ & $39.8^{+2.5}_{-2.2}$ & $41.7\pm2.3$\\
\hspace{1em}All isolated -- KS probability & $0.0001$ & $0.01$ & $0.0006$ & $0.17$\\
\hspace{2em}-- mean & $42.6\pm0.7$ &
  $42.1\pm1.2$ & $41.3\pm1.2$ & $43.6\pm0.9$\\
\hline
\end{tabular}
\end{table*}

Our measured alignment must be a lower limit on the intrinsic
three-dimensional alignment for several reasons.
Firstly, images of triaxial elliptical galaxies contain isophotal
twists due to projection effects, which introduces scatter
between the isophotal PA we use and the intrinsic 3D major axis.
Secondly, galaxies seen closer to their symmetry axis have their anisotropic
signal diluted. Finally, interlopers may dilute the signal.

\begin{figure}
\includegraphics[scale=0.48]{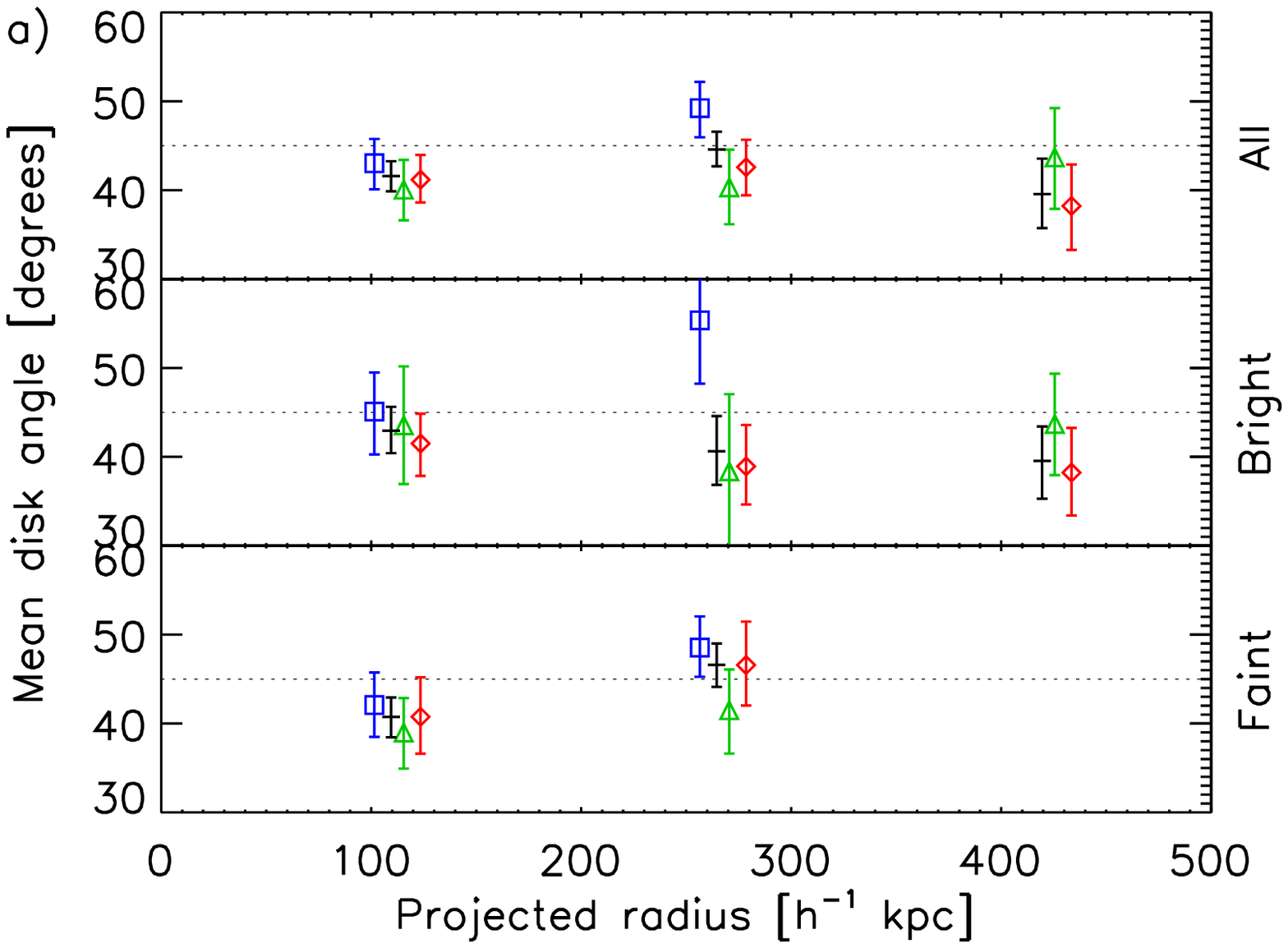}
\includegraphics[scale=0.28]{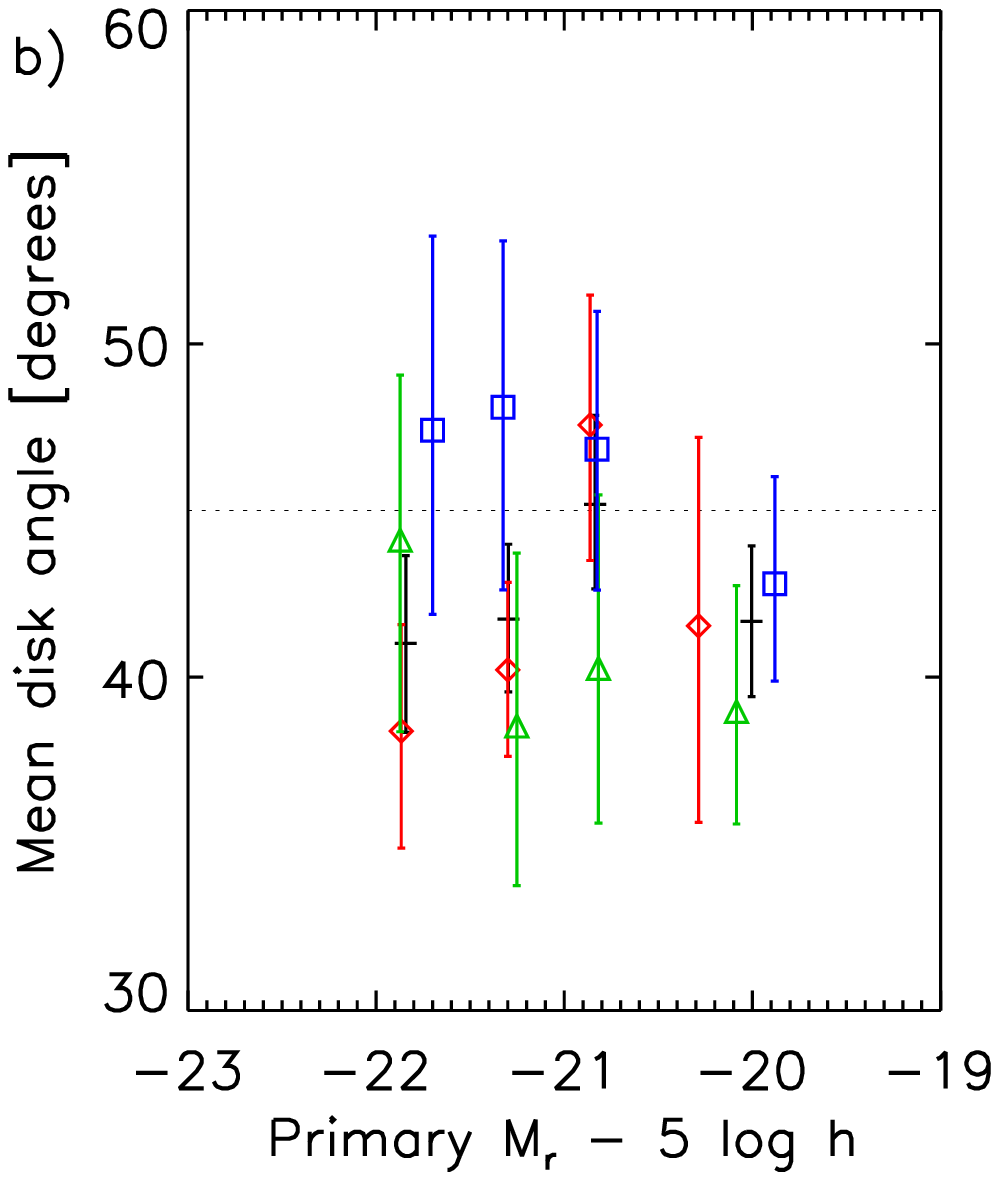}%
\includegraphics[scale=0.28]{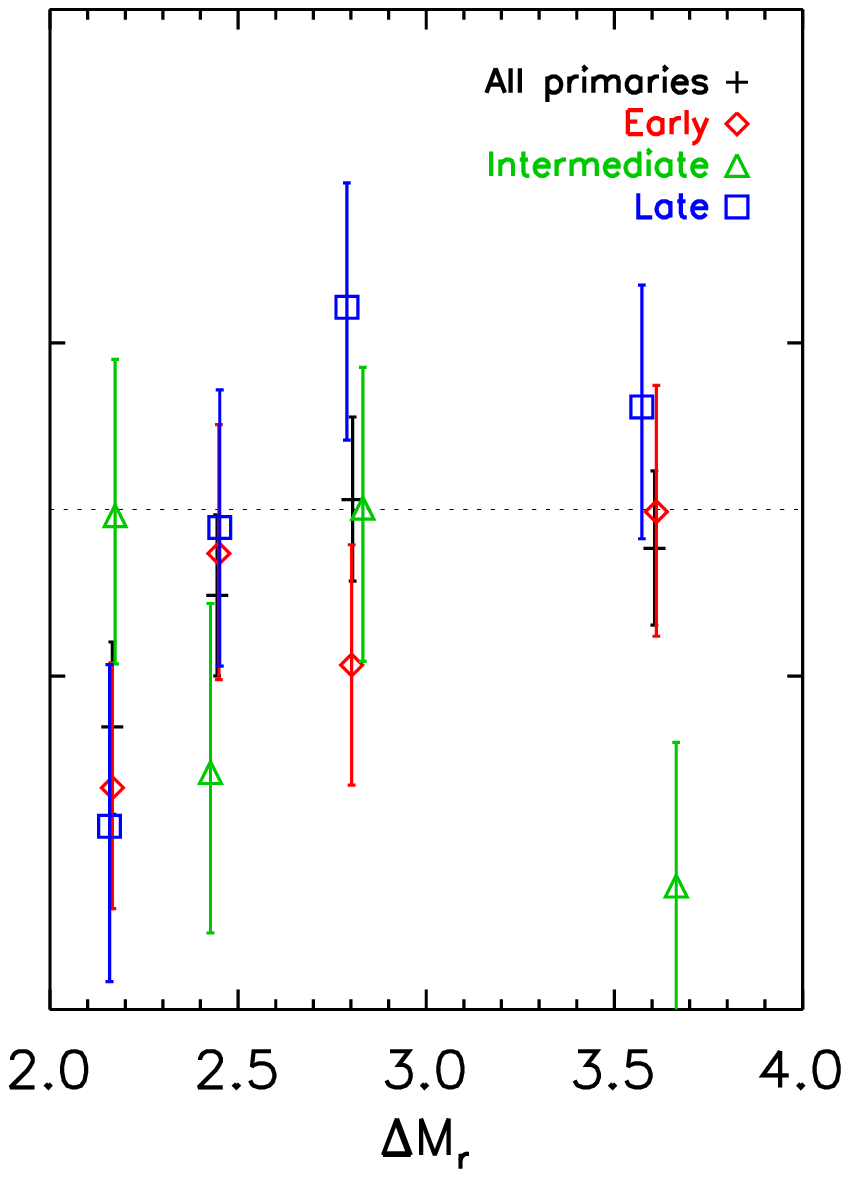}%
\includegraphics[scale=0.28]{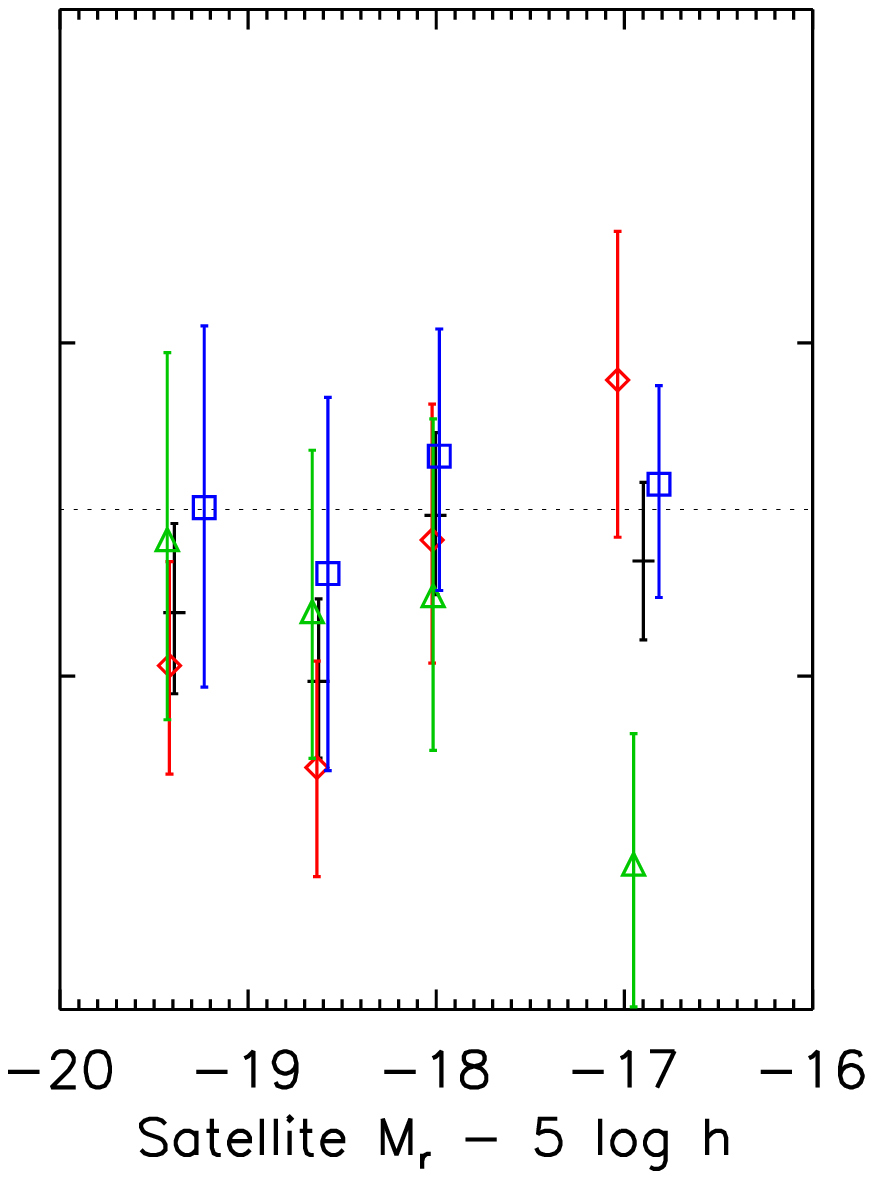}
\caption{%
\textit{(a)}
Mean disc angle of satellites as a function of projected
separation from their primary.
Black plusses, red diamonds, green trianges and blue squares
refer to satellites of all primaries, early-type primaries,
intermediate-type primaries, and late-type primaries respectively.
Radial bins are $155~\hikpc$ wide and are plotted at the central
bin radius with different symbols offset slightly for clarity.
The top panel contains all satellites while the middle and bottom
panels contain satellites of primaries with $M_r - 5 \log h < -21.1$
and $M_r - 5 \log h > -21.1$ respectively.
\textit{(b)} Mean disc angle of satellites as a function
of the absolute magnitude of the primary (left), the magnitude
difference between the primary and satellite (middle),
and the absolute magnitude of the satellite (right).
Bins are chosen to have approximately equal number of satellites
per bin and are plotted at the mean magnitude.
\label{diskangle vs radius mag figure}}
\end{figure}

In Figure~\ref{diskangle vs radius mag figure}a, we plot the mean disc
angle of satellites as a function of projected separation
from their primary.
Satellites are binned in 3 annuli spaced evenly in radius
from $35\,\hikpc$ to $500\,\hikpc$, and are separated into
those more luminous and less luminous than the median
$M_r - 5 \log h = -21.1$. There is no significant
trend with radius.
Our intermediate-separation bin covers approximately
the same radii as the outermost bin of \citetalias{ab07}, in which
they detect minor-axis alignment around late-type primaries;
our mean disc angle around late-types at these radii is also greater
than $45\degr$, but not at a statistically significant level.
While we do not detect
the major-axis alignment that they see at small radii,
a large number of systems in their innermost radial bin fall
within the $35~\hikpc$ region that we exclude to avoid contamination
from H\,\textsc{ii} regions in the outskirts of the parent galaxy;
if such regions are mistakenly included as
satellites, they will bias the result towards major-axis alignment.

In Figure~\ref{diskangle vs radius mag figure}b, we compare the anisotropy
as a function of the luminosity of the primary,
luminosity of the satellite, and of the difference
in magnitude between the primary and satellite.
The widths of the bins are chosen such that there are similar
numbers of satellites in each bin; in all panels, the symbols are
plotted at the mean luminosity or mean $\Delta M_r$ of the galaxies
in the bin.
No clear trend is apparent as a function of primary luminosity.
The anisotropy shows a weak dependence on the magnitude
difference between the primary and satellite, although the
sense of the trend differs for different primary types.
Around early-type primaries, those satellites that are not much
fainter than their primaries show a stronger major-axis alignment
than the satellites that are much fainter, while around
intermediate-type primaries the opposite trend holds.
Around late-type primaries,
those satellites that are bright relative to their primary show $\sim 2\sigma$
major-axis alignment, while the faintest $50\%$ of the satellites
relative to their primary do not (or, if anything, show a \textit{polar}
distribution, although not at a statistically significant level).
These trends could either reflect a dependence on the degree to which the
primary dominates, i.e.~it could truly depend on the magnitude
difference, or it could reflect a dependence on the luminosity of the
satellite. The right panel of Figure~\ref{diskangle vs radius mag figure}b
reveals that the anisotropy around intermediate-type galaxies
can equally well be explained as being a function of satellite luminosity,
while the angular
distributions around early- and late-type primaries are not; therefore,
for these galaxies,
the trends seen in the middle panel truly reflect a dependence on the
relative dominance of the primary.

\begin{figure}
\includegraphics[scale=0.5]{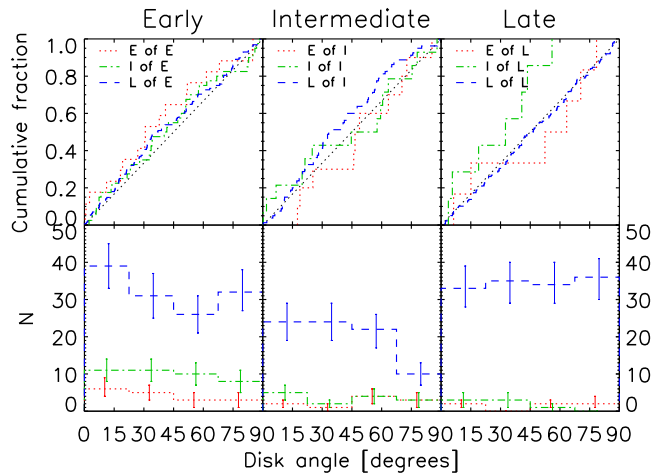}
\caption{\label{disc angle vs satellite type figure}%
Cumulative \textit{(top)} and differential \textit{(bottom)}
distributions of disc angle as a function of morphological type
of both the primary and satellite.
The three panels specify the type of the primary, while line colours
and styles denote the type of the satellite.%
}
\end{figure}

\citet{kg06} found that only the early-type satellites of M31 have a
polar alignment.
\citetalias{yang-etal06} and \citet{faltenbacher-etal07}
found that the red satellites of red primaries show stronger
major-axis alignment than the blue satellites,
and \citetalias{sl04} found that quiescent
satellites show stronger anisotropy than star-forming satellites.
To determine if these signals are evident in our sample, we split the sample by
the galaxy type of both the primary and satellite and plot the
distributions in Figure~\ref{disc angle vs satellite type figure}.
Mean disc angles as a function of satellite type
are given in Rows~5 through 7 of
Table~\ref{satellite anisotropy stats table}.
Although we see small deviations for particular subsamples
(e.g.~the early-type vs. late-type satellites of intermediate-type
primaries, or the intermediate-type satellites of late-type primaries),
none are statistically significant due to the small number of
early- and intermediate-type satellites.

\begin{figure}
\includegraphics[scale=0.5]{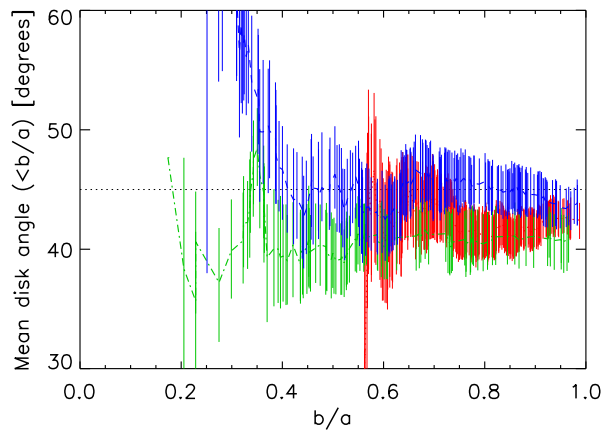}
\caption{%
Mean disc angle for satellites of primaries with isophotal
axis ratios less than or equal to the plotted abscissa.
One error bar is plotted at the location of each primary galaxy.
Colors/line styles are as in Figure~\ref{disc angle distribution figure}.%
\label{disc angle vs axis ratio}}
\end{figure}

In \S~\ref{the sample section},
we excluded primary galaxies with isophotal axis ratios
$b/a > 0.8$.
Our choice of cutoff is motivated by
Figure~\ref{disc angle vs axis ratio}, where we plot the mean disc
angle around primaries with axis ratios less than or equal to
the plotted abscissa.
A cutoff at $b/a=0.8$ provides enough systems that the results have
converged, while avoiding the dilution in the signal seen at higher
values of $b/a$.
For a circular disc galaxy of intrinsic thickness $0.2$, this corresponds
to an inclination of $43\degr$.

\begin{figure*}
$\begin{array}{c@{\hspace{.1in}}c}
\includegraphics[scale=0.5]{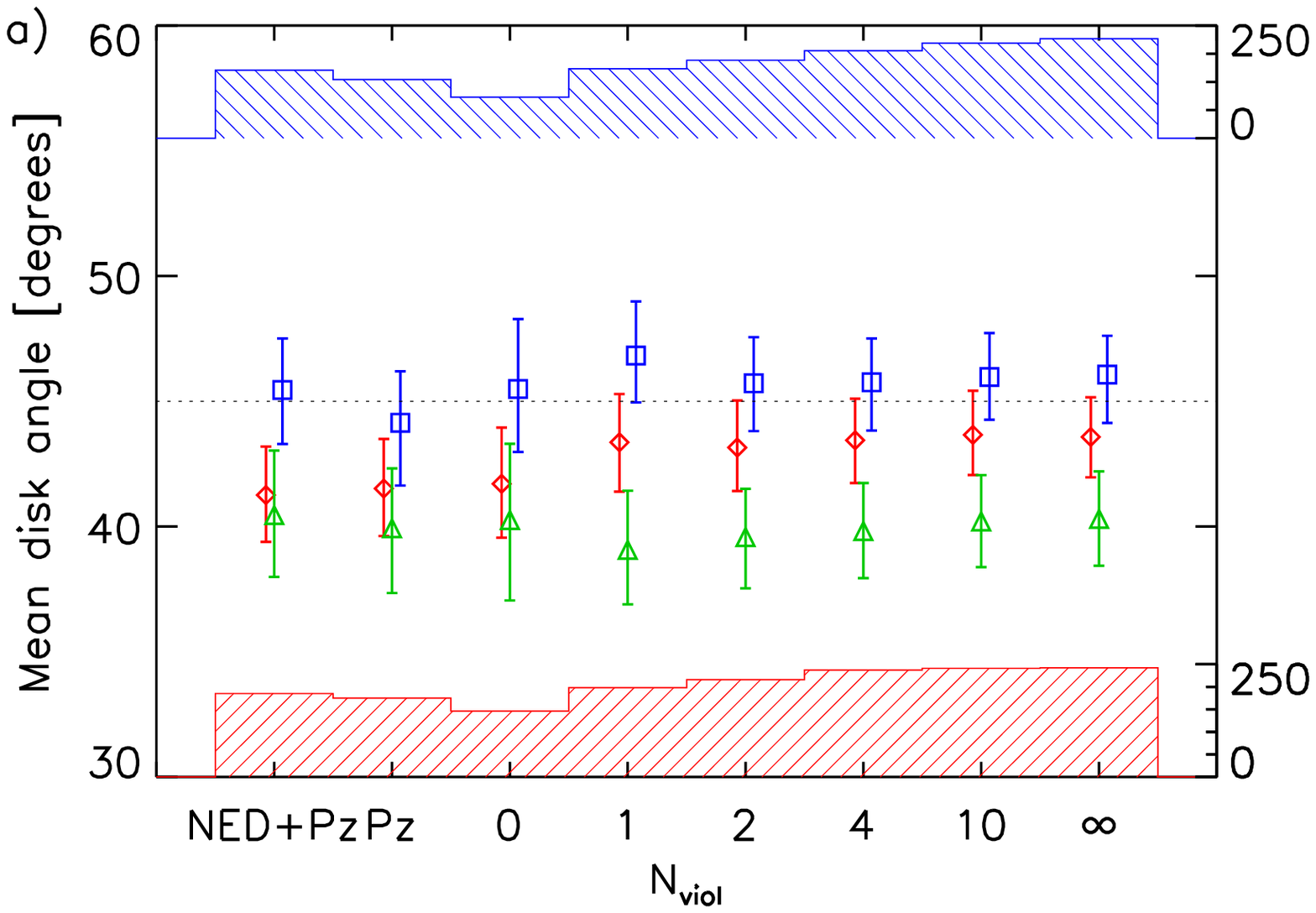} &
\includegraphics[scale=0.5]{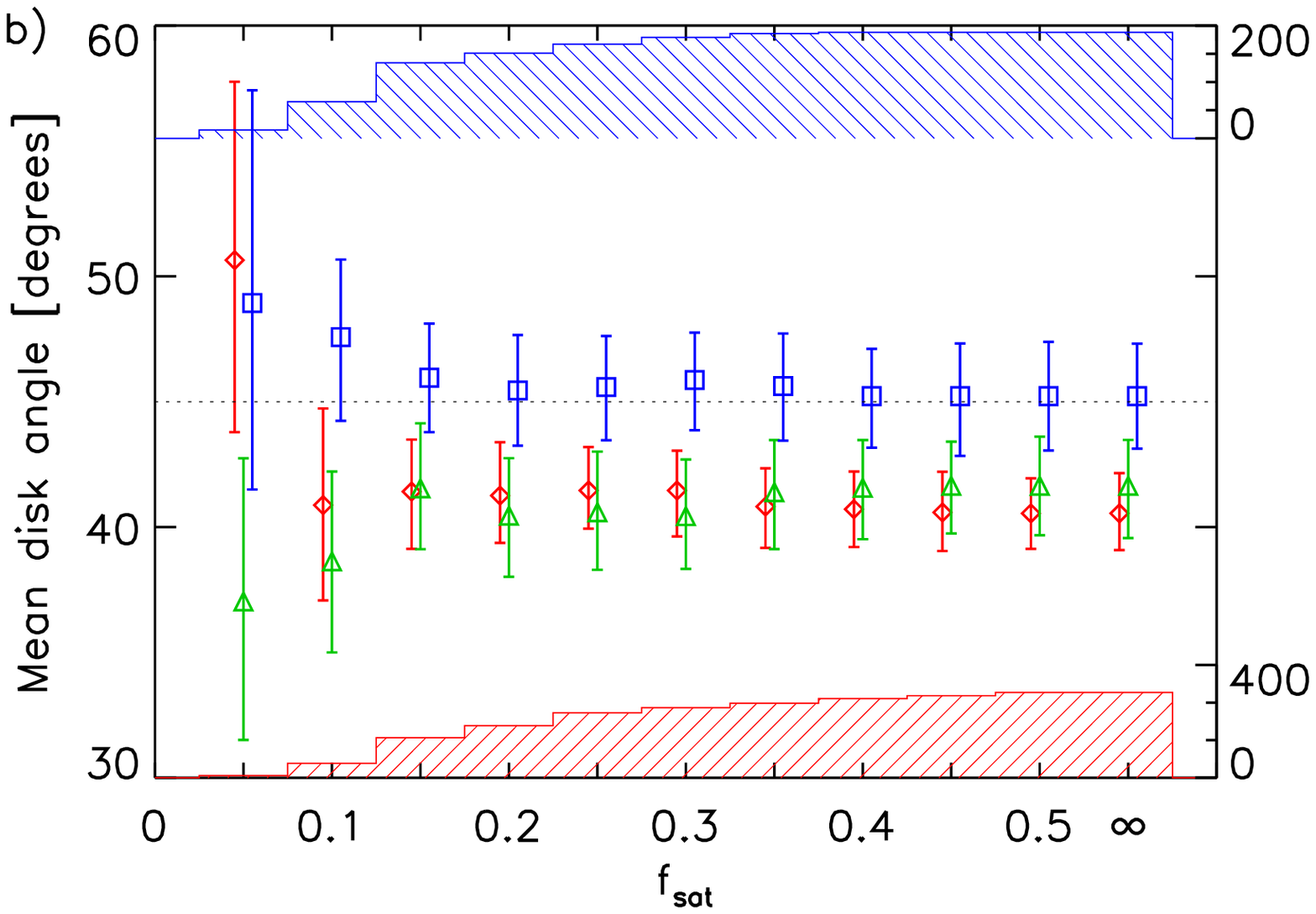}\\
\includegraphics[scale=0.5]{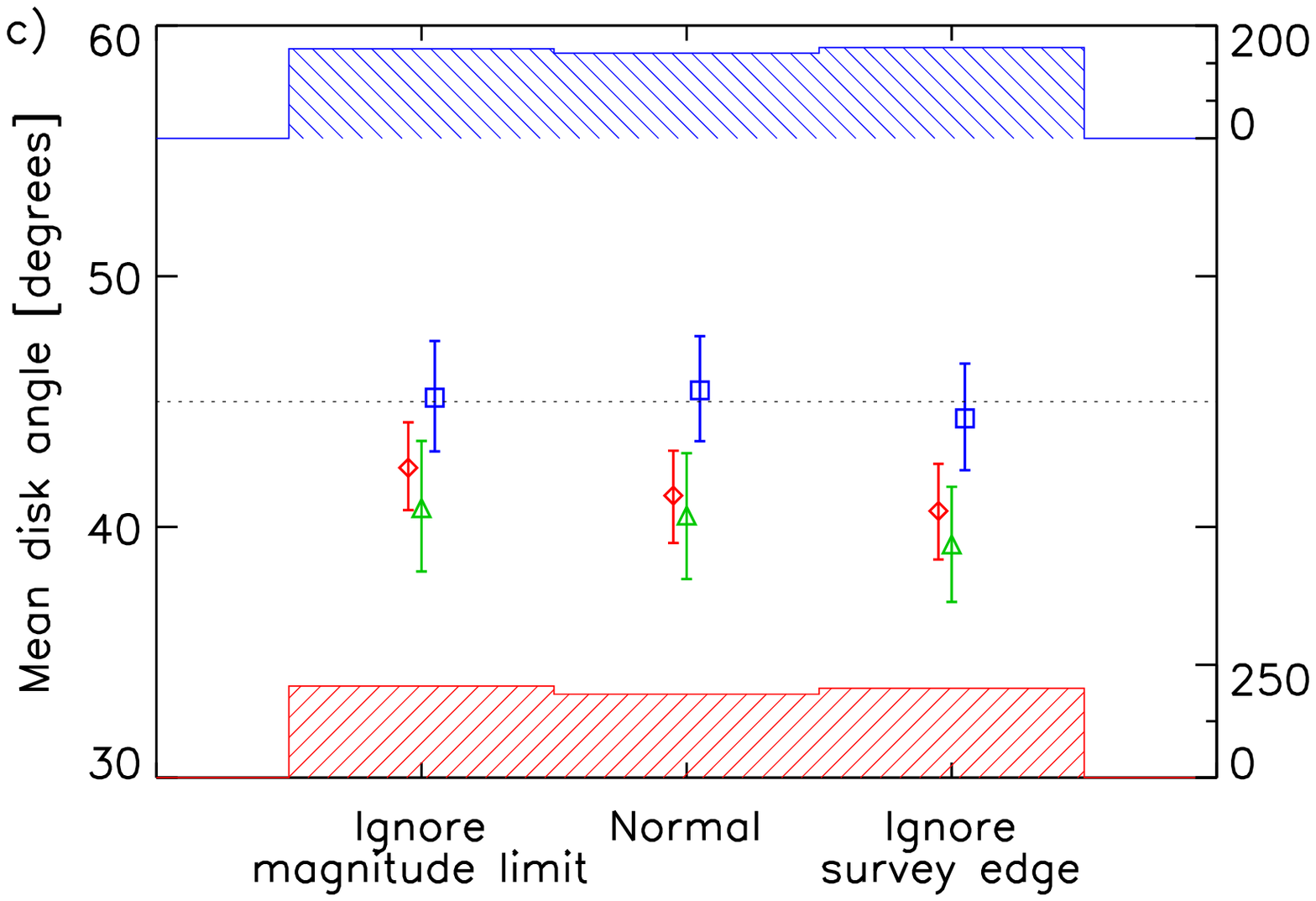} &
\includegraphics[scale=0.5]{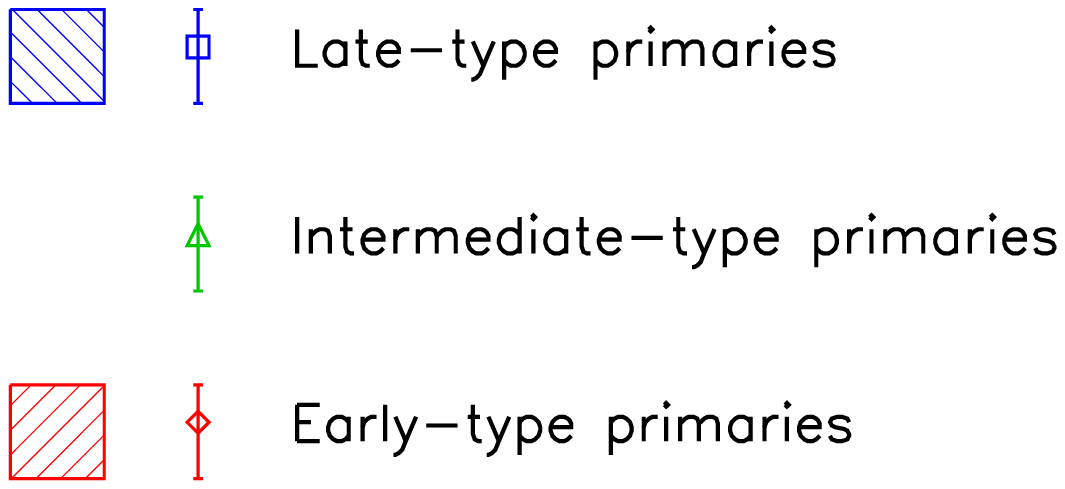}\\
\end{array}$
\caption{Effects of varying the selection parameters on the
results and sample size. In each plot, the mean disc angle for
early/intermediate/late-type primaries is shown as the red
diamond/green triangle/blue square symbols with error bars,
while the sample size for the early- and late-type samples are
shown as histograms below and above. The scale for each histogram
is shown on the right side of each plot. The histograms for the
intermediate-type samples show identical trends.
\textit{(a)} The effects of replacing
the fiducial NED plus photometric redshift (``NED+Pz'') method for
dealing with violators with either a pure photometric redshift method
(``Pz'') or with a cut on \Nviol.
\textit{(b)} The effects of changing \fsatlum.
\textit{(c)} The effects of limiting primaries to be at least
\magin\ magnitudes brighter than the survey limit and of limiting
primaries to be at least projected radii \rout\ from the edge of the
photometric survey footprint and \rsat\ from the edge of the
spectroscopic survey footprint.
\label{results for different keywords plot}}
\end{figure*}

As discussed in \S~\ref{constructing mock grs section},
it is difficult to assess the effects of some of the selection
parameters using the mock catalogues.
Therefore, we now empirically investigate the effects of
varying all of the selection parameters, including several
parameters that have been neglected by many previous studies.
In Figure~\ref{results for different keywords plot},
we plot the mean disc angle and the sample size as we adjust
\Nviol, \fsatlum, and whether or not the survey magnitude limit
or survey edge are taken into account.
The mean disc angles are shown as symbols with error bars while the
early and late-type sample sizes are shown as histograms above and below
(the intermediate-type sample sizes show identical trends).

In panel (a), we compare our combined NED and photometric redshift-based
method of dealing with criteria violators, a method that only uses
the photometric redshift, and methods based on a cut at various
values of \Nviol.
More restrictive values of \Nviol\ lead to smaller
sample sizes, particularly for $\Nviol \le 4$.
The measured anisotropy is relatively constant as a function of \Nviol.
Our combined NED+photometric redshift method produces a sample size equivalent
to using $\Nviol=1$, and therefore recovers a reasonable fraction
of the sample available from ignoring the presence of violators,
while conservatively excluding any systems that could contaminate
the sample.
In panel (b), we plot the effects of \fsatlum.
The measured anisotropy rises as \fsatlum\ is reduced to very
low values; this is accompanied by a sharp decrease in sample size.
As indicated by Figure~\ref{diskangle vs radius mag figure}b,
this indicates that in many cases the anisotropy is related to the
dominance of the primary galaxy, and samples selected
with large values of \fsatlum\ may be significantly contaminated.
In panel (c), we show the effects of ignoring the survey magnitude
limit or of ignoring the survey edge. Both samples are marginally larger
with no significant change in the measured anisotropy.

Although the exact strength of the anisotropy can vary with
some of these oft-neglected parameters, the qualitative result,
that the satellites of early- and
possibly intermediate-type galaxies show a major-axis
distribution while the satellites of late-type galaxies are
isotropically distributed, is not dependent
on the value of any one of these parameters.
Varying the other parameters from
Table~\ref{selection criteria parameters table} has very little
effect on the measured anisotropy,
as anticipated by the results of \citetalias{brainerd05},
who found identical results in three samples with quite different
values of these parameters.

\subsection{Satellite and Primary Distribution Relative to Large
Scale Structure}\label{lss angle section}

If satellites are accreted from filaments, then the most recently-accreted
satellites will be aligned preferentially with the surrounding
filamentary large scale structure (LSS).
We test this expectation by determining the axis of the LSS surrounding
each primary galaxy. To determine this axis, we use all spectroscopic 
galaxies with projected radii between $1000$ and $3000\,\hikpc$ (thereby
explicitly ensuring that there is no overlap between the galaxies used 
to determine the orientation of the LSS and those used to evaluate the 
isolatedness of the primary or the satellites themselves)
with velocities that differ from that of the primary by no more than 
$400\,\kms$. The velocity dimension of this cylinder is significantly smaller
than the cylinder used to select isolated galaxies and satellites.
This is because filaments are not virialised structures and their intrinsic
velocity dispersion about the Hubble flow is much lower than that inside a halo
\citep[for example, the scatter about the Hubble flow among the galaxies
surrounding the Local Group is a mere $85~\kms$, or as low as $40~\kms$
if galaxies inside virialised groups are excluded;][]{karachentsev-etal03},
and therefore a much smaller additional velocity is required to account
for peculiar velocities on top of the Hubble component of $300\,\kms$
corresponding to the radial dimension of the cylinder.
We calculate the PA and axial ratio of the distribution
on the sky of these surrounding galaxies by diagonalising the moment
of inertia tensor relative to the primary galaxy.
We have used the mock catalogues to confirm that this procedure
reliably recovers the three-dimensional PA of the LSS surrounding
the primary (see Appendix~\ref{appendix:lss pa}).

\begin{figure}
\includegraphics[scale=0.45]{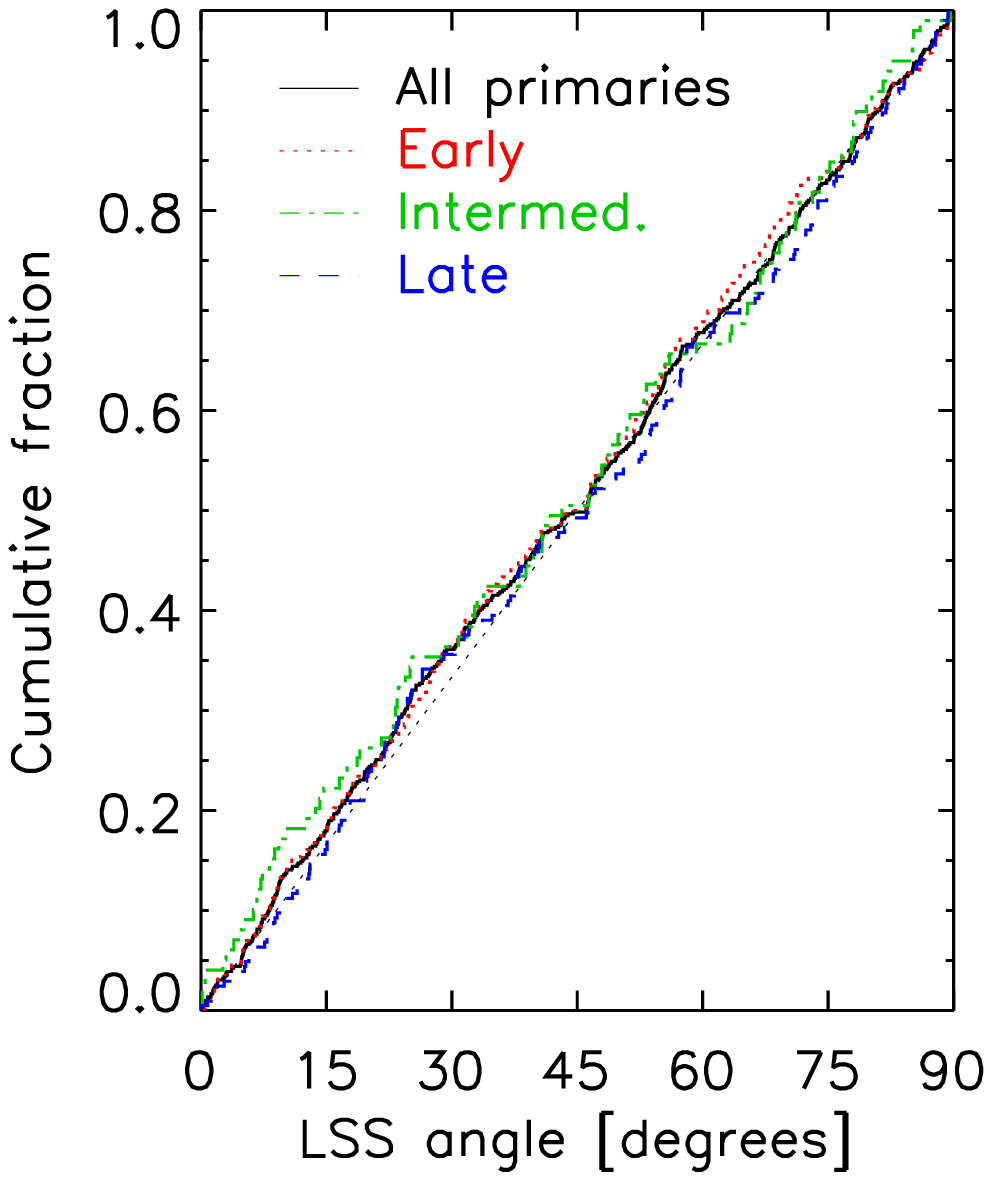}%
\includegraphics[scale=0.45]{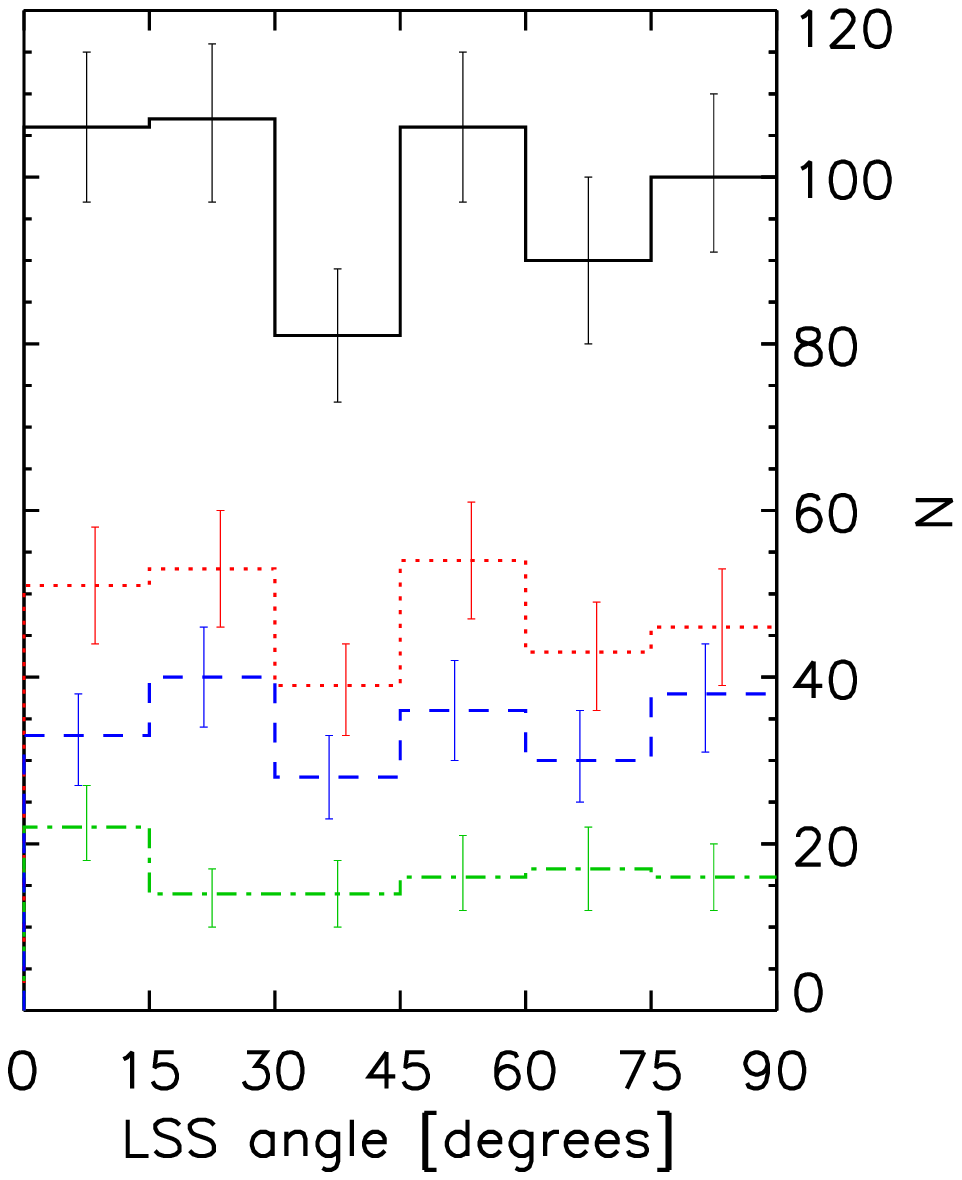}
\caption{%
\textit{(Left)} Cumulative distribution of angle between
the axis of the local LSS and the location of the satellite
(``LSS angle''). The thick black solid/red dotted/green dot-dashed/blue dashed
lines refer to the distribution of satellites
around all primaries, early-type primaries,
intermediate-type primaries and late-type primaries
respectively. The thin dotted line shows the distribution expected
if satellites are isotropically distributed.
\textit{(Right)} Differential distribution of the LSS angles.
The error bars are determined by bootstrap resampling
of all primary galaxies.%
\label{LSS angle distribution figure}}
\end{figure}

\begin{figure}
\includegraphics[scale=0.5]{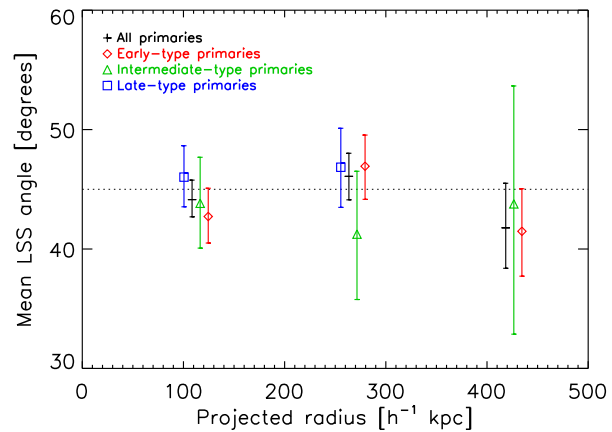}
\caption{%
Mean LSS angle of satellites as a function of their projected
separation from the primary.
Black plusses, red diamonds, green triangles and blue squares refer to
satellites of all primaries, early-type primaries,
intermediate-type primaries and late-type
primaries respectively.
Symbols are offset in radius slightly for clarity.
\label{lss angle vs radius figure}}
\end{figure}

\begin{figure}
\includegraphics[scale=0.45]{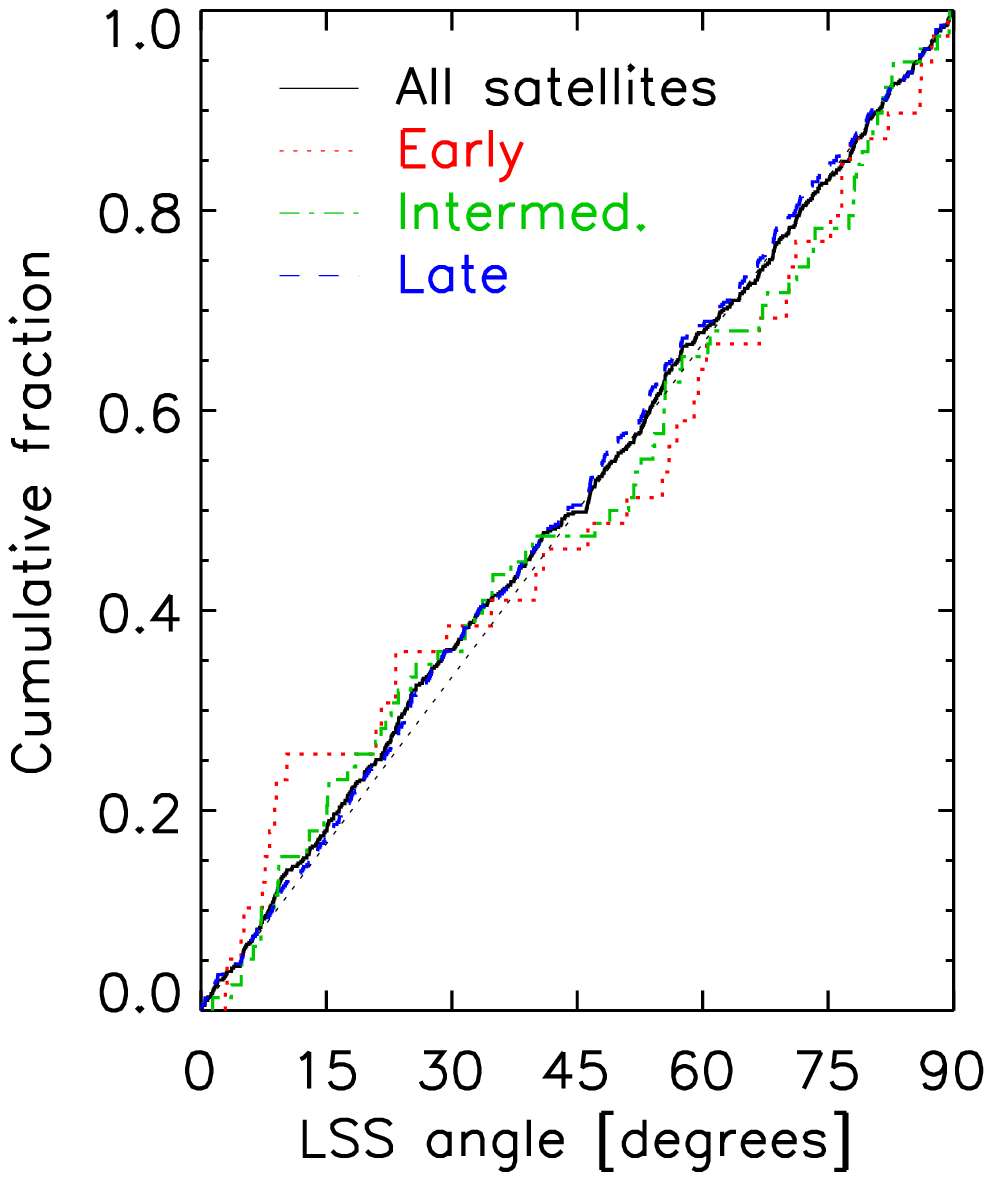}%
\includegraphics[scale=0.45]{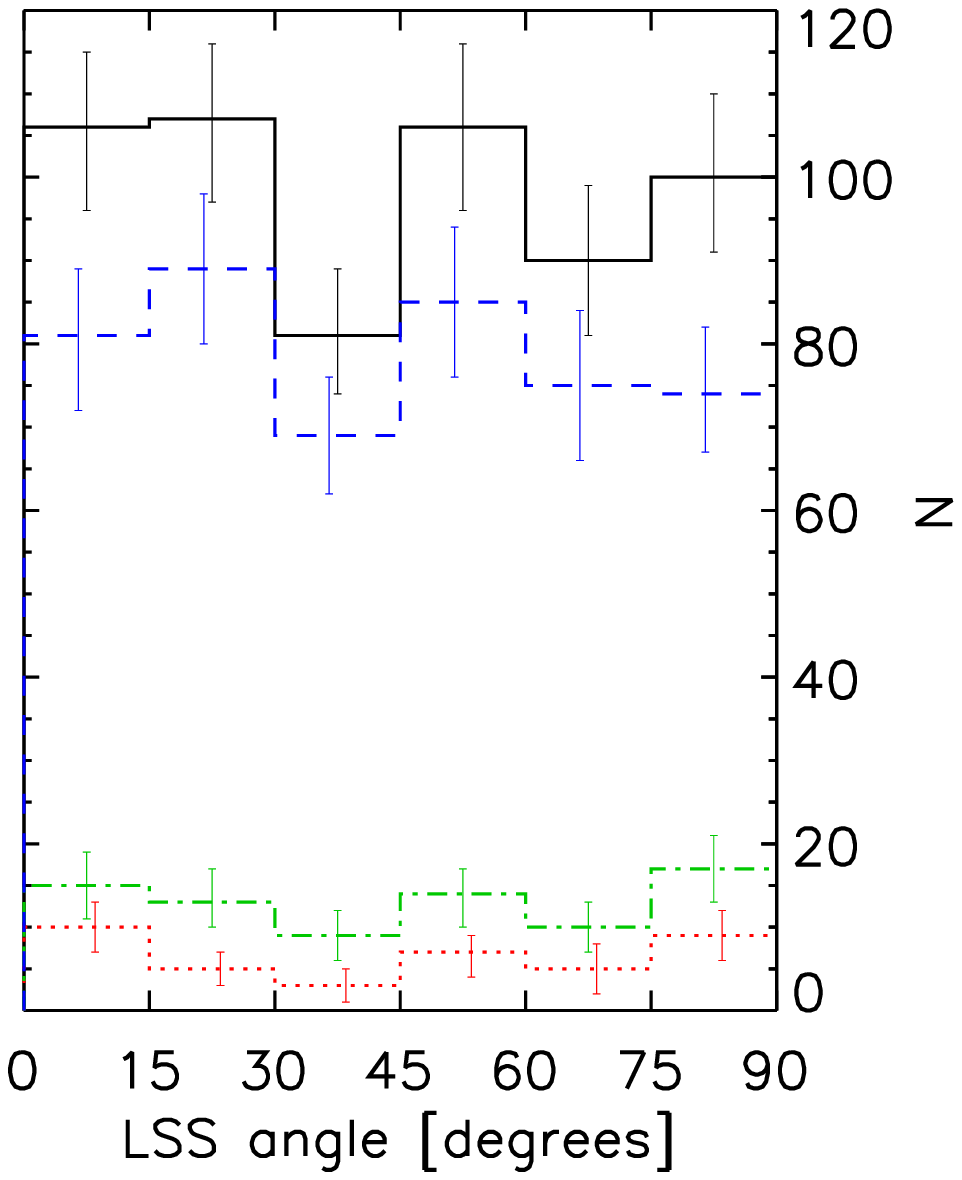}
\caption{%
As in Figure~\ref{LSS angle distribution figure}, but separated
by the classification of the satellite galaxy.%
\label{LSS angle bysat figure}}
\end{figure}

\begin{figure}
\includegraphics[scale=0.5]{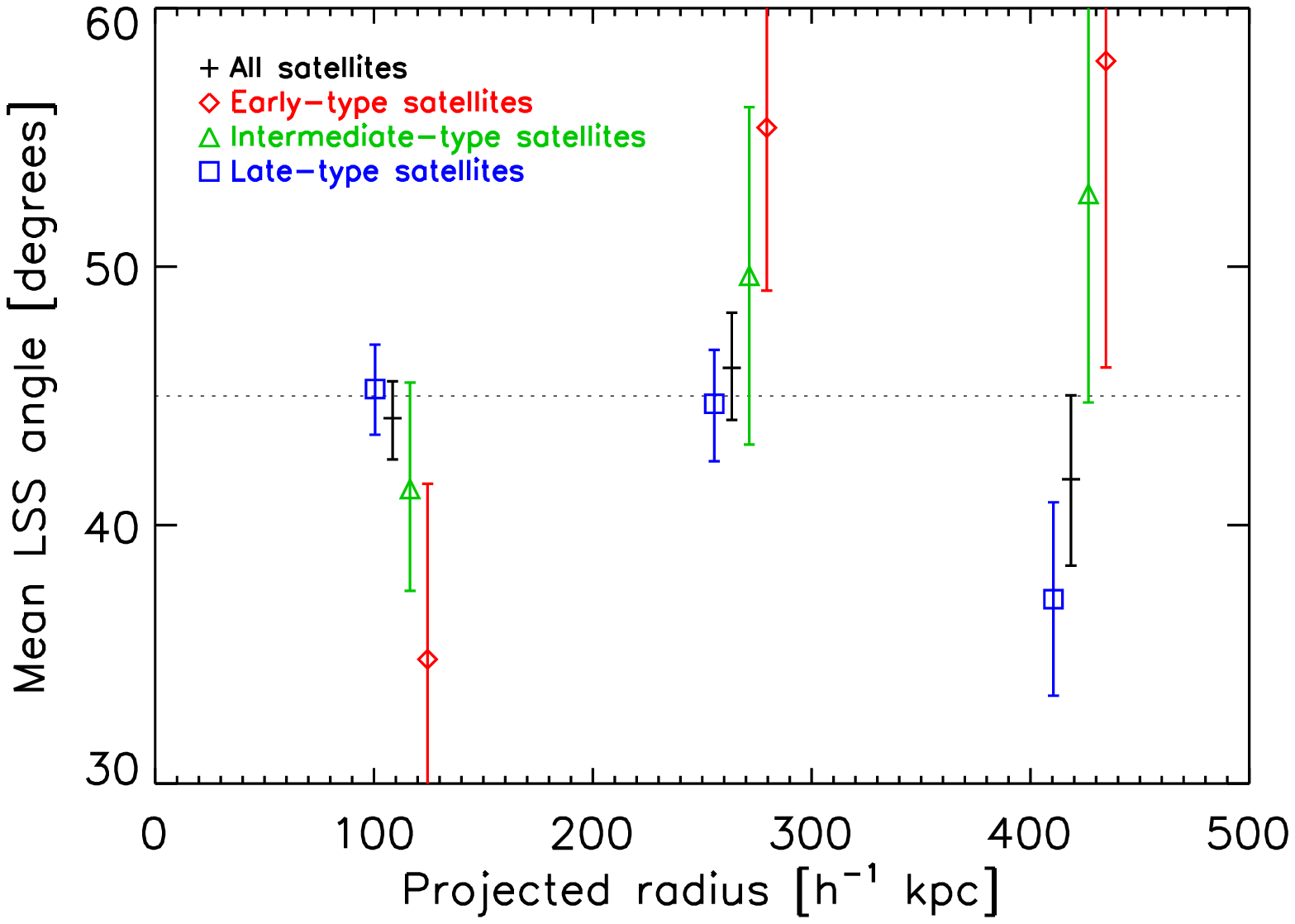}
\caption{%
As in Figure~\ref{lss angle vs radius figure}, but separated
by the classification of the satellite galaxy.%
\label{lss angle bysat radius figure}}
\end{figure}

Figure~\ref{LSS angle distribution figure} shows the distribution of
angles between the PA of the great circle connecting the primary
and the satellite and the PA of the LSS surrounding the primary
(``LSS angles''; as with the disc angles, these
are folded into the range $0$--$90\degr$).
We present the results from KS tests and the mean LSS angles in
Table~\ref{satellite anisotropy stats table}.

The satellites of each population of galaxies are consistent
with being isotropically distributed with respect to the LSS. However,
all samples except that around late-type primaries
have mean LSS angles less than $45\degr$. Further data are
required to determine if this hint of an alignment is real.
In Figure~\ref{lss angle vs radius figure}, we plot the mean
LSS angle as a function of radial separation from the primary.
Although each individual point is consistent with isotropy,
the satellites at large separations all have mean LSS
angles less than $45\degr$ (as there are by definition no satellites
of late-types in this bin, this may explain why the late-types also
show no hint of LSS alignment). Further information may be gained by
plotting the LSS alignment as a function of satellite type
(Figures~\ref{LSS angle bysat figure} and \ref{lss angle bysat radius figure}).
Although the number of early- and intermediate-type satellites is
too small to draw any conclusions, the late-type satellites that
are found at large radius are aligned with the LSS at the $2\sigma$
level.

\begin{figure}
\includegraphics[scale=0.45]{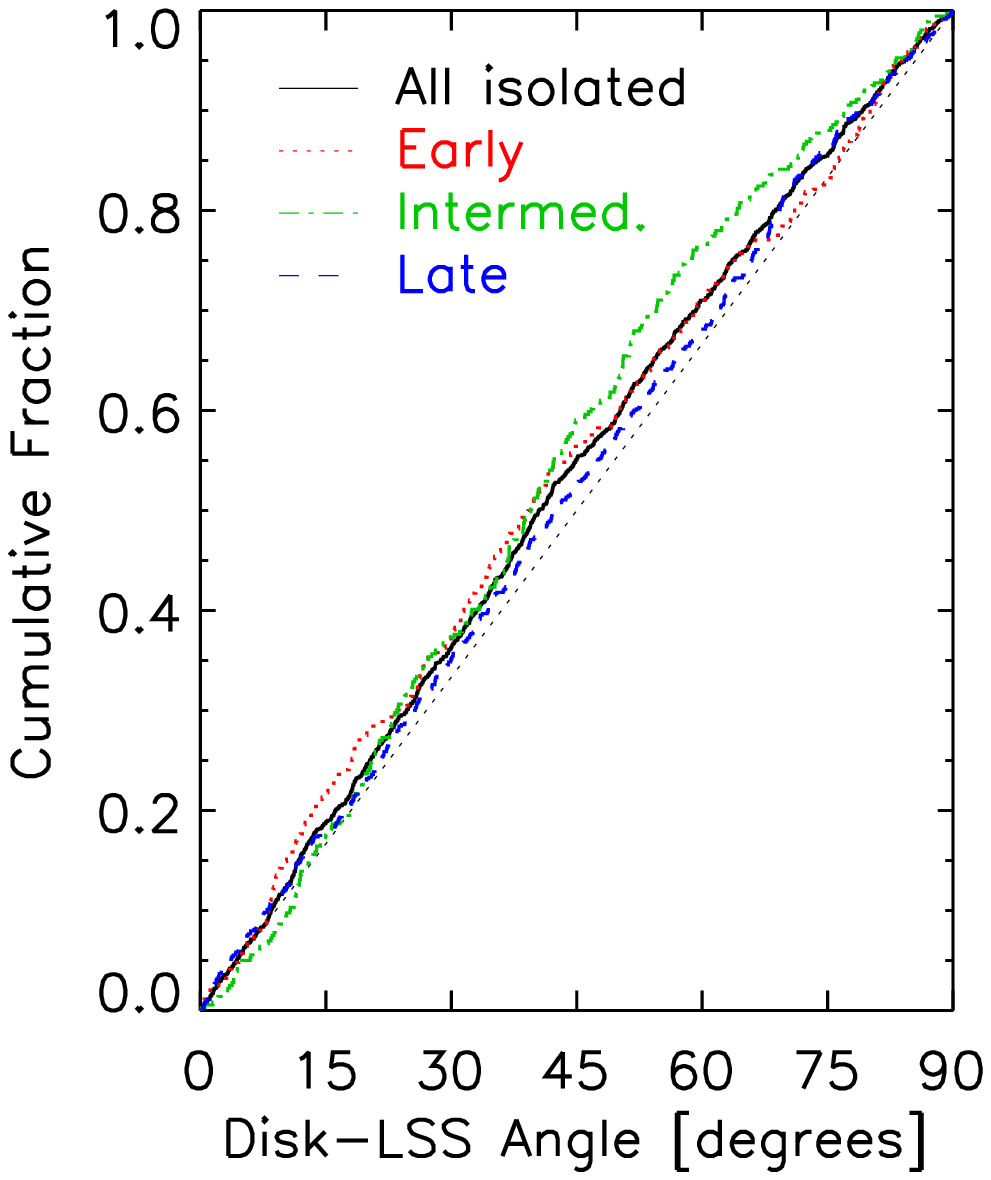}%
\includegraphics[scale=0.45]{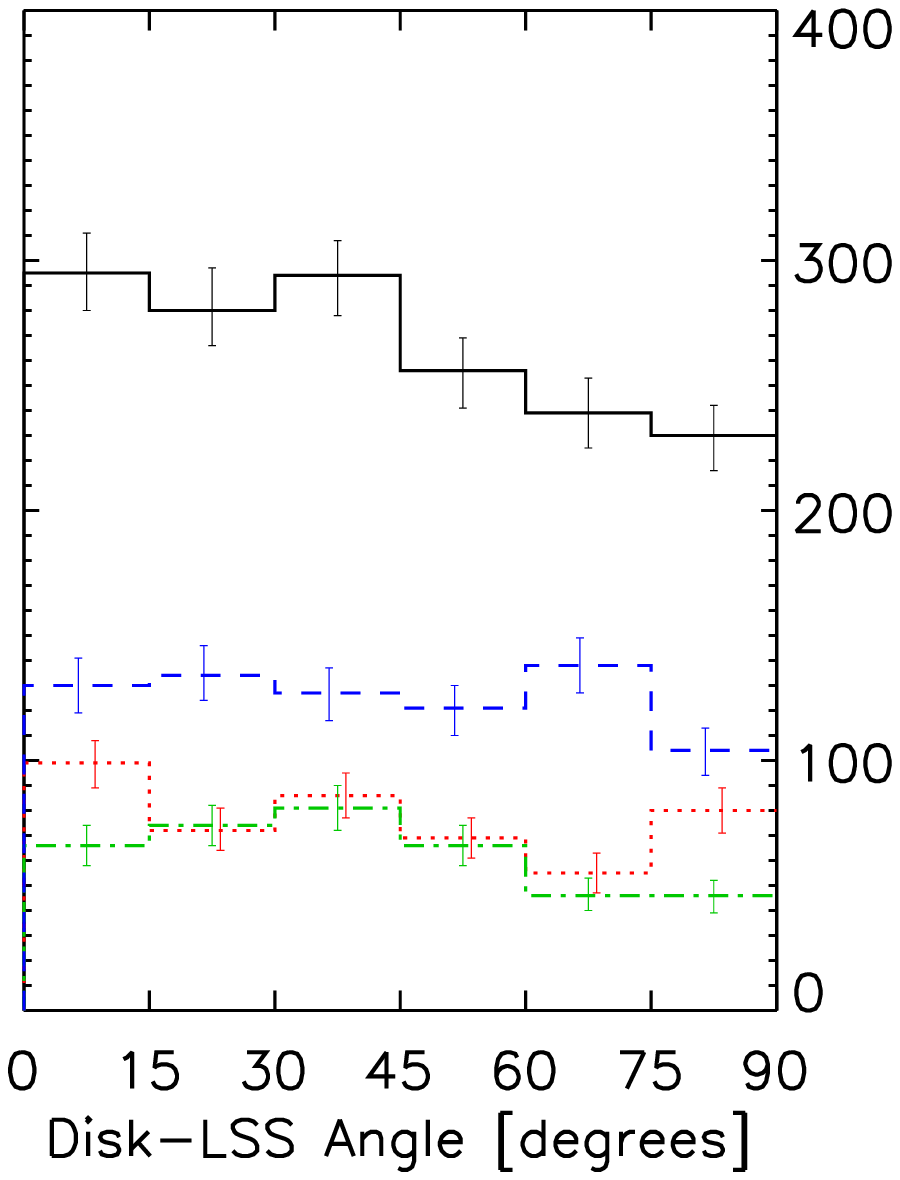}
\caption{\label{disc vs lss figure}%
Cumulative \textit{(left)} and differential \textit{(right)} distributions
of angles between the major axes of isolated galaxies and their
surrounding LSS. The sample contains all isolated galaxies
that pass the ``Disc'' and ``LSS'' sample quality cuts, regardless
of whether they host any satellite galaxies.
Red/dotted lines refer to early-type galaxies while blue/dashed lines
refer to late-type galaxies.}
\end{figure}

Given the orientation of the primary galaxy and the LSS, 
we now investigate their relative alignment.
The orientation of a disc galaxy is determined by its angular
momentum, which originates from tidal torques due to the surrounding
material.
Analytic arguments and cosmological simulations suggest that
this angular momentum (and therefore the disc spin axis)
aligns with the intermediate axis of the surrounding
mass distribution, and
such alignment has been measured for disc galaxies in the supergalactic
plane \citep{nas04}
and for galaxies in SDSS and 2dFGRS on the surfaces of
voids \citep{tcp06}.
The orientation of an early-type galaxy is determined
by its anisotropic velocity ellipsoid, as is that of its halo; therefore,
the two are expected to be aligned, and preferentially aligned with
the large scale structure \citep{bs05-alignment}.
This has been measured for Brightest Cluster Galaxies (BCGs)
at low \citep{argyres-etal86,lgp88b,ml89} and high \citep{dol06} redshift,
but not for field early-types.
We directly compared the PA of our primary galaxies to that of their
local LSS. For this comparison, we use all isolated galaxies
that pass both the ``Disc'' and ``LSS'' quality cuts, regardless
of whether they host satellite galaxies; our results are unchanged
if we only include those that host satellite galaxies.
The distributions are shown in Figure~\ref{disc vs lss figure},
and the associated mean angles and KS test probabilities of
being drawn from an isotropic distribution are given in
Table~\ref{satellite anisotropy stats table}.
There is a detection of alignment between the orientation of
isolated early-type galaxies and the surrounding LSS
at $99\%$ confidence,
a strong alignment for isolated intermediate-type galaxies
at $99.94\%$ confidence.
We do not detect a significant alignment for isolated late-type galaxies.
The samples containing all isolated galaxies and only those with
satellites are consistent with each other.

\subsection{Satellites in the Local Group}\label{LG}
Many satellites of the two dominant galaxies in the Local Group,
the Milky Way and M31, appear to lie on planes that are highly inclined
to their parent discs
\citep[and references therein]{mkj07}.
In order to determine how the anisotropy of
the SDSS galaxies compares to the anisotropy around the Local
Group spirals, we have determined the signal we would have recovered
around both the Milky Way and M31 if they had fallen into our sample%
\footnote{In fact, neither the Milky Way nor M31 would fall into
our sample if they were observed in a redshift survey because the presence
of each would violate the isolation criteria around the other.
Our isolation criteria are required to be so strict in order that systems
intrinsically less isolated than Local Group galaxies are not
mistakenly included.}.
Based on the distribution of primary and satellite magnitudes
in our sample (Figure~\ref{sample distribution figure}a),
we use all satellites with absolute magnitudes
within $5$ magnitudes of their primary;
for the Milky Way this consists of the LMC and SMC,
and for M31 this consists of M33, IC~10, M32, and NGC~205.
The 3D locations of these satellites with respect to their
parent galactic disc are taken from
\citet{mkj07},
using the \citet{mi06} parameters for the M31 satellites.
We select $20000$ random viewing directions isotropically distributed
about each galaxy and calculate the disc angle for each satellite from
each viewing direction.
We calculate the mean disc angle of the satellites
averaged over all viewing angles where the
projected axis ratio of the
parent disc is less than $0.8$, to provide a direct comparison
to the SDSS sample.

The resulting anisotropies are listed
in Table~\ref{local group anisotropy table}.
The satellites of the Local Group spirals show polar distributions
with mean disc angles of $49.1\degr$ and $54.0\degr$ for the satellites
of the Milky Way and M31 respectively.
We do not measure such a polar distribution around late-type disc galaxies
in SDSS.
However, we do measure a minor-axis alignment of this magnitude
among those satellites much fainter than their primary
(see Figure~\ref{diskangle vs radius mag figure}b) and most
of the Local Group satellites above meet that description. Therefore,
perhaps the dependence of satellite anisotropy on degree of 
primary dominance may explain the discrepancy between the satellites
of the Local Group spirals and the results from galaxy redshift surveys,
or perhaps we are simply victims of coincidence and small number statistics.

\begin{table}
\caption{\label{local group anisotropy table}%
Anisotropy of Local Group Satellites}
\begin{tabular}{lcc}
\hline
{Parameter} & {Milky Way} & {M31}\\
\hline
\hspace{1em}Mean disc angle [\degr] & $49.1$ & $54.0$\\
\hspace{1em}Median disc angle [\degr] & $47.7$ & $53.4$\\
\hspace{1em}Polar fraction & $0.57$ & $0.60$\\
\hline
\end{tabular}
\end{table}

\section{Comparison with Previous Results: The Effects of Environment}
\label{comparison with previous results section}

Our results indicate that the satellites of isolated early-type
galaxies show a preference for lying near the major axis
of the primary, while those of intermediate-type galaxies may lie
near the disc plane and those of late-type galaxies are isotropically
distributed.
The result of combining
these populations gives a distribution that is purely a function of the
morphological mixture in the sample of isolated galaxies;
in our case, the early- and intermediate-type galaxies dominate
over the late-type galaxies and therefore we find that
satellites tend to exhibit a net major axis alignment.

It is interesting to compare our results to those of previous
studies.
Our results for late-type galaxies agree with previous
detections of an isotropic satellite distribution
\citepalias{sl04,azpk06,appz07,yang-etal06},
and our results for early-type galaxies agree with previous
detections of major-axis alignment
\citepalias{sl04,yang-etal06,appz07,ab07}.
Our intermediate-type galaxies would have been identified as part
of the ``early''-type sample by those studies that used colour
to classify galaxies, and would have been split between
apparent ellipticals and apparent disc galaxies depending
on inclination by those studies that classified galaxies by eye
\citep[see][]{bailinharris08-classification}. As the
anisotropy around intermediate-type galaxies and early-type galaxies
is similar, the former situation would not have affected the
measured distribution, while in the latter case the relatively
small number of contaminating intermediate-types would not have
strongly affected the measured distribution around late-types.
Finally, our results for the full sample agree with \citetalias{sl04},
\citetalias{brainerd05}, \citetalias{yang-etal06} and \citetalias{ab07},
who all found that the full sample shows major-axis alignment.

The main disagreement between our results and previous studies are
with those studies that found a polar alignment of satellites around
late-type galaxies (\citealp{holmberg-effect}; \citetalias{zsfw97-holmberg}).
However, there are physical sub-classes of systems for which our results
are consistent with a polar alignment (i.e. although our results in
these regimes are consistent with isotropy and we therefore do not
claim detection of a polar alignment, the mean disc angles are sufficiently
larger than $45\degr$ that they are also consistent with a polar
alignment):
satellites at intermediate separation from their primary,
and those that are much fainter than their primary.
The polar alignment found by these studies may be explained if they were
dominated by such satellites; indeed,
\citetalias{zsfw97-holmberg} 
detected their polar alignment for satellites with
similar projected separations as our intermediate separation bin,
and dominance by relatively faint satellites may explain the
alignment seen around the Milky~Way and M31 (see \S~\ref{LG}).

\begin{figure*}
\includegraphics{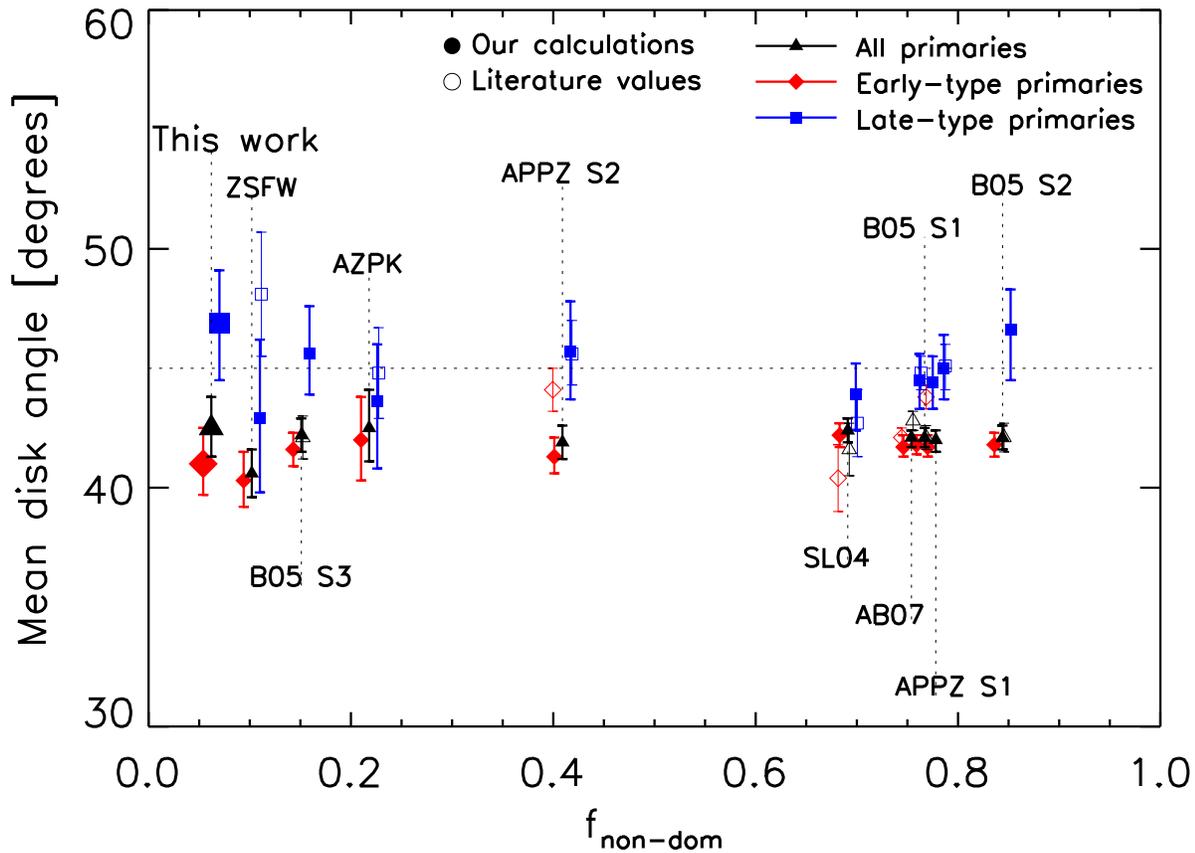}
\caption{%
Mean disc angle of satellites selected using each
set of criteria vs. \fnondom, the fraction of satellites
selected using each criteria that lie around
non-isolated primaries
(see Table~\ref{selections on mocks table}).
Colors/symbols are as in Figure~\ref{diskangle vs radius mag figure},
where we have used location on the CMD to separate early from
late-type galaxies in all cases.
Filled symbols with thick error bars indicate our calculations, while
open symbols with thin error bars indicate the values given
in each previous study.
As the degree of agreement is excellent, the filled and open symbols
typically lie almost overtop each other.
Early and late-type symbols are offset in \fnondom\ for clarity.%
\label{diskangle vs criteria figure}}
\end{figure*}

To determine the effects of the different selection criteria on
the measured anisotropy, we adopt the criteria from the previous
studies (as given in Table~\ref{selection criteria parameters table})
to select corresponding samples of galaxies from SDSS~DR6 and to
measure the disc angle distribution. The mean value of the disc
angle determined by the previous studies and the value we derive
using identical selection criteria are shown in
Figure~\ref{diskangle vs criteria figure}
as a function of \fnondom, the fraction of satellites estimated
to lie around non-isolated primaries according to the mock
catalogue analysis (see Table~\ref{selections on mocks table}).
For simplicity, in this Figure only we separate galaxies into two classes,
using the location on the CMD. More details
about the comparison are presented in Appendix~\ref{criteriacomp-appendix}.
This Figure confirms that if we select galaxies according to the
criteria used in each previous study, we recover their results.

The angular distribution found using each sample is quite similar.
However, the selected galaxies lie in very different environments.
Our selection criteria have been fine-tuned using mock catalogues to
select satellites of isolated primaries. While most previous authors
have also implemented selection criteria aimed at identifying isolated
primaries, the results of \S~\ref{mock catalogue section} indicate
that they have had variable success. In particular, the fraction
of satellites around non-isolated primaries is over 50\%\ in several 
of the previous studies. The satellites in these systems should be
considered group members rather than satellites of isolated galaxies,
as should those of \citetalias{yang-etal06},
\citet{faltenbacher-etal07}, and \citet{wang-etal08}, whose selection
criteria were tuned to find associated galaxies 
with no constraints on whether the largest galaxy in each group is isolated.

The combination of our results and those of
previous studies constrain the environmental dependence
(or lack thereof) of the satellite distribution.
Satellites surrounding spheroidal galaxies
show the same major-axis alignment regardless of whether that spheroid
is isolated or is at the centre of a group.
Similarly, the satellites surrounding isolated late-type galaxies
are as isotropically distributed as the members of groups
that surround late-type galaxies. Although previous studies have not identified
intermediate-type galaxies, we can select galaxies using their
criteria and examine the distribution of those satellites around
intermediate-type galaxies.
For example, when using the sample generated by the \citetalias{brainerd05}~S2
criteria, which lies at the far right
side of Figure~\ref{diskangle vs criteria figure} and contains almost
$85\%$ group members,
the satellites of intermediate-type galaxies have a mean disc angle of
$41\fdg5\pm0\fdg9$, in good agreement with the planar alignment
tentatively detected around
isolated intermediate-type galaxies using our fiducial criteria.
Therefore, we conclude that group-specific processes are not
responsible for the angular distributions of their member galaxies,
but rather processes that also apply to the satellites of isolated
galaxies.

\section{Summary and Discussion}\label{summary discussion section}

We summarise our results as follows:

\begin{itemize}

\item Satellites of isolated early-type spheroidal galaxies lie preferentially
  along the major axis of the galaxy. The degree of alignment 
  increases slightly for satellites that are brighter relative to their primary.

\item Satellites of isolated disc galaxies appear to have different angular
  distributions depending on the colour of the disc. Satellites of
  \textit{red} discs (intermediate-type galaxies) show hints of
  lying preferentially near the disc plane, with intrinsically or relatively
  fainter satellites showing a
  tendancy to stronger anisotropy. Satellites of
  \textit{blue} discs (late-type galaxies) are distributed isotropically.

\item Late-type satellites that are found far from their primary
  show preferential alignment with the surrounding LSS (i.e.~filaments)
  (Figure~\ref{lss angle bysat radius figure}).
  
\item Isolated early- and intermediate-type galaxies show an alignment
  with the surrounding LSS.
  This alignment is strongest for intermediate-type galaxies (with a
  KS test significance of $99.94\%$).

\item The angular distribution of group members about the BGG is
  very similar to the distribution of satellites around
  an isolated galaxy of the same type.

\item Great care must be taken in order to select truly isolated galaxies
  and their satellites in galaxy redshift surveys. Unless the region
  immediately surrounding the primary is devoid of galaxies too large to
  be considered satellites (\textit{whether or not those galaxies have been
  observed spectroscopically}), the sample will be dominated by group
  members rather than isolated galaxies.

\end{itemize}

\noindent
These results provide us with the foundations on which we can build 
our understanding of the mass distribution in and around
galaxies and
raise a number of interesting questions. What 
can we learn about the role of dynamical effects in driving preferential
alignments of satellites? What is the nature of these effects and
does it depend on the morphological type
and history of the galaxy? What role is
played by the host dark matter halo? What role is played by the larger
scale environment? Our results allow us to begin to address these 
questions.\\

The most straightforward interpretation of the preferential alignment 
of satellites at large projected radii with the surrounding LSS is that
this is a signature of the anisotropic infall of satellite galaxies along 
filaments. This interpretation is favoured by the following observations:

\begin{itemize}
\item the mean angle between the projected positions of satellites and
  the surrounding LSS (i.e. the LSS angle) is similar around different
  types of galaxies;
\item the alignment is present for \textit{only} the more recently
  accreted satellites, i.e. those that are most distant and those blue
  late-type satellites for which halo-specific transformation
  mechanisms have not yet had time to operate; and
\item the alignment is distinctive when compared to the
  alignment of the satellites with respect to their primary.
\end{itemize}

\noindent 
This evidence argues in favour of an origin that is imposed by 
the larger scale environment rather than one driven by the primary
galaxy. In other words, it is improbable that the dynamics of the 
most recently accreted satellites will be significantly affected by any 
process internal to the primary's dark matter halo. This is in
agreement with the results of cosmological simulations 
\citep{knebe-etal04,zentner-etal05,libeskind-etal05}.\\

The relationship between the orientation of a galaxy and its
surrounding LSS can be understood in terms of the relationship between
the orientation of the galaxy's dark matter halo and its surrounding
LSS \textit{and} the galaxy's orientation within its dark matter halo. 
Cosmological $N$-body simulations predict that dark matter haloes are 
strongly triaxial systems in the absence of baryons
\citep[e.g.][]{allgood-etal06}, and these haloes tend to
align with their major axes along the large scale filaments and 
their minor axes perpendicular to filaments \citep{bs05-alignment}. 

However, the presence of baryons can have a dramatic effect on the shapes
and internal alignments of dark matter haloes, with interesting
consequences for the haloes of disc galaxies. A number of 
studies have shown that cooling baryons at the centre of a dark
matter halo tend to circularise the orbits of dark matter particles,
modifying the halo's inner mass profile \cite[e.g.][]{gfsl06}
and reducing the ellipticity of the halo's 
isodensity surfaces \citep[e.g.][]{dubinski94,kazantzidis-etal04}.
\citet{bailin-etal05-diskhalo} examined the structure of the host
haloes of several disc galaxies that formed in high resolution
cosmological $N$-body hydrodynamical simulations, and discovered that
haloes consisted of two distinct regions. The inner halo is
flattened along the disk axis, while the orientation of the outer 
halo is unrelated to that of the inner halo and is unaffected by the
presence of the luminous galaxy. 

The \citet{bailin-etal05-diskhalo} result is interesting because it 
implies that the major axis of the inner halo around a typical disc 
galaxy is aligned with the major axis of the light distribution, while 
the major axis in the outer region is independent of the light but is 
aligned with the LSS. If satellite galaxies are more common along the 
halo major axis, then this ``twisting'' of the halo should be evident 
in the distribution of satellites around disc galaxies.
Indeed, we observe that it is the outermost satellites around all types of
galaxies that are preferentially aligned with the LSS, while
it is the inner satellites
around intermediate-type disc galaxies (which correspond most closely to
the relatively red
concentrated simulated discs studied by \citet{bailin-etal05-diskhalo})
that are found preferentially in the disc plane.
However, the significant relative alignment between intermediate-type
galaxies and the LSS argues that some residual halo-LSS alignment
remains.

In the case of early-type galaxies, the orientation of the galaxy 
with respect to its dark matter halo has not been studied explicitly 
in a cosmological context. However, we would expect that the shapes 
of both their stellar and dark matter components are supported by their 
anisotropic velocity ellipsoid. Consequently, we would expect that both 
the galaxy and its dark matter halo will share the same orientation,
and therefore the galaxy will tend to align with the surrounding LSS
\citep{bs05-alignment},
as is observed.\\

How should we interpret the preferential alignments of satellites
around isolated galaxies? Can we determine whether
the alignments are imprinted by the dynamical effects of the galaxy or
the host dark matter halo? 

\citet{ab06-satdist}, \citetalias{ab07} and \citet{kang-etal07} studied
the angular distribution of satellite galaxies in cosmological
simulations selected according to the criteria of \citetalias{brainerd05},
\citetalias{ab07} and \citetalias{yang-etal06} respectively
(note that all of these criteria select samples dominated by groups).
The orientation of a mock galaxy must be assumed and so these authors explored
different assumptions about how primary galaxies are oriented with respect
to their dark matter haloes and the larger scale environment. They found
that if the primary galaxy is a spheroid whose principal axes are
perfectly aligned with those of its dark matter halo, then satellites in
these systems tend to show a major-axis anisotropy that is stronger than
observed in groups whose BGG is an early-type galaxy. However, if there is
a small offset between the principal axes of the galaxy and its halo, as
may arise if the galaxy aligns with the halo's angular momentum rather
than its minor axis, then the anisotropy is of the same order that is
observed. Because the alignment of satellites relative to isolated
early-types is identical to that seen in galaxy groups whose BGG is an
early-type, we may therefore conclude that the principal axes of isolated
early-types are similarly well aligned with those of their dark matter
haloes ($\sim20\degr$). As we argued above, this is in accord with our
expectation that the dynamics of baryons and dark matter in collisionless
ellipsoidal systems are similar.

The theoretical situation around discs is less
clear. If the disc is oriented perpendicular to its halo's angular
momentum, it is simply a special case of an oblate spheroid, and therefore
these studies predict that its satellites will exhibit major-axis
alignment. This is consistent with what we observe around red discs,
but in stark contrast to what is seen around blue discs.
If, on the other hand, the
angular momenta of galaxy discs align with the intermediate axis of the
surrounding mass distribution, as seen in the simulations of
\citet{nas04}, then satellites in these systems show no preferential
alignments, as we observe around blue discs.
If we interpret disc colour as a measure of how long the
baryonic material has been part of the luminous galaxy, then
perhaps red discs have had more time to come to equilibrium with their
halo, while most of the material in blue discs has been acquired
more recently and retains a memory of the external
tidal torques.

However, these explanations are still largely speculative.
A theoretical analysis that takes into account
the detailed dependences of the satellite distribution,
the alignment of satellites with the LSS,
and the differences between early-type spheroidal galaxies, intermediate-type
red disc galaxies, and late-type blue disc galaxies
must be performed to determine whether the
distribution of satellites is determined
predominantly by the orientation of the halo
or if dynamical processes within the
halo are important. We are in the process of performing such an
analysis (Power et~al., in preparation).

\section*{Acknowledgements}

JB thanks Mike Blanton for writing KCORRECT, the SDSS help desk for
their help with the Catalog Archive Server,
St\'ephane Herbert-Fort, Pavel Kroupa and
Manuel Metz for useful discussions, and
John and Lexi Moustakas for writing the RED IDL cosmological
routines.
Early stages of this research benefitted from three-dimensional
visualisation conducted with the S2PLOT
programming library \citep{s2plot}.

JB acknowledges the financial assistance of the
Australian Research Council.
CP and BKG gratefully acknowledge the support of the Australian
Research Council supported ``Commonwealth Cosmology Initiative'',
DP 0665574.
During the course of this work, PN acknowledges funding from a Zwicky
Fellowship at ETH and a PPARC PDRA Fellowship at the IfA.
DZ acknowledges support from NASA LTSA award NNG5-GE82G,
NSF grant AST-0307482, and a Guggenheim fellowship, and thanks
the NYU Physics department and Centre for Cosmology and
Particle Physics for their generous support during his sabbatical there.

    Funding for the SDSS and SDSS-II has been provided by the Alfred P. Sloan Foundation, the Participating Institutions, the National Science Foundation, the U.S. Department of Energy, the National Aeronautics and Space Administration, the Japanese Monbukagakusho, the Max Planck Society, and the Higher Education Funding Council for England. The SDSS Web Site is http://www.sdss.org/.

    The SDSS is managed by the Astrophysical Research Consortium for the Participating Institutions. The Participating Institutions are the American Museum of Natural History, Astrophysical Institute Potsdam, University of Basel, University of Cambridge, Case Western Reserve University, University of Chicago, Drexel University, Fermilab, the Institute for Advanced Study, the Japan Participation Group, Johns Hopkins University, the Joint Institute for Nuclear Astrophysics, the Kavli Institute for Particle Astrophysics and Cosmology, the Korean Scientist Group, the Chinese Academy of Sciences (LAMOST), Los Alamos National Laboratory, the Max-Planck-Institute for Astronomy (MPIA), the Max-Planck-Institute for Astrophysics (MPA), New Mexico State University, Ohio State University, University of Pittsburgh, University of Portsmouth, Princeton University, the United States Naval Observatory, and the University of Washington.

This research has made use of the NASA/IPAC Extragalactic Database (NED) which is operated by the Jet Propulsion Laboratory, California Institute of Technology, under contract with the National Aeronautics and Space Administration.

\bibliography{../../masterref.bib}
\appendix

\section{The Conditional Luminosity Function : Parameters}

\label{appendix:details}

\citet{yang-etal03} deduced a functional form for the variation of
the mass-to-light ratio with dark matter halo mass by comparing the 
\citet{st99} dark matter halo mass function with the Schechter luminosity
function \citep{schechter76}. They noted that the mass-to-light ratio must 
increase (decrease) with decreasing (increasing) halo mass, and 
proposed a parameterisation for the variation of the average total 
mass-to-light ratio with halo mass,
\noindent
\begin{equation}
  \label{eq:mtol}
	{\frac{\left<M\right>}{\left<L\right>}(M) = \frac{1}{2} 
	  \left(\frac{M}{L}\right)_0 
	  \left[ \left(\frac{M}{M_1}\right)^{-\gamma_1} +
	    \left(\frac{M}{M_1}\right)^{\gamma_2}\right] .}
\end{equation}

\noindent
Here the free parameters correspond to $M_1$, the characteristic mass
for which the mass-to-light ratio in $b_J$ is equal to $(M/L)_0$, and 
$\gamma_1$ and $\gamma_2$ which determine the behaviour at the low- and 
high-mass ends of the of the mass function respectively. We follow
\citet{yang-etal04} (\citetalias{yang-etal04}) in adopting 
$M_1$=$10^{10.94} h^{-1} {\rm M_{\odot}}$, 
$(M/L)_0$=$124\,h\,{\rm (M/L)_{\odot}}$ in $b_J$, $\gamma_1$=2.02, 
and $\gamma_2$=0.30.

The characteristic luminosity $\tilde{L}^{\ast}$ is parameterised in a
similar manner;

\begin{equation}
  \label{eq:lstar}
	{\frac{M}{\tilde{L}^{\ast}} = \frac{1}{2} \left(\frac{M}{L}\right)_0 f(\tilde{\alpha}) \left[ \left(\frac{M}{M_1}\right)^{-\gamma_1} + \left(\frac{M}{M_2}\right)^{\gamma_3}\right],}
\end{equation}
\noindent
where $M_2$ is a characteristic mass and $\gamma_3$ determines the behaviour 
at the high-mass end of the mass function; $\tilde{\alpha}$ follows

\begin{equation}
  \label{eq:alpha}
  {\tilde{\alpha} = \alpha_{15} + \eta \log (M_{15}),}
\end{equation}
\noindent
where $M_{15}$ is the mass of the halo in units of $10^{15} h^{-1}\,M_{\odot}$
and $\alpha_{15}$ and $\eta$ are free parameters. We follow 
\citetalias{yang-etal04} in adopting $M_2$=$10^{12.04} h^{-1} {\rm
  M_{\odot}}$, $\gamma_3$=0.72, $\eta$=-0.22, and $\alpha_{15}$=-1.1.\\

Expression~\ref{eq:mtol} and ~\ref{eq:lstar} allow an expression for
$\left<L\right>/\left<M\right>$ to be derived, from which 
$\tilde{\Phi}^{\ast}$ is deduced;

\begin{equation}
  \label{eq:mtol_alt}
	{\frac{\left<L\right>}{\left<M\right>}(M) = \int_0^\infty \Phi(L|M) \frac{L}{M}
	  dL = \Phi^{\ast} \frac{\tilde{L}^{\ast}}{M} \Gamma(\tilde{\alpha+2})}
\end{equation}
\noindent leads to

\begin{equation}
  \label{eq:phistar}
	{\tilde{\Phi}^{\ast}(M) = \frac{1}{\Gamma(\tilde{\alpha}+1,1)} \frac{\left[ \left(M/M_1\right)^{-\gamma_1} + \left(M/M_2\right)^{\gamma_3}\right]}{\left[ \left(M/M_1\right)^{-\gamma_1} + \left(M/M_1\right)^{\gamma_2}\right]} }
\end{equation}

\noindent Here $\Gamma(x)$ and $\Gamma(x,a)$ are the Gamma and
Incomplete Gamma functions respectively; formally these are expressed
as
\begin{equation}
  \label{eq:gamma}
  \Gamma(x)=\int^\infty_0 t^{x-1} \exp(-t) dt
\end{equation}
\noindent and 
\begin{equation}
  \label{eq:incgamma}
  \Gamma(x,a)=\int^\infty_a t^{x-1} \exp(-t) dt
\end{equation}

\noindent We note that \citetalias{yang-etal04} denote the average 
mass-to-light ratio by $\left<M/L\right>$ (see their equation 2). We
prefer $\left<M\right>/\left<L\right>$ because the meaning is clear -- 
the average luminosity associated with a halo of mass $M$ is
$\left<L\right>$ and so the average mass-to-light ratio is 
$\left<M\right>/\left<L\right>$. If we adopt $\left<M/L\right>$, 
this means that
\begin{equation}
  \label{eq:mtol_wrong}
	{\left<\frac{M}{L}\right>(M) = \int_0^\infty
	  \Phi(L|M) \frac{M}{L} dL = 
	  \Phi^{\ast} \frac{M}{\tilde{L}^{\ast}} \Gamma(\tilde{\alpha}).}
\end{equation}
This produces an expression for $\tilde{\Phi}^{\ast}$ that is
quite different from equation~\ref{eq:phistar}, and which does not
recover the correct behaviour of quantities such as $\left<N\right>(M)$.\\

Having deduced the form of $\tilde{\Phi}^{\ast}$, we can compute the
``conditional luminosity function'' $\Phi(L|M)$,
\[
  { \Phi(L|M)dL =
    \frac{\tilde{\Phi}^{\ast}}{\tilde{L}^{\ast}}\left(\frac{L}{\tilde{L}^{\ast}}\right)^{\tilde{\alpha}}
    \exp(-L/\tilde{L}^{\ast})dL .}
  \]

\begin{figure*} 
  \includegraphics[scale=0.4]{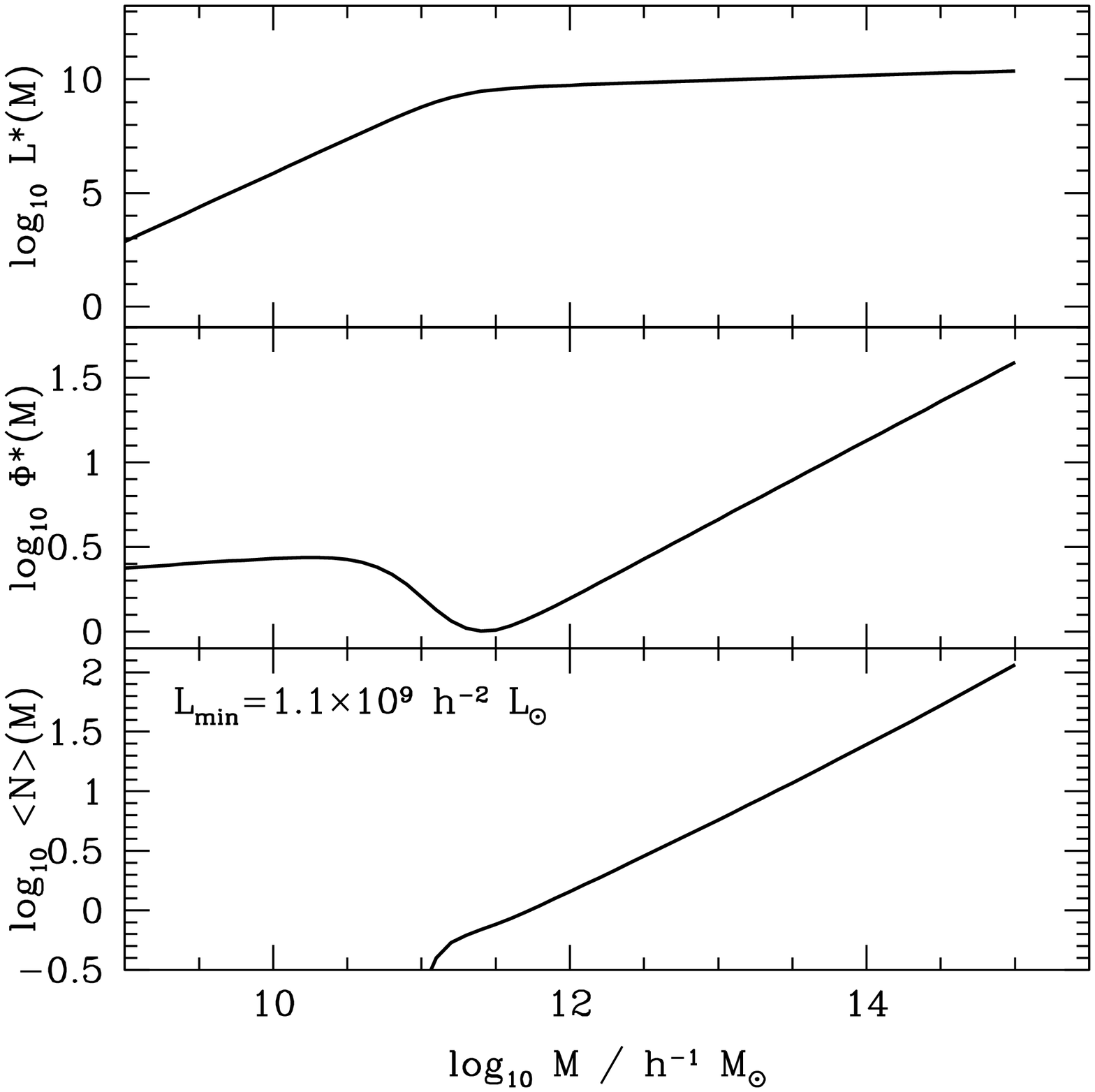}\includegraphics[scale=0.4]{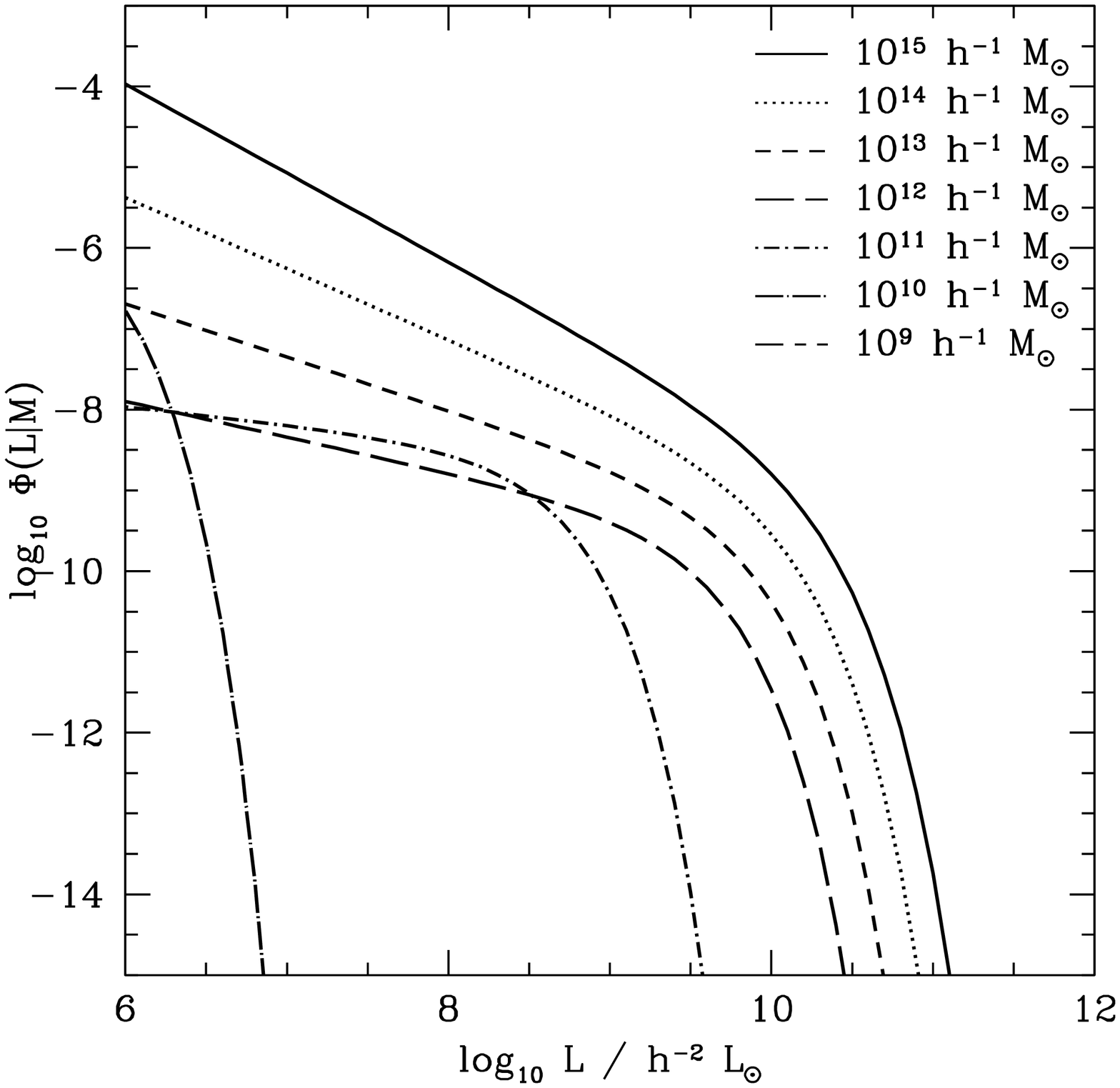}
  \caption{%
    \textit{(LEFT)} Variation of the conditional luminosity function parameters
    $\tilde{L}^{\ast}$ (upper panel) and $\tilde{\Phi}^{\ast}$ (middle
    panel), and the halo occupation number $\left<N\right>(M)$ (bottom
    panel) with halo mass. We adopt the \citetalias{yang-etal04} choice
    of 2dFGRS parameters and a minimum luminosity $L_{\rm min}=1.1
    \times 10^9 h^{-2} \,{\rm L_{\odot}}$.
    \textit{(RIGHT)} Variation of the conditional luminosity function with
    luminosity at fixed halo masses, for the \citetalias{yang-etal04} 
    parameters.
  \label{fig:clf_vars}}
\end{figure*}

The upper left-hand panels of Figure~\ref{fig:clf_vars} show how
$\tilde{L}^{\ast}$ and $\tilde{\Phi}^{\ast}$ vary with halo
mass, while the right-hand panel shows the variation of $\Phi(L|M)$
with luminosity at a fixed halo mass for the \citetalias{yang-etal04} 
choice of 2dFGRS parameters.

\section{Populating Dark Matter Haloes with Galaxies : Details}
\label{appendix:mocks}
We perform a suite of cosmological $N$-body simulations and
constructed catalogues of dark matter haloes at $z$=0. Dark matter 
haloes are identified using a \emph{friends-of-friends} (FOF)
algorithm with a linking length of $b=0.2$ times the mean interparticle
separation. For each of the groups identified in this way we compute
the virial mass $M_{180}$, defined as the mass of the spherical
overdensity that is $180$ times the critical density of the Universe
at $z$=0. In the following discussion, we define a halo's mass $M$ to
be its virial mass $M_{180}$ rather than $M_{\rm FOF}$,
the mass of the FOF group; this is required by the \citetalias{yang-etal04}
prescription. \\

The minimum halo mass $M_{\rm min}$ that is ``reliably''
recovered in each of the simulations governs the minimum luminosity, 
$L_{\rm min}$, that is used in constructing the mock catalogues.
$L_{\rm min}$ defines the threshold luminosity fainter than which there
are no galaxies. $M_{\rm min}$ is the halo mass above which we expect 
the mass function to be unaffected by finite numerical resolution;
below this threshold the number density of haloes tends to be suppressed
relative to the number density they would have in the limit of infinite
numerical resolution. Previous studies have examined how the mass
function is affected by finite mass and force resolution, time-stepping 
accuracy and starting redshift, as well as the influence of the 
group-finding algorithm used to identify dark matter haloes
\citep[e.g.][]{jenkins-etal01,lukic-etal07}. \citet{jenkins-etal01}
performed careful convergence tests and found that mass functions 
constructed from FOF groups are adversely affected by numerical effects
below a halo mass equivalent to $\sim 20$ particles. In this work we 
adopt a more conservative lower mass limit of 50 particles, to ensure
that the mass function of haloes in higher density regions is 
converged \citep[see Chapter 3 of][]{power-phd}; this gives 
$M_{\rm min}$=50$\,m_{\rm part}$ in Table~\ref{table:simulations}.

Having determined $M_{\rm min}$, we estimate $L_{\rm min}$ using the 
``conditional probability distribution'' $P(M|L)$ (see right panel 
of Figure 1, \citetalias{yang-etal04}). $L_{\rm min}$ is a critical 
parameter because it fixes the halo occupation number, the average 
number of galaxies per halo of mass $M$; 
\begin{equation}
  \label{eq:nav}
	{\left<N\right>(M) = \tilde{\Phi}^{\ast} \Gamma(\tilde{\alpha}+1,L_{\rm min}/\tilde{L}^{\ast}).}
\end{equation}

\noindent The variation of $\left<N\right>(M)$ with halo mass for the
\citetalias{yang-etal04} 2dFGRS parameters and $L_{\rm min}=1.1
\times 10^9 h^{-2} \,{\rm L_{\odot}}$ is shown in the bottom left-hand
panel of Figure~\ref{fig:clf_vars}. This $L_{\rm min}$ is appropriate
for the $M_{\rm min}$ in the mock catalogues A to E. 

Note the important role played by the ratio $L_{\rm min}/\tilde{L}^{\ast}$
in equation~\ref{eq:nav}, which controls the number of galaxies
per halo. At fixed $L_{\rm min}$ it increases dramatically as halo mass 
decreases, leading to low mass haloes containing one ``central'' galaxy 
at most, and as $L_{\rm min}$ decreases, the number of galaxies per
halo increases. The number of galaxies per halo of mass $M$ is Poisson 
distributed with a mean of $\left<N\right>(M)$.

We note also that \citet{yang-etal03} introduced a ``hard'' lower mass 
cut-off of $M_{\rm min}=10^9 h^{-1} \rm M_{\odot}$ below which haloes 
cannot host galaxies -- galaxy formation is suppressed in these haloes 
following cosmological reionisation.\\

Having determined the number of galaxies hosted by a halo, we must
assign luminosities. We follow \citetalias{yang-etal04} and give special status
to the central galaxy by assuming that it is the brightest in the halo,
with an average luminosity
\begin{equation}
  \label{eq:lc}
	{\left<L_c\right> = \tilde{\Phi}^{\ast} \tilde{L}^{\ast} \Gamma(\tilde{\alpha}+2,L_1/\tilde{L}^{\ast}).}
\end{equation}
\noindent
The luminosity $L_1$ is a function of halo mass and is chosen such that 
\begin{equation}
  \label{eq:l1}
  {\tilde{\Phi}^{\ast} \Gamma(\tilde{\alpha}+1,L_1/\tilde{L}^{\ast})=1;}
\end{equation}
\noindent
when choosing the central galaxy luminosity, we assume that $L_c$ is a random
variable drawn from $\Phi(L|M)$ for the range of luminosities $L>L_1$. The 
remaining $N-1$ galaxies within the halo are assigned luminosities in the 
range $L_{\rm min} < L < L_1$, drawn at random from the luminosity function
(the ``intermediate'' approach of \citetalias{yang-etal04}).\\

The penultimate step involves assigning morphological types to each
mock galaxy; this is done by defining a function $f_{\rm late}(L,M)$
that specifies the fraction of galaxies with luminosity $L$ in haloes 
of mass $M$ that are late-type. This function can be expressed as the 
product of functions, 
\begin{equation}
  \label{eq:late}
	{ f_{\rm late}(L,M) = g(L) h(M) q(L,M),}
\end{equation}
where
\begin{equation}
  \label{qlm}
  q(L,M) = \left\{
  \begin{array}{lll}
    1                      & \mbox{if $g(L) \, h(M) \leq 1$} \\
    {1 \over g(L) \, h(M)} & \mbox{if $g(L) \, h(M) > 1$}
  \end{array} \right.,
\end{equation}

\begin{equation}
  \label{gl}
  g(L) = {\hat{\Phi}_{\rm late}(L) \over \hat{\Phi}(L)}
  {\int_{0}^{\infty} \Phi(L \vert M) \, n(M) \, {\rm d}M \over
    \int_{0}^{\infty} \Phi(L \vert M) \, h(M) \, n(M) \, {\rm d}M},
\end{equation}
and
\begin{equation}
\label{hm}
h(M) = \max \left( 0, \min\left[ 1, \left({{\rm log}(M/M_a)
\over {\rm log}(M_b/M_a)} \right) \right] \right).
\end{equation}

\noindent
Here $n(M)$ is the halo mass function \citep{st99};
$\hat{\Phi}_{\rm late}(L)$ and $\hat{\Phi}(L)$ correspond to the
{\it observed} luminosity functions of the late-type and entire galaxy  
samples respectively; and $M_a$ and $M_b$ are free parameters defined as
the masses at which $h(M)$ takes on the values $0$ and $1$ respectively.
\citet{vdb-etal03} demonstrated that this parameterisation allowed the 
galaxy population to be split into early- and late-types such that the
respective luminosity functions and clustering properties could be recovered.
We follow \citetalias{yang-etal04} in adopting 
$M_{\rm a}=10^{17.26} h^{-1} {\rm M_{\odot}}$ and 
$M_{\rm b}=10^{10.86} h^{-1} {\rm M_{\odot}}$.
Formally we assign morphological type by drawing a random number $R$
that is uniformly distributed between $[0,1]$ and comparing it to 
$f_{\rm late}(L,M)$. If $R < f_{\rm late}(L,M)$, the galaxy is designated 
late-type, otherwise it is early-type.\\

The final step involves assigning phase space coordinates
(i.e. positions and velocities) to each of the $N$ galaxies within the
halo. The brightest central galaxy is associated with the most bound
particle of the halo and is assigned its position and velocity.
The remaining $N$-1 galaxies can be treated in a variety of ways.
For the purposes of this study, in which our main concern is testing
the reliability of our selection criteria, we follow \citetalias{yang-etal04}
in randomly sampling dark matter particles from the FOF
group (their ``FOF approach''). More sophisticated approaches, in which
we explicitly track the merging history of individual haloes, will be
essential for future work, especially with regards to kinematics (Power
et al., in preparation).

\section{Galaxy Classification}\label{appendix:classification}

Our primary classification method is that of
\citet{bailinharris08-classification}, which has been validated
using high-quality imaging from the Millennium Galaxy
Catalogue \citep{mgc-imaging}.
However, given the qualitative difference between our results
for the different subpopulations, it is important
to confirm that the difference seen is not
an artefact of the galaxy classification scheme.
We examine how the anisotropy of the satellite
distribution varies using other methods:
\begin{enumerate}
  \item Inclination-corrected location on the colour-magnitude diagram (\cmdfo),
    which is strongly bimodal \citep{bailinharris08-classification}.
    Galaxies are considered ``early'' if they are redder than
    $\cmdfo = -0.05$,
    and ``late'' if they are bluer.
  \item Spectroscopic Principal Component Analysis (PCA) eClass parameter.
    Galaxies are considered ``early'' if they have $\mathrm{eClass} < -0.07$,
    otherwise they are considered ``late''.
  \item The inclination-corrected global concentration of the light
    profile, \cnorm. The distribution of galaxy concentrations is
    trimodal \citep{bailinharris08-trimodality}.
    We label the ``Elliptical'' (high-\cnorm\ and high-$b/a$) region
    from \citet{bailinharris08-trimodality}
    as ``early'', their ``Disk''
    (low-\cnorm) region as ``late'', and all other galaxies
    as ``intermediate''.
\end{enumerate}
It should be noted that these measurements are completely independent:
the CMD location is based on global photometry, the PCA analysis is based
on spectroscopy, and the concentration is based
on the distribution of the light profile.

\begin{figure*}
\scalebox{0.6}[0.65]{\includegraphics{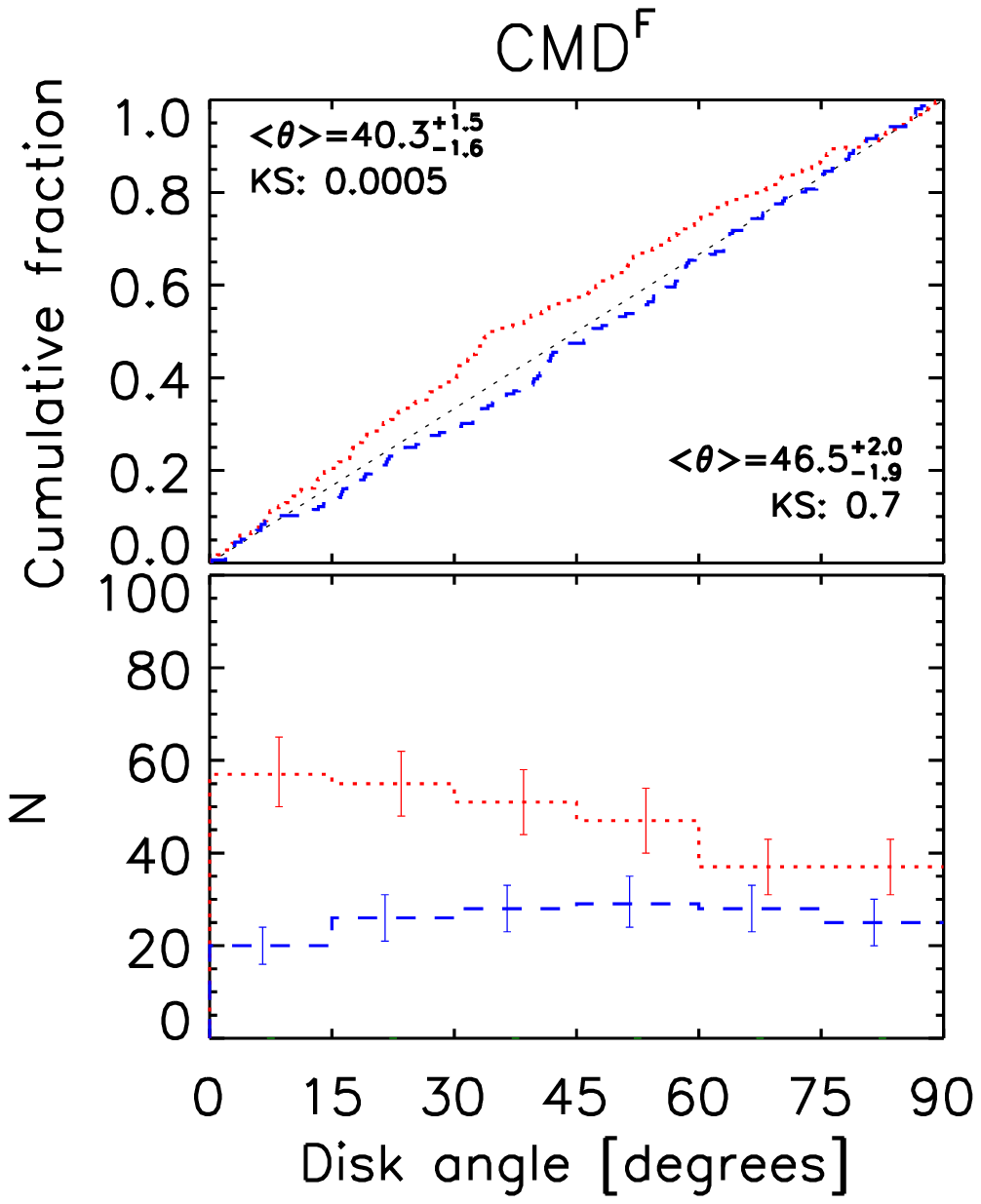}}%
\scalebox{0.6}[0.65]{\includegraphics{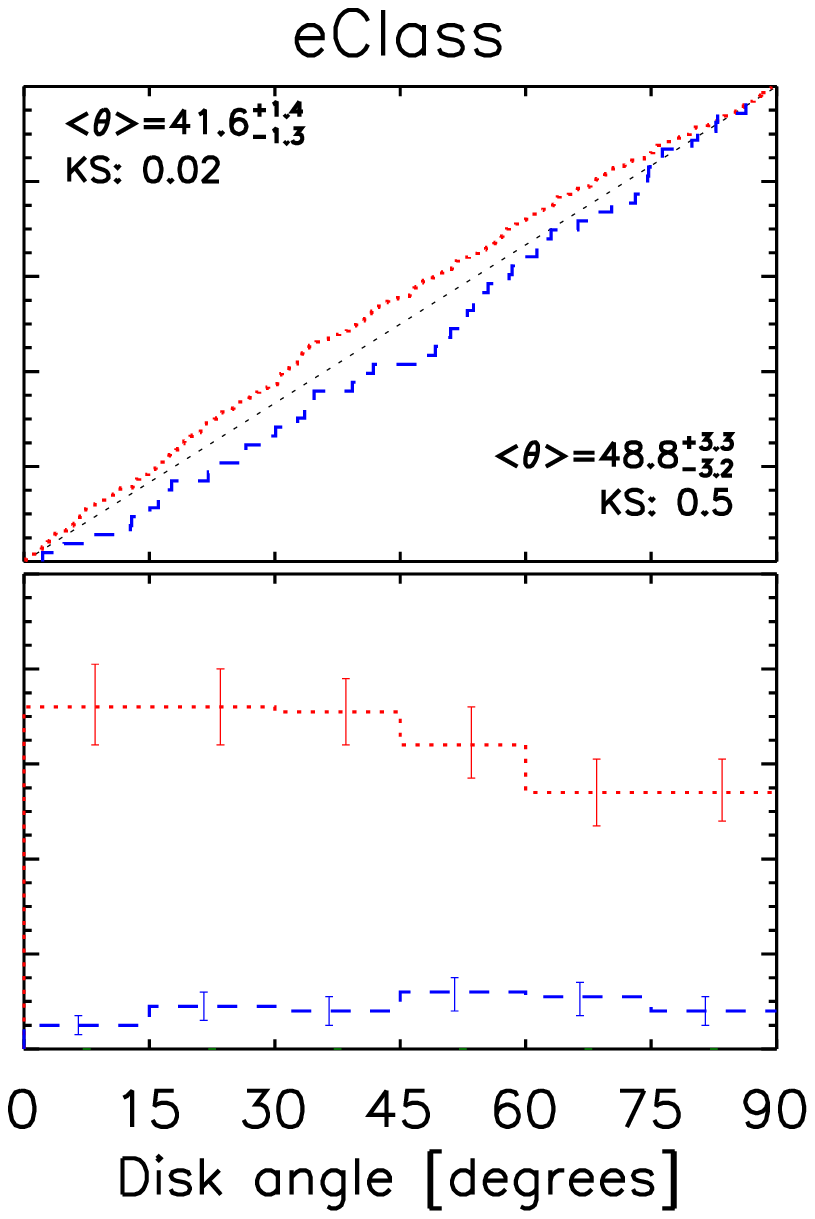}}%
\scalebox{0.6}[0.65]{\includegraphics{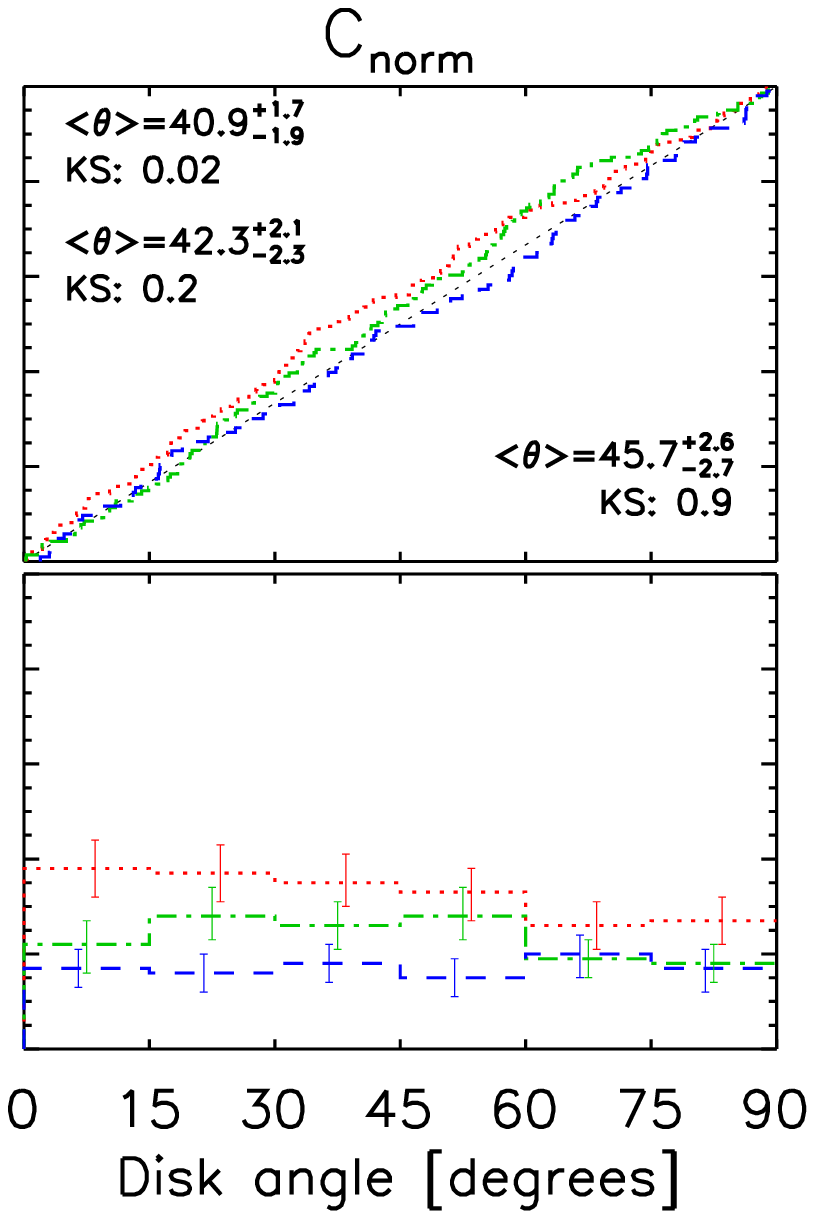}}%
\caption{Cumulative \textit{(top panels)} and
differential \textit{(bottom panels)}
distributions of satellite disc angles of
primary galaxies classified using the
following schemes (left to right): \cmdfo, spectroscopic eClass
parameter, and inclination-corrected Petrosian concentration parameter \cnorm.
Colours/line styles are as in
Figure~\ref{disc angle distribution figure}.
Mean disc angles and KS test probabilities that the samples
are drawn from a uniform distribution are given in the top-left
(bottom-right) corners of the cumulative plots for the early-type
(late-type) samples. The statistics for the intermediate-type samples
are given below the early-type statistics for \cnorm.%
\label{disc angle galaxy classification plot}}
\end{figure*}

The results using these alternative classification schemes are shown in
Figure~\ref{disc angle galaxy classification plot}.
The mean disc angle and KS test probability that each sample
is drawn from an isotropic distribution are also given.
The major-axis distribution around early-type galaxies
and a distribution consistent with isotropy around late-type galaxies is seen
using every method.
Intermediate-type galaxies are red with intermediate concentrations;
the galaxies with red \cmdfo\ and with intermediate \cnorm\ show the
same major-axis distribution as around the intermediate-type
galaxies of \citet{bailinharris08-classification}.
Therefore, although the magnitude of the measured anisotropy
varies at a $\sim 1\sigma$ level,
the detected anisotropy cannot be simply a galaxy classification
artefact: satellites of early-type and late-type galaxies have
different angular distributions.

We note that the classification of some of our galaxies is uncertain.
Because the galaxies that constitute our primary sample
are typically more luminous and more isolated than typical SDSS
spectroscopic galaxies, they provide a biased sample of parameter space.
In particular, several of our primary galaxies have
$\cnorm < 1$ (i.e. they have low concentrations) but have
$\cmdfo > -0.05$ (i.e. they are red): $4.3\%$ of all primaries and
$12.0\%$ of primaries classified as Late-type fall into this region of
parameter space, compared to just $1.8\%$ of the visually-classified
galaxies in \citet{bailinharris08-classification} and $3.6\%$ of those
classified as Late. Given that the anisotropy of the satellite
distribution differs between galaxy classes, and shows the
strongest difference between the Late and Intermediate types,
examining the anisotropy around the galaxies in this region of
parameter space can provide insight into their nature.

If we separate our Late types into red and blue sub-classes
(divided at $\cmdfo=-0.05$, as above), we find that the mean
disc angle around the blue subclass is $46.7\pm2.2\degr$,
consistent with isotropy and with the results from the full Late-type
sample. However, the mean disc angle around the red subclass
is $37\fdg3^{+5\fdg0}_{-5\fdg5}$,
exhibiting major-axis
alignment consistent with the results from the Intermediate-type
sample. This suggests that the red low-concentration galaxies
more properly belong in the Intermediate classification.

\section{Angular Distribution of Satellites in the Mock Catalogues}%
\label{appendix:mockdist}

\begin{figure*}
\scalebox{0.4}[0.5]{\includegraphics{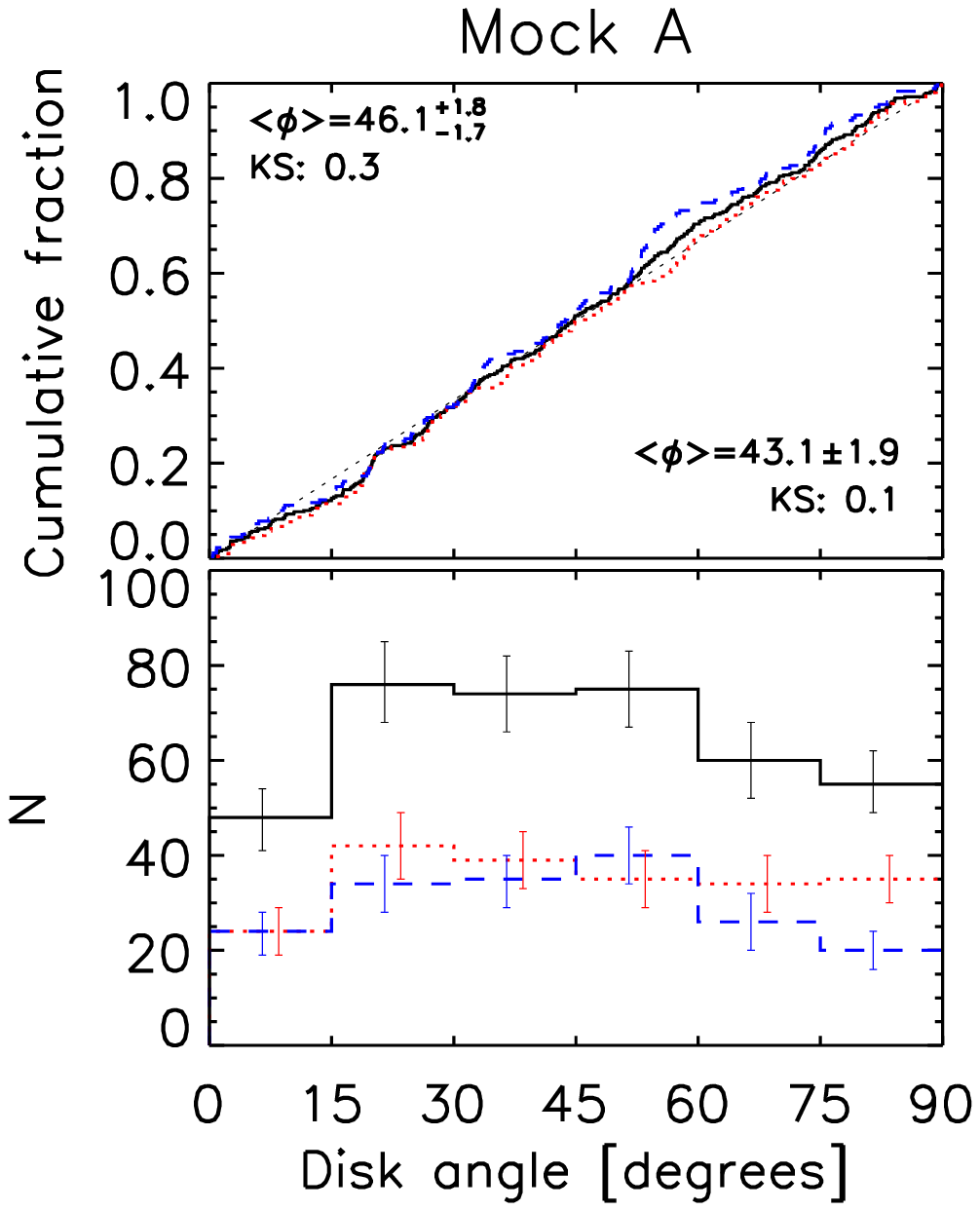}}%
\scalebox{0.4}[0.5]{\includegraphics{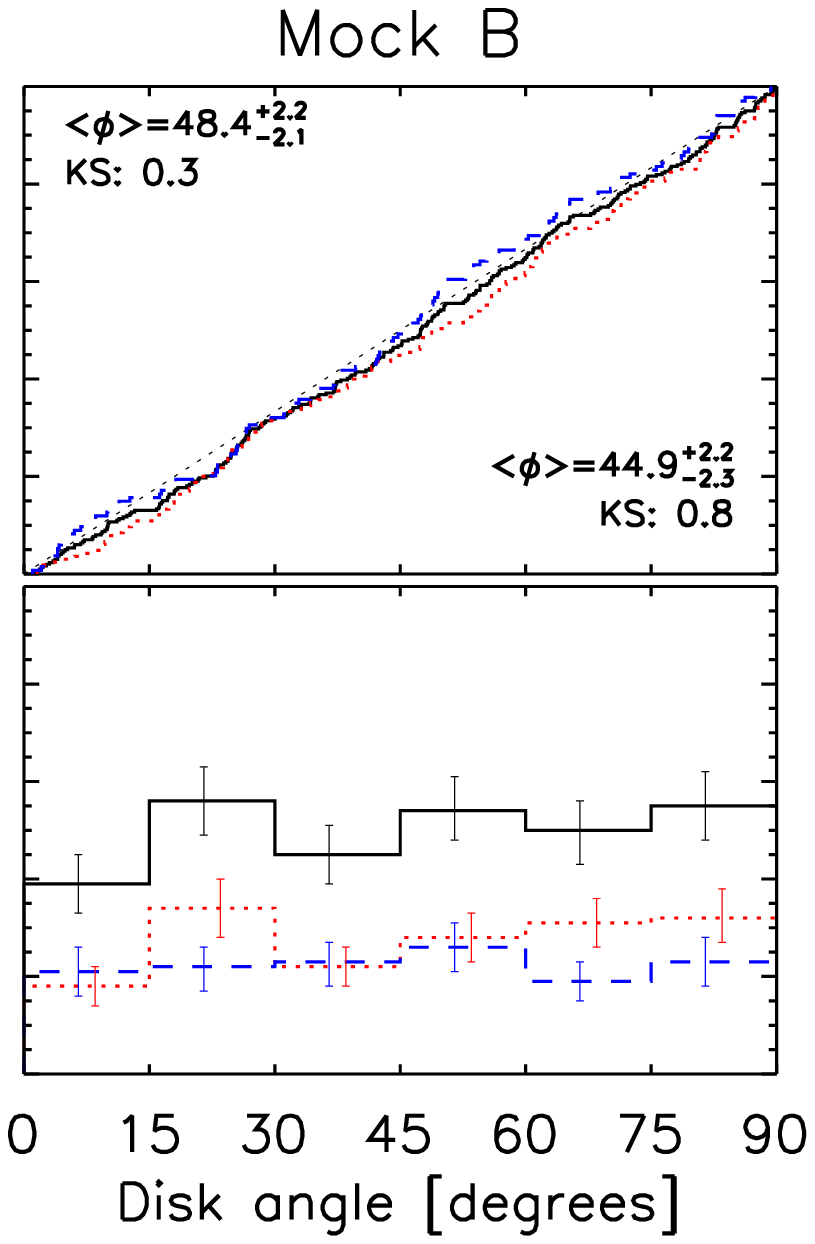}}%
\scalebox{0.4}[0.5]{\includegraphics{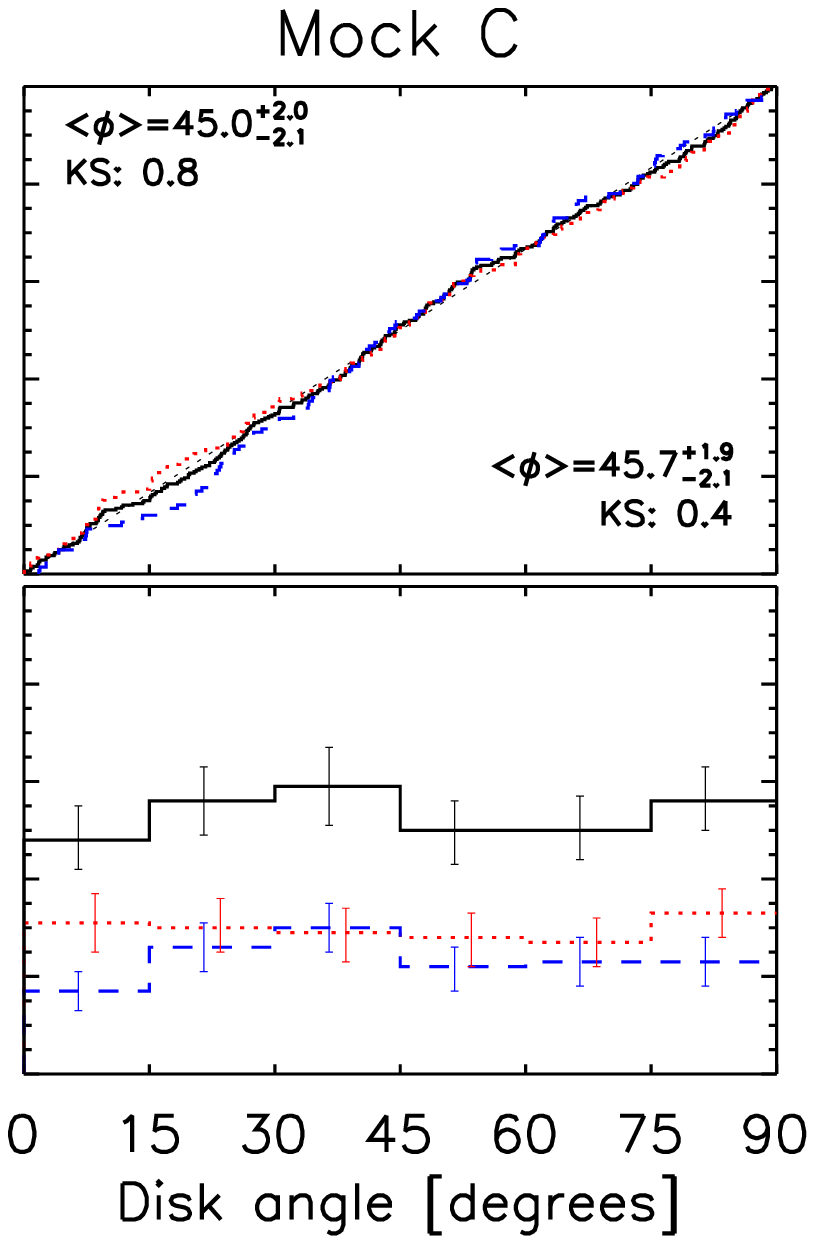}}%
\scalebox{0.4}[0.5]{\includegraphics{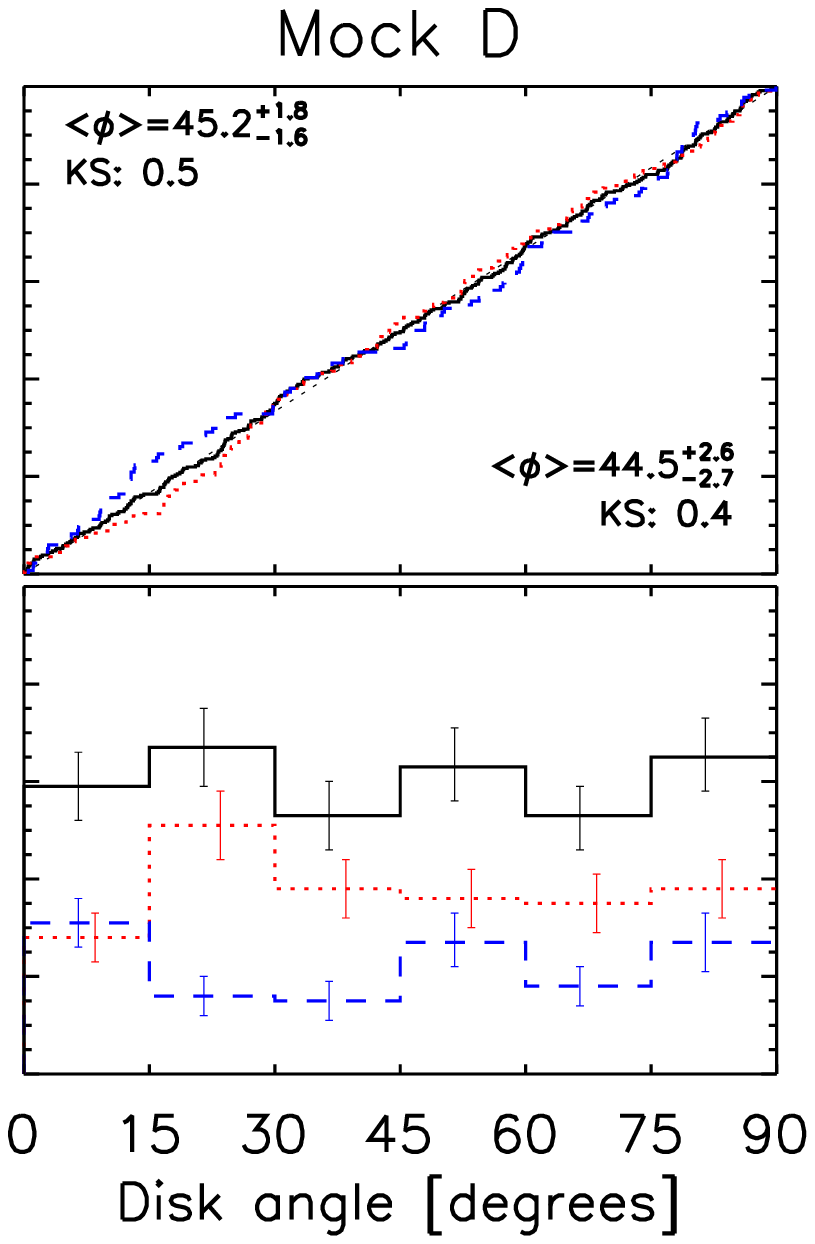}}%
\scalebox{0.4}[0.5]{\includegraphics{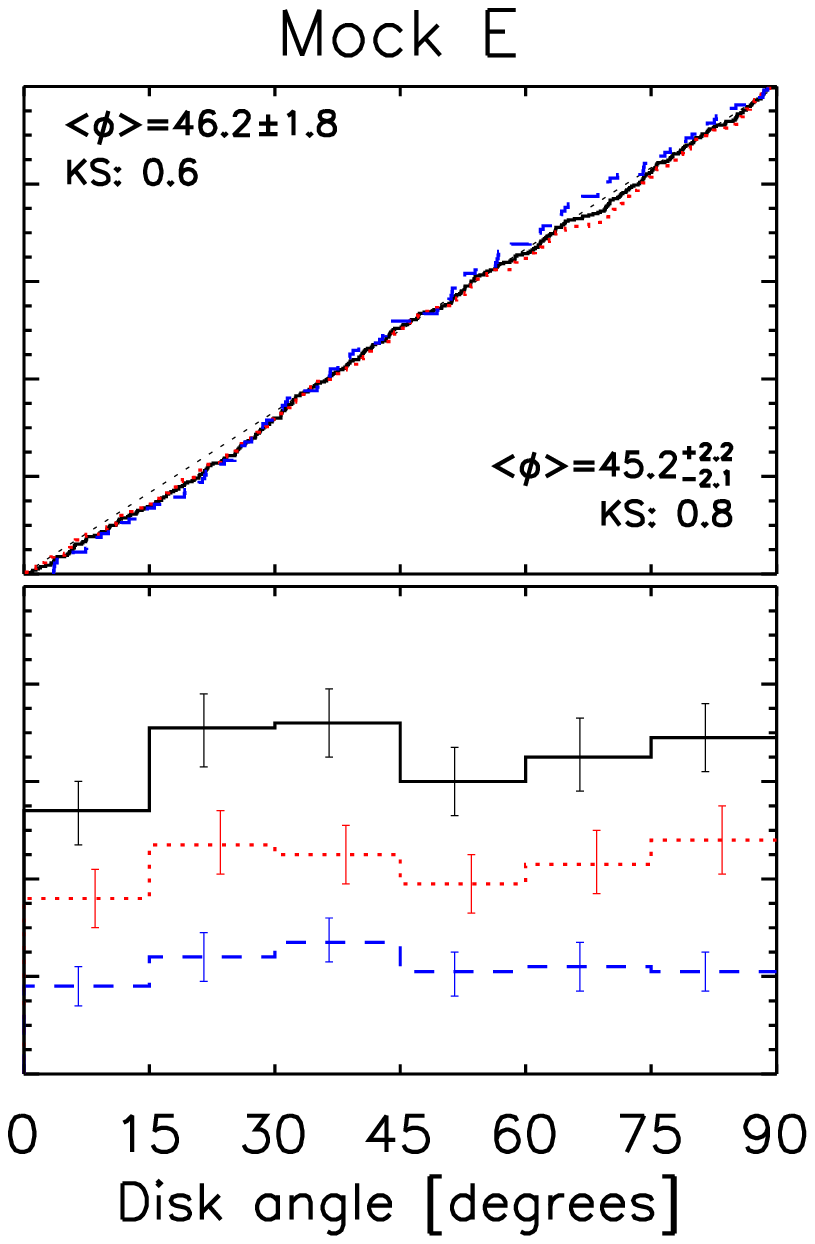}}
\caption{Cumulative \textit{(top panels)} and
differential \textit{(bottom panels)}
distributions of satellite disc angles
in the five mock catalogues.
Colours/line styles are as in
Figure~\ref{disc angle distribution figure}.
Mean disc angles and KS test probabilities that the samples
are drawn from a uniform distribution are given in the top-left
(bottom-right) corners of the cumulative plots for the early-type
(late-type) samples.%
\label{disc angle mock plot}}
\end{figure*}

To confirm that our measurement of an anisotropic distribution of
satellite galaxies is due to an intrinsic anisotropy rather than
an artefact of our method, we have performed an identical analysis
on the mock catalogues, whose satellite distributions are,
by construction, isotropic.
We use mock catalogues generated from the five independent
$100~h^{-1}~\mathrm{Mpc}$ simulations
in order to account for cosmic variance.
The distributions are plotted in
Figure~\ref{disc angle mock plot} and the statistical measures
of anisotropy are listed in Table~\ref{mock anisotropy stats table}.

The level of anisotropy that we measure from isotropically-distributed
satellites in the mock catalogues is small.
Even when the isotropic KS test probabilities in the mock samples are
low, the deviations are not systematic: the mean disc angle
almost always deviates from $45\degr$ by less than $2\degr$
the polar fraction never deviates from $0.5$ by more than
$4\%$.
These are much smaller than the anisotropies that we detect
around early-type galaxies in the
observational sample, confirming that our detection of anisotropy
cannot be explained by intrinsically isotropically distributed satellites.

\begin{table*}
\caption{\label{mock anisotropy stats table}%
Anisotropy of Satellites in Mock Catalogues}
\begin{tabular}{lccccc}
\hline
{Parameter} & {Mock A} & {Mock B} &
  {Mock C} & {Mock D} & {Mock E}\\
\hline
\multicolumn{4}{l}{Full sample:}\\
\hspace{1em}KS probability & $0.28$ & $0.41$ &
  $0.95$ & $0.94$ & $0.65$\\
\hspace{1em}Mean disc angle [\degr] & $44.7^{+1.4}_{-1.2}$ & $46.8\pm1.5$ &
  $45.3\pm1.5$ & $44.9^{+1.5}_{-1.4}$ & $45.9^{+1.5}_{-1.4}$\\
\hspace{1em}Median disc angle [\degr] & $44.7^{+2.1}_{-2.0}$ & $47.4^{+1.7}_{-2.2}$ &
  $44.4^{+2.9}_{-1.7}$ & $45.4^{+2.5}_{-2.7}$ & $44.1^{+3.0}_{-1.6}$\\
\hspace{1em}Polar fraction & $0.49\pm0.03$ & $0.53\pm0.03$ &
  $0.49\pm0.03$ & $0.50\pm0.03$ & $0.50\pm0.03$\\
\multicolumn{4}{l}{Early-type primaries:}\\
\hspace{1em}KS probability & $0.30$ & $0.33$ &
  $0.76$ & $0.46$ & $0.64$\\
\hspace{1em}Mean disc angle [\degr] & $46.1^{+2.0}_{-1.7}$ & $48.4\pm2.2$ &
  $45.0^{+2.2}_{-1.9}$ & $45.2\pm1.7$ & $46.2^{+1.8}_{-1.7}$\\
\hspace{1em}Median disc angle [\degr] & $44.9^{+4.2}_{-1.9}$ & $48.8^{+5.8}_{-1.4}$ &
  $44.8^{+3.2}_{-2.1}$ & $44.0^{+3.4}_{-1.8}$ & $45.2^{+1.9}_{-3.5}$\\
\hspace{1em}Polar fraction & $0.50^{+0.03}_{-0.04}$ & $0.55\pm0.04$ &
  $0.49\pm0.04$ & $0.48\pm0.03$ & $0.50\pm0.03$\\
\multicolumn{4}{l}{Late-type primaries:}\\
\hspace{1em}KS probability & $0.10$ & $0.83$ &
  $0.45$ & $0.40$ & $0.80$\\
\hspace{1em}Mean disc angle [\degr] & $43.1^{+2.0}_{-1.8}$ & $44.9^{+2.3}_{-2.2}$ &
  $45.7^{+2.0}_{-2.2}$ & $44.5^{+2.6}_{-2.4}$ & $45.2\pm2.2$\\
\hspace{1em}Median disc angle [\degr] & $43.7^{+2.4}_{-2.7}$ & $45.2^{+2.9}_{-2.2}$ &
  $43.7^{+4.1}_{-3.2}$ & $47.9^{+5.1}_{-9.4}$ & $43.4^{+6.6}_{-4.0}$\\
\hspace{1em}Polar fraction & $0.48\pm0.04$ & $0.51\pm0.04$ &
  $0.48^{+0.5}_{-0.4}$ & $0.54\pm0.05$ & $0.48\pm0.05$\\
\hline
\end{tabular}
\end{table*}

\section{Determination of the Large Scale Structure Axis}
\label{appendix:lss pa}

The physical environment of a galaxy is best described by the region
in which the presence of the galaxy predicts the presence of other
matter;
the radial extent of this region
is characterised by the correlation length $r_0$.
For the global $\sim L^*$ galaxy population,
$r_0 \sim 4$--$6~\hiMpc$ \citep{norberg-etal02}.
However, the isolated galaxies
that constitute our sample are, by construction, much less clustered
than average; for example, HI-selected galaxies, which are much
less likely than average to have large nearby neighbours, have
a much smaller
$r_0 \sim 3.3~\hiMpc$ \citep{basilakos-etal07,meyer-etal07}.
For our very isolated sample,
$3~\hiMpc$ is a reasonable radius in which to characterise the
large-scale environment of each galaxy.

We therefore determine the PA of the large scale structure around each
primary galaxy by diagonalising
the moment of inertia tensor of the projected positions of all spectroscopic
galaxies with projected radii of between $1000$ and $3000\,\hikpc$ (thereby
explicitly ensuring that there is no overlap between the galaxies used 
to determine the orientation of the LSS and those used to evaluate the 
isolatedness of the primary or the satellites themselves)
and with velocities that differ from that of the primary by no more than 
$400\,\kms$ \citep[this is larger than the $300\,\kms$ Hubble flow component
 in order to account for the peculiar velocities of galaxies, which have a
dispersion of $85~\kms$ in the Local Volume;][]{karachentsev-etal03}.

\begin{figure}
\includegraphics[scale=0.45]{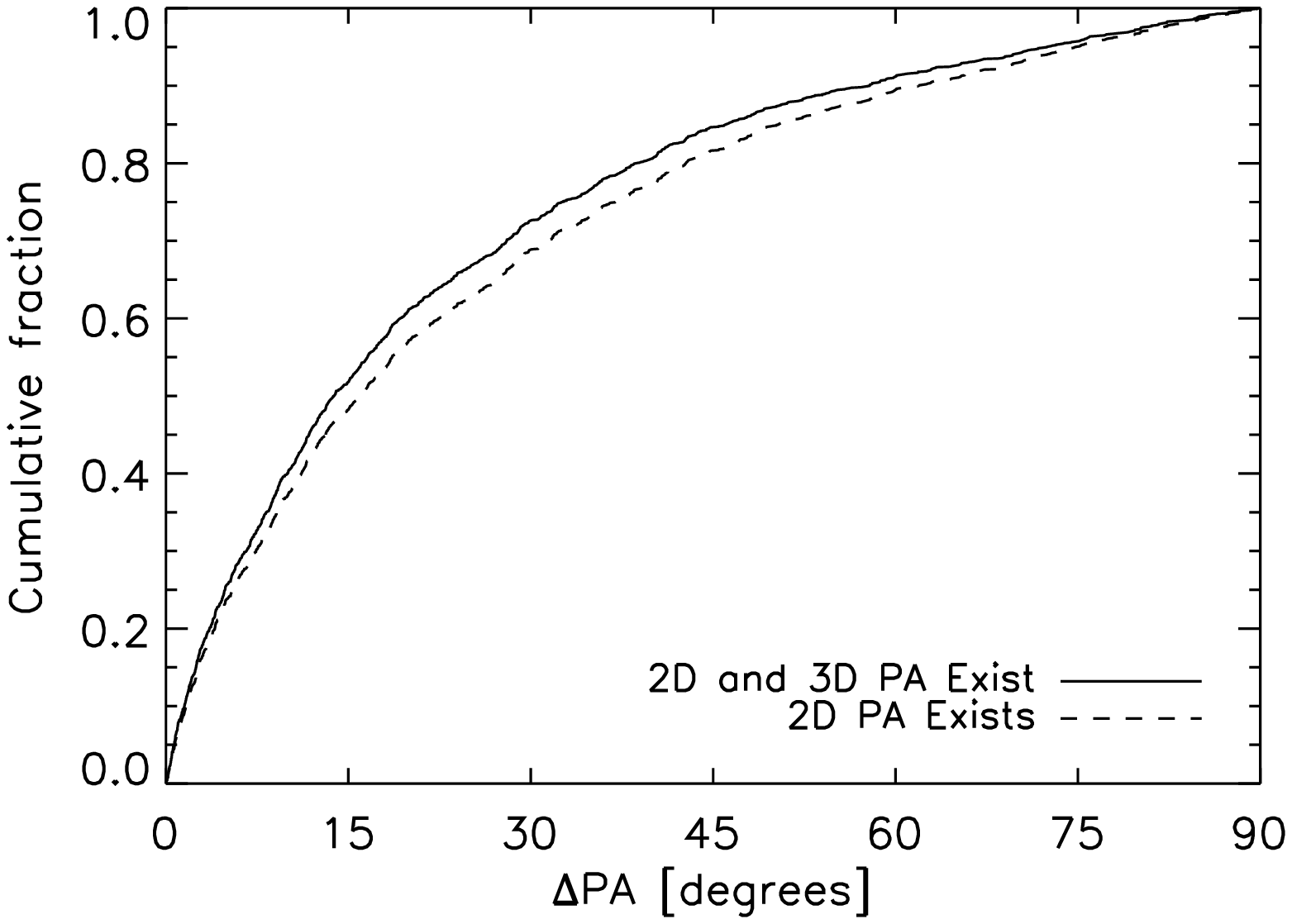}
\caption{\label{lss pa verification figure}%
Cumulative distribution of the difference between the PA of the LSS
measured around isolated mock galaxies using the ``observed'' 2D galaxy
positions and redshifts versus that measured using the
known three-dimensional positions of surrounding haloes.
The solid line indicates galaxies for which both the 2D and 3D
PA is well-defined, while the dashed line also includes haloes
for which only the 2D PA is well-defined.}
\end{figure}

We have used the mock catalogues to confirm that this procedure
reliably recovers the three-dimensional PA of the LSS surrounding the primary.
We have taken the known three-dimensional positions of all haloes
within a spherical volume of $3000~\hikpc$ around the halo of each primary
galaxy in the mock catalogues,
constructed and diagonalized their inertia tensor, and projected the major
axis onto the plane of the sky. The PA of this axis is then compared to the
two-dimensional PA inferred from the ``observed'' galaxies in the mock
catalogues.
The cumulative distribution of the misalignment between the PA determined
using 3D positions and that inferred
from the 2D observables is plotted in Figure~\ref{lss pa verification figure}.
To improve the statistics, we have included in this plot
all galaxies that match the isolation criteria, regardless of whether
they have satellites;
however, the results are consistent if we restrict
the sample to just those with satellites.
The solid line indicates the relative alignment when both the
3D LSS PA and the inferred 2D are well-defined
(i.e.~that contain galaxies within the defining sphere or cylinder),
and has a median misalignment of $13.8\degr$.
A small fraction of galaxies have well-defined 2D LSS PAs,
and would therefore be included in the observational analysis,
but no well-defined 3D LSS axis because none of the galaxies that lie
within the redshift-space cylinder lie within the 3D sphere.
We account for these cases by assuming that their intrinsic 3D LSS PAs
are isotropically distributed and indicate the alignment
of the full sample including them as
the dashed line in Figure~\ref{lss pa verification figure}.
Half of the LSS PAs are aligned to within $15.8\degr$;
therefore, this procedure successfully recovers the PA of the LSS.

\section{Comparisons using different selection criteria}%
\label{criteriacomp-appendix}

In Figure~\ref{diskangle vs criteria figure}, we have compared the mean
disc angle determined by previous studies of satellite anisotropy
to the values we derive using identical selection criteria.

No previous study has identified intermediate-type galaxies as a
separate class; we therefore adopt a simple binary classification based
on the location of the galaxy on the CMD:
a galaxy is considered Early-type if its \grkc\ colour is redder than
\begin{equation}
  \grkc = 0.70 - 0.0325 (M_r - 5 \log h + 19)
\end{equation}
and Late-type if it is bluer than this threshold.
The results of Appendix~\ref{appendix:classification} indicate
that different methods of classifying galaxies may introduce $\sim 1\sigma$
differences in the measured anisotropy.

We split each sample by the galaxy type of its primary
although \citetalias{zsfw97-holmberg}
and \citetalias{azpk06} only studied late-types, and
\citetalias{brainerd05} did not separate the sample by type.
\citetalias{brainerd05} suggested that her samples were dominated
by systems with late-type primaries; in contrast, 
we find that $46\%$/$58\%$/$37\%$
of the primaries we select using her criteria are classified as
early types according to \citet{bailinharris08-classification},
containing $52\%$/$64\%$/$43\%$
of the satellites for samples 1, 2, and 3 respectively. Using
the location on the CMD, the fraction of early-type primaries is
even higher: $85\%$/$90\%$/$83\%$ containing $89\%$/$94\%$/$86\%$
of the satellites.

\citetalias{sl04} do not quote a mean disc angle; rather, they fit
the distribution of disc angles $\theta$ to the form
\begin{equation}
f(\theta) = A\cos(2\theta) + B
\end{equation}
and quote the values of $A$ and errors $\sigma_A$.
To enable a more direct comparison with other studies, we calculate
the mean disc angle of the associated distribution as
\begin{equation}
\left<\theta\right> = \frac{\pi}{4} - \frac{A}{2}
\end{equation}
with uncertainty
\begin{equation}
\sigma_{\left<\theta\right>} = \frac{\sigma_A}{2},
\end{equation}
in radians.
\citetalias{appz07} also do not quote a mean disc angle. We have
derived the mean and the error of the distributions from their plotted
histograms, assuming that all satellites lie at the central value
of the bin they fall in. The errors are calculated by bootstrap
resampling.
\citetalias{zsfw97-holmberg} do not quote a mean disc angle, but it
can be derived from the data in Table~2 of \citet{zsfw97-moresats}.
The errors are calculated by bootstrap resampling.

The only cases where our results deviate from the previous
results by more than $2\sigma$ are the early-type samples of
\citetalias{appz07}. Even in these cases,
the sense of the observed anisotropy is the same, only the magnitude
is different. These are cases where there is no mean disc
angle quoted by the authors, and therefore we have used indirect
methods to determine the appropriate mean;
they also use different methods to classify early-type galaxies.
We conclude that the
numerical differences in these cases are not significant.

\end{document}